\documentclass{aa}  
\usepackage{multirow}
\usepackage{ulem}
\usepackage{threeparttable}
\usepackage{color}
\usepackage{arydshln}
\usepackage{txfonts}
\usepackage{natbib}
\usepackage{ulem}
\usepackage{subfig}
\bibpunct{(}{)}{;}{a}{}{,}
\def\ltsima{$\; \buildrel < \over \sim \;$}
\def\simlt{\lower.5ex\hbox{\ltsima}}
\def\gtsima{$\; \buildrel > \over \sim \;$}
\def\simgt{\lower.5ex\hbox{\gtsima}}

\begin{document} 

   \title{Testing the universality of the star formation efficiency \\in dense molecular gas}
   \subtitle{}

   \author{
    Y. Shimajiri \inst{1},
    Ph. Andr$\acute{\rm e}$ \inst{1}, 
    J. Braine \inst{2},
    V. K$\ddot{\rm o}$nyves \inst{1},
    N. Schneider \inst{3}, 
    S. Bontemps \inst{2},
    B. Ladjelate \inst{1},
    A. Roy \inst{1},
    Y. Gao \inst{4,5},
 and
    H. Chen \inst{6,7,8}
}

   \institute{$^{1}$Laboratoire AIM, CEA/DSM-CNRS-Universit$\acute{\rm e}$ Paris Diderot, IRFU/Service d'Astrophysique, CEA Saclay, F-91191 Gif-sur-Yvette, France
              \email{Yoshito.Shimajiri@cea.fr} \\
    $^{2}$Laboratoire d'Astrophysique de Bordeaux, Univ. Bordeaux, CNRS, B18N, all\'ee Geoffroy Saint-Hilaire, 33615 Pessac, France\\
    $^{3}$I. Physik. Institut, University of Cologne, Z$\ddot{\rm u}$lpicher Str. 77, 50937 Koeln, German \\
    $^{4}$Purple Mountain Observatory, Chinese Academy of Sciences,
2 West Beijing Road, Nanjing 210008, P. R. China\\
    $^{5}$Key Laboratory of Radio Astronomy, Chinese Academy of
Sciences, Nanjing 210008, P. R. China\\
     $^{6}$ School of Astronomy and Space Science, Nanjing University,
Nanjing 210093, P. R. China\\
     $^{7}$Key Laboratory of Modern Astronomy and Astrophysics,
Nanjing University, Nanjing 210093, P. R. China\\
     $^{8}$Collaborative Innovation Center of Modern Astronomy and
Space Exploration Nanjing, 210093, P.R. China
             }

   \date{Received \today; accepted }

% \abstract{}{}{}{}{} 
% 5 {} token are mandatory
 
  \abstract
  % context heading (optional)
  % {} leave it empty if necessary  
{Recent studies with, e.g., {\it Spitzer} and {\it Herschel} have suggested that star formation in dense molecular gas may be governed by essentially the same ``law'' in Galactic clouds and external galaxies. This conclusion remains controversial, however, in large part because different tracers have been used to probe the mass of dense molecular gas in Galactic and extragalactic studies.
}
  % aims heading (mandatory)
 {
We aimed to calibrate the HCN and HCO$^+$ lines commonly used as dense gas tracers in extragalactic studies and to test the possible universality of the star formation efficiency in dense gas {($\simgt$10$^4$ cm$^{-3}$)}, SFE$_{\rm dense}$.
 }
  % methods heading (mandatory)
{ 
We conducted wide-field mapping of the Aquila, Ophiuchus, and Orion B clouds at $\sim$0.04 pc  resolution in the $J$=1--0 transition of HCN, HCO$^+$, and their isotopomers. For each cloud, we derived a reference estimate of the dense gas mass $M_{\rm Herschel}^{A_{\rm V}>8}$,
as well as the strength of the local far-ultraviolet (FUV) radiation field, using {\it Herschel} Gould Belt survey data products, and estimated the star formation rate from direct counting of the number of  {\it Spitzer} young stellar objects.
}
  % results heading (mandatory)
{ 
The H$^{13}$CO$^+$(1--0) and H$^{13}$CN(1--0) lines were observed to be good tracers of the dense star-forming filaments detected with {\it Herschel}. Comparing the luminosities $L_{\rm HCN}$ and $L_{\rm HCO^+}$ measured in the HCN and HCO$^+$ lines with the reference masses $M_{\rm Herschel}^{A_{\rm V}>8}$, the empirical conversion factors $\alpha_{\rm Herschel-HCN}$ (= $M_{\rm Herschel}^{A_{\rm V}>8}$/$L_{\rm HCN}$) and  $\alpha_{\rm Herschel-HCO^+}$ (= $M_{\rm Herschel}^{A_{\rm V}>8}$/$L_{\rm HCO^+}$) were found to be significantly anti-correlated with the local FUV strength. In agreement with \citet{Pety16}, the HCN and HCO$^+$ lines were also found to trace gas down to $A_{\rm V}$ $\simgt$ 2. As a result, published extragalactic HCN studies must be tracing all of the moderate density gas down to $n_{\rm H_2}  \simlt 10^3\, \rm{cm}^{-3}$. Estimating the contribution of this moderate density gas from the typical column density PDFs in nearby clouds, we obtained the following $G_0$-dependent HCN conversion factor for external galaxies: $\alpha_{\rm Herschel-HCN}^{\rm fit^\prime} = 64\times G_{0}^{-0.34}$. Re-estimating the dense gas masses in external galaxies with $\alpha_{\rm Herschel-HCN}^{\rm fit^\prime}(G_0)$, we found that SFE$_{\rm dense}$ is remarkably constant, with a scatter of less than 1.5 orders of magnitude around $4.5\times10^{-8}\, {\rm yr}^{-1} $, over 8 orders of magnitude in dense gas mass.
}
  % conclusions heading (optional), leave it empty if necessary 
{ 
Our results confirm that SFE$_{\rm dense}$ of galaxies is quasi-universal on a wide range of scales from $\sim 1$--10 pc to $> 10\,$kpc. Based on the tight link between star formation and filamentary structure found in {\it Herschel} studies of nearby clouds, we argue that SFE$_{\rm dense}$ is primarily set by the ``microphysics'' of core/star formation along  filaments.
}

\titlerunning{Relationship between SFR and $M_{\rm dense}$}
\authorrunning{Shimajiri et al.}
\maketitle
%
%________________________________________________________________

%%%%%%%%%%%%%%%%%%%%%%%%%
% Introduction
%%%%%%%%%%%%%%%%%%%%%%%%%
\section{Introduction}\label{sect:Intro}

A close connection between dense molecular gas (with $n_{\rm H_2}  > 10^4\, {\rm cm}^{-3} $) 
and star formation has been established for quite some time on 
both Galactic and extragalactic scales.
On small scales, individual stars of low to intermediate masses are known to form from the collapse of prestellar dense cores \citep[e.g.][]{Myers83,Ward-Thompson94,Andre00}, themselves often embedded in dense cluster-forming gas clumps 
within molecular clouds \citep[e.g.][]{Lada92,Myers98}.
On galaxy-wide scales, the global star formation rate (SFR) is linearly correlated with the total amount of dense molecular gas as traced by HCN observations \citep{Gao04b,Gao04a}, while the correlation between the SFR and the amount of either atomic (HI) or low-density molecular (CO) gas 
is non linear and not as tight \citep[e.g.][]{Kennicutt89}.
Moreover, as pointed out by \citet{Lada12}, essentially the same relation between 
SFR and mass of dense gas 
$M_{\rm dense}$ is found in nearby Galactic clouds 
[${\rm SFR} = 4.6 \times 10^{-8}\,  M_\odot \, {\rm yr}^{-1}\, \times \left(M_{\rm dense}/M_\odot \right) $ -- \citet{Lada10}]
and external galaxies 
[${\rm SFR} = 2 \times 10^{-8}\,  M_\odot \, {\rm yr}^{-1}\, \times \left(M_{\rm dense}/M_\odot \right) $ -- \citet{Gao04a}]. 
{The only exception seems to be the extreme star-forming environment of the central molecular zone (CMZ) of our Galaxy, 
where a very low star formation efficiency in dense gas has been reported \citep{Longmore13}.}
Investigating the nature and origin of this quasi-universal ``star formation law'' in the dense molecular gas of galaxies 
is of fundamental importance for our understanding of star formation and galaxy evolution in the Universe.

Key insight is provided by the results of recent submillimeter imaging surveys of Galactic molecular clouds with 
the {\it Herschel} Space Observatory, which emphasize 
the role of interstellar filaments in the star formation process \citep[e.g.][]{Andre10, Molinari10}. 
The presence of filamentary structures in molecular clouds was already known long before {\it Herschel} 
\citep[e.g.][]{SchneiderElmegreen79,Myers09}, but 
{\it Herschel} observations now demonstrate 
that molecular filaments are truly ubiquitous, 
make up a dominant fraction of the dense gas in molecular clouds,  
and present a high degree of universality in their properties 
\citep[e.g.][]{Arzoumanian11,Hill11,Schisano14,Konyves15}. 
A detailed analysis of the radial column density profiles 
of nearby {\it Herschel} filaments shows that they are characterized 
by a narrow distribution of central widths with a typical 
full width at half maximum (FWHM) value of $\sim 0.1$~pc and a dispersion of less than a factor of 2 \citep[][]{Arzoumanian11}. 
Another major result from the {\it Herschel} Gould Belt survey \citep[HGBS --][]{Andre10}  
is that the vast majority of prestellar cores 
are found in dense, ``supercritical''  filaments for which 
the mass per unit length exceeds the critical line mass of nearly isothermal, long cylinders  
\citep[e.g.][]{Inutsuka97}, $M_{\rm line, crit} = 2\, c_{\rm s}^2/G \sim 16\, M_\odot$ pc$^{-1}$,  
where $c_{\rm s} \sim 0.2$~km s$^{-1}$ is the isothermal sound speed for molecular gas at $T \sim 10$~K 
\citep[e.g.][]{Konyves15}. 
These {\it Herschel} findings in nearby Galactic clouds support a scenario of star formation 
in two main steps \citep[cf.][]{Andre14}: 
First, large-scale compression of interstellar material in supersonic 
flows (turbulent or not) generates a quasi-universal web of 
filaments  in the cold interstellar medium (ISM); 
second, the densest filaments fragment into 
prestellar cores (and subsequently protostars) by gravitational instability above $M_{\rm line, crit} $.

The realization that, at least in nearby clouds,  
prestellar core formation occurs primarily along dense filaments of roughly constant inner width 
has potential implications for our understanding of star formation on global galactic scales. 
Remarkably, the critical line mass   
of a filament, $M_{\rm line, crit} = 2\, c_{\rm s}^2/G$, depends only on gas temperature 
(i.e., $T \sim 10$~K for the bulk of Galactic molecular clouds, away from the immediate vicinity of massive stars). 
Given the common filament width $W_{\rm fil} \sim $~0.1~pc \citep[][]{Arzoumanian11}
this may set a quasi-universal threshold for core/star formation in the giant molecular clouds (GMCs) 
of galaxies at 
$M_{\rm line, crit} \sim 16\, M_\odot \, {\rm pc}^{-1} $ 
in terms of filament mass per unit length, or $M_{\rm line, crit}/W_{\rm fil} \sim 160\, M_\odot \, {\rm pc}^{-2} $ in terms of gas surface 
density (corresponding to a visual extinction $A_{\rm V} \sim 8$), 
or $M_{\rm line, crit}/W_{\rm fil}^2 \sim 1600\, M_\odot \, {\rm pc}^{-3} $ in terms of gas density 
(i.e., a number density $n_{\rm H_2} \sim 2.3 \times 10^4\, {\rm cm}^{-3} $). 
Indeed,  independent {\it Spitzer} infrared studies of the SFR as a function of gas surface density in nearby cloud complexes 
\citep[e.g.][]{Heiderman10,Lada10,Evans14} show that the SFR tends to be linearly proportional to the 
mass of dense gas above a surface density threshold $\Sigma_{\rm gas}^{\rm th} \sim $~130~$M_\odot \, {\rm pc}^{-2} $ 
and drops to negligible values below $\Sigma_{\rm gas}^{\rm th} $. 
The observed star formation threshold $\Sigma_{\rm gas}^{\rm th} $ corresponds to within a factor of $<< 2$ to the 
line-mass threshold above which interstellar filaments are expected to be gravitationally unstable.

\begin{table*}
\centering
\begin{threeparttable}
\caption{Observations\label{obs_parameters}}
\begin{tabular}{lccc}
\hline
Region & Aquila &  Ophiuchus   & Orion B  \\
\hline
Distance & 260 pc\tnote{*}  & 139 pc\tnote{$\dag$}  & 400 pc\tnote{$\ddag$} \\
Telescope & IRAM 30m & MOPRA 22m   & Nobeyama 45m  \\
Receiver & EMIR & 3mm & TZ  \\
Correlator & FTS50 & MOPS  & SAM45  \\
Obs. period &  18 -- 29 Dec 2014 & 26 July -- 2 Aug 2015 & 7--21 May 2015  \\
                    &  2--9 Sep 2015     & &   \\
Mapping area &  0.42 deg$^2$ ($\sim$8.7 pc$^2$) & 0.21 deg$^2$ ($\sim$1.2 pc$^2$) & 0.14 deg$^2$ ($\sim$6.8 pc$^2$)  \\
$\theta_{\rm ant}$ at 86 GHz &  28$\arcsec$.6 ($\sim$0.04 pc) &39$\arcsec$.0 ($\sim$0.03 pc) & 19$\arcsec$.1 ($\sim$0.04 pc)   \\
$dV$ at 86 GHz &  0.15 km/s & 0.10 km/s & 0.023 km/s   \\
\hline
\end{tabular}
\begin{tablenotes}
\item[*] The distance to the Aquila molecular complex is  under debate \citep{Drew97, Dzib10,Ortiz-Leon16}. In this paper, we adopt a distance of 260 pc for Aquila according to \citet{Maury11} and \citet{Konyves15}. 
\item[$\dag$] See \citet{Mamajek08}.
\item[$\ddag$] See \citet{Gibb08}. 
\end{tablenotes}
\end{threeparttable}
\end{table*}
 % Table 1

While the observational results summarized above are very encouraging and tentatively point to a unified picture for star formation 
on GMC scales in both Galactic clouds and external galaxies, there are at least two caveats.  
First, direct comparison between Galactic \citep[e.g.][]{Lada10} 
and extragalactic \citep[e.g.][]{Gao04a} studies of the dense gas -- star formation connection is difficult 
at this stage because different tracers have been used to probe dense gas in Galactic and extragalactic situations so far.
For instance, \citet{Lada10} 
used column density maps from near-infrared 
extinction 
and derived the total mass $M^{A_{\rm V} > 8}_{\rm dense} $ above the extinction/surface density threshold mentioned 
earlier (i.e.,  $A_{\rm V}$ $>$ 8), while \citet{Gao04a} used HCN (1--0) data and estimated the mass of dense gas $M_{\rm dense}$ above 
the effective density $\sim$3$\times$10$^{4}$ cm$^{-3}$ of the HCN (1--0) transition from the HCN (1--0) line luminosity, i.e., 
$M^{\rm HCN}_{\rm dense}$ = $\alpha_{\rm GS04-HCN}$ $L_{\rm HCN}$ with 
$\alpha_{\rm GS04-HCN}$ $\sim$10 $M_{\odot}$ (K km s$^{-1}$ pc$^2$)$^{-1}$ \citep[see also][]{Wu05}. 
While the effective density of the HCN(1--0) transition (cf. Evans 1999) turns out to be close to the critical threshold density 
$\sim 2.3 \times 10^4\, \rm{cm}^{-3} $ quoted  above for $\sim 0.1$-pc-wide supercritical filaments, 
the relation between $M^{A_{\rm V} > 8}_{\rm dense} $ and $M^{\rm HCN}_{\rm dense} $ remains 
to be properly calibrated in nearby Galactic clouds.
Second, significant variations in the apparent star formation efficiency in dense gas SFE$^{\rm extragal}_{\rm dense} \equiv L_{\rm FIR}/L_{\rm HCN}$ 
as a function of stellar surface density or galactocentric radius 
have been found in resolved observations of the disks of several nearby galaxies \citep[][]{Usero15,Chen15,Leroy16, Bigiel16}. 
To confirm that there is a universal 
star formation law converting the dense molecular gas of GMCs
into stars, wide-field line mapping observations of Galactic clouds in the same dense gas tracers as used in extragalactic work
and at a spatial resolution high enough to resolve $\sim$0.1-pc-wide molecular filaments 
are crucially needed.

With the advent of sensitive heterodyne receivers and wideband spectrometers on millimeter-wave telescopes, wide-field mapping observations in 
several lines simultaneously are now feasible at an angular resolution down to $\sim$10--30$\arcsec$. 
In this paper, we present wide-field imaging data in the 
HCN (1--0), H$^{13}$CN (1--0), HCO$^{+}$ (1--0), and H$^{13}$CO$^+$ (1--0) transitions toward three nearby star-forming clouds, Aquila, Ophiuchus, and Orion B. 
The paper is organized as follows. In Sect.~2, we describe our IRAM 30m, MOPRA 22m, and Nobeyama 45m observations. In Sect.~3, 
we present the results of the HCN, H$^{13}$CN, HCO$^{+}$, and H$^{13}$CO$^+$ mappings and estimate the far ultraviolet (FUV) field strength 
from {\it Herschel} HGBS 70 $\mu$m and 100 $\mu$m data. In Sect.~4, we discuss evidence of significant variations in the 
conversion factor $\alpha_{\rm HCN}$ between HCN luminosity and mass of dense gas, 
and, in particular, the dependence of $\alpha_{\rm HCN}$ on the FUV field strength. 
We then revisit the question of the universality of the star formation efficiency in the dense molecular gas of galaxies 
and propose an interpretation of this universality in terms of the filamentary structure of GMCs.
Our conclusions are summarized in Sect.~5.

%%%%%%%%%%%%%%%%%%%%%%%%%
% Observations
%%%%%%%%%%%%%%%%%%%%%%%%%
\section{Millimeter line observations}

We carried out observations in the HCN ($J$=1--0, 88.6318473 GHz), H$^{13}$CN ($J$=1--0, 86.340167 GHz), HCO$^+$ ($J$=1--0, 89.188526 GHz), and H$^{13}$CO$^+$ ($J$=1--0, 86.75433 GHz) transitions toward three nearby star-forming regions, Aquila, Ophiuchus, and Orion B using the IRAM 30m, MOPRA 22m, and Nobeyama 45m telescopes. The effective excitation densities\footnote{The effective excitation density of a molecular transition at a given gas kinetic temperature 
is defined as the density which results in a spectral line 
with an integrated intensity of 1 K km s$^{-1}$ \citep{Shirley15}.} of the HCN (1--0), H$^{13}$CN (1--0), HCO$^+$ (1--0), and H$^{13}$CO$^+$ (1--0) lines at 10 K are 8.4$\times$10$^3$ cm$^{-3}$, 3.5$\times$10$^5$ cm$^{-3}$, 9.5$\times$10$^2$ cm$^{-3}$, and 3.9$\times$10$^4$ cm$^{-3}$, respectively \citep{Shirley15}. Table \ref{obs_parameters} shows a summary of our molecular line observations. We describe the details of each observation below.

%Figure 1
\begin{figure*}
\begin{center}
\includegraphics[width=180mm, angle=0]{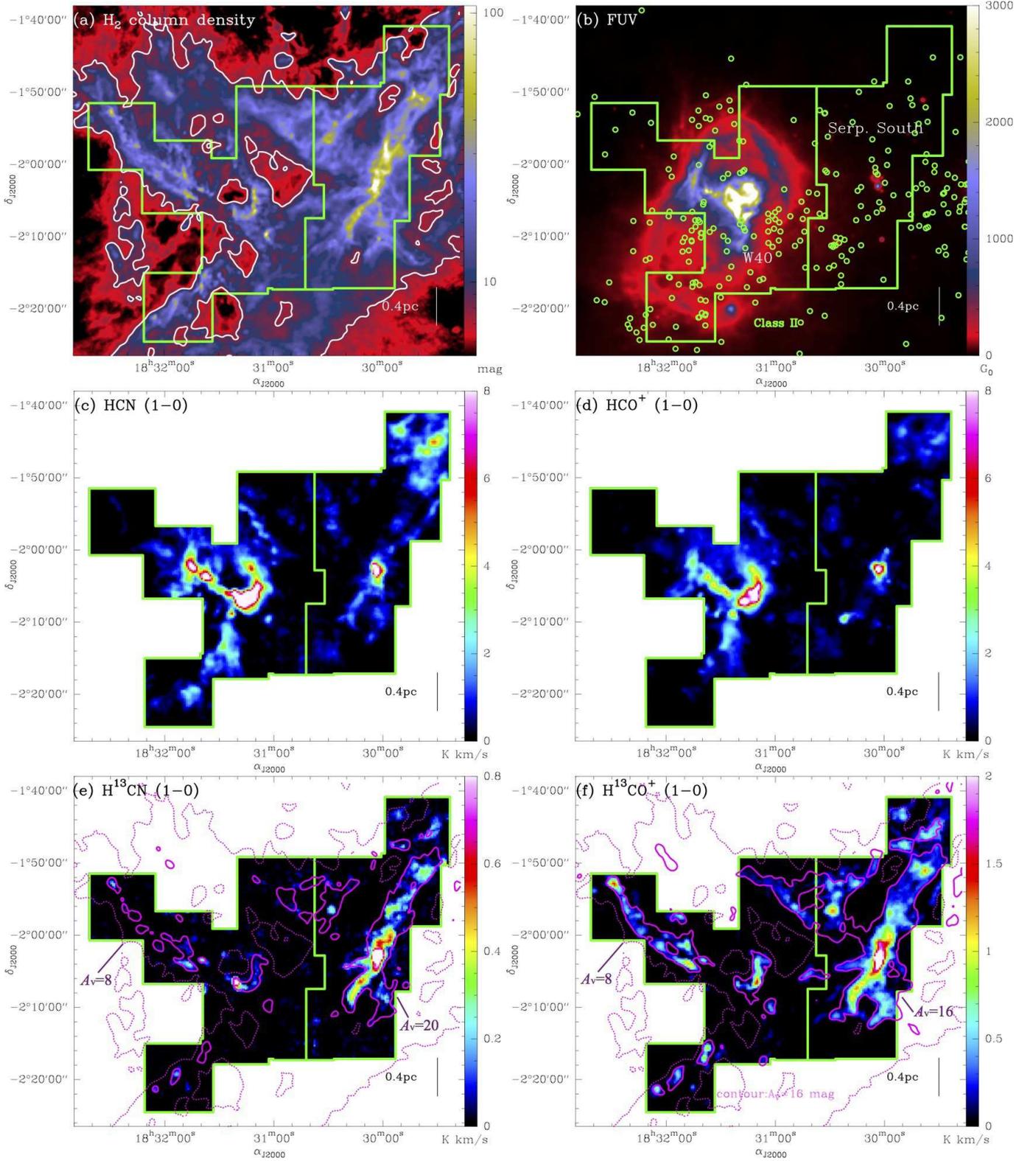} 
\caption{a) Column density map of the Aquila region derived from {\it Herschel} Gould Belt survey (HGBS) data \citep{Andre10, Konyves15} 
at an angular resolution of 18.2$\arcsec$ and in units of {$A_{\rm V}$}. 
b) FUV field strength map derived from HGBS 70 $\mu$m and 100 $\mu$m data smoothed to an angular resolution of 18.2$\arcsec$ in Habing units, 
and integrated intensity maps of c) HCN(1--0), d) HCO$^{+}$(1--0), {e) H$^{13}$CN(1--0) , f) H$^{13}$CO$^+$(1--0)} in units of K km s$^{-1}$ ($T_{\rm MB}$). 
The angular resolutions of the HCN, HCO$^{+}$,  {H$^{13}$CN, and H$^{13}$CO$^+$} maps are 40$\arcsec$, 40$\arcsec$, 40$\arcsec$, and 50$\arcsec$, respectively. 
In each panel, a green polygon outlines the field observed in molecular lines. 
The HCN and H$^{13}$CN integrated intensity includes all components of the hyperfine structure (HFS). 
{The white contour in panel (a) and the magenta dotted contours in panels (e) and (f) show} the $A_{\rm V}$ = 8 level obtained after smoothing the {\it Herschel} column density map 
to 40$\arcsec$ resolution. 
In panel (b), green open circles indicate the positions of the Class~II objects identified by \citet{Dunham15}. 
{In panel (e) and (f), the magenta solid contour indicates the rough $A_{\rm V}$ column density level 
above which significant line emission is detected, 
i.e., $A_{\rm V}$ = 20 for H$^{13}$CN (1--0) and  $A_{\rm V}$ = 16 for H$^{13}$CO$^+$ (1--0).}}
\label{fig1}
\end{center}
\end{figure*}

\subsection{IRAM 30m observations toward the Aquila cloud}
During two observing runs (18 December 2014 - 29 December 2014 and 2 September 2015 -- 9 September 2015), we carried out mapping observations toward a 0.4 deg$^{2}$ region in the Aquila cloud, including the three subregions Aquila/W40, Aquila/Serp. South, and Aquila/cold (see Figs. \ref{fig1} and \ref{fig1_cold}), with the Eight MIxer Receiver (EMIR) receiver on the IRAM-30m telescope. All molecular line data were obtained simultaneously. At 86 GHz, the 30 m telescope has a beam size of 28.6$\arcsec$ (HPBW) and {the forward and main beam efficiencies ($F_{\rm eff}$ and $B_{\rm eff}$) are 95\% and 81\%}.  As backend, we used the FTS50 spectrometer, providing a bandwidth of 1820 MHz and a frequency resolution of 50 kHz. The latter corresponds to a velocity resolution of $\sim$0.15 km s$^{-1}$ at 86 GHz.  
The standard chopper wheel method was used to convert the observed signal to the antenna temperature $T_{\rm A}^*$ in units of K, corrected for the atmospheric attenuation. The data are given in terms of the main beam brightness temperature corresponding to {$T_{\rm MB} = T_{\rm A}^* \times F_{\rm eff}/B_{\rm eff}$. }
During the  observations, the system noise temperatures ranged from 80 K to 330 K. 
The telescope pointing was checked every hour by observing the quasar source 1741-038 and was found to be better than 3$\arcsec$ throughout the two runs.  
Our mapping observations were made with the on-the-fly (OTF) mapping technique. 
We chose the positions (RA$_{\rm J2000}$, DEC$_{\rm J2000}$) = (16:29:06.4, -24:26:57.0), (16:25:34.5, -24:36:44.0), and (16:26:14.5, -24:02:00.0) 
as our reference (off) positions. 
We decomposed the target field into a series of 10$\arcmin$$\times$10$\arcmin$ subfields and took pairs of 
OTF maps toward each subfield using two perpendicular scanning directions (along the RA or Dec axes).
Combining such pairs of OTF maps reduces scanning artifacts.
We smoothed the data spatially with a Gaussian function resulting in an effective beam size of 40$\arcsec$. 
The 1$\sigma$ noise level of the final mosaiced data cube at an effective angular resolution of 40$\arcsec$ 
and a velocity resolution of $\sim$0.15 km s$^{-1}$ is 0.07 K in $T_{\rm MB}$.

\subsection{MOPRA observations toward the Ophiuchus cloud}
Between 26 July 2015 and 2 August 2015, we carried out mapping observations toward a 0.21 deg$^2$ region in the Ophiuchus cloud, {including the two subregions Oph/main (L1688) and Oph/cold (see Fig. \ref{fig_oph_maps})}, with the 3 mm receiver installed on the MOPRA-22m telescope. All molecular line data were obtained simultaneously.  At 86 GHz, the telescope has a beam size of 39$\arcsec$ (HPBW) and a main beam efficiency $\eta_{\rm MB}$ of 49\% \citep{Ladd05}. As backend, we used the Mopra spectrometer (MOPS) in zoom mode, providing a bandwidth of 137.5 MHz and a frequency resolution of 33.8 kHz. The latter corresponds to a velocity resolution of $\sim$0.1 km s$^{-1}$ at 86 GHz. The standard chopper wheel method was used to convert the observed signal to the antenna temperature $T_{\rm A}^*$ in units of K, corrected for atmospheric attenuation. 
The data are given in main beam brightness temperature, $T_{\rm MB}$ = $T_{\rm A}^*$/$\eta_{\rm MB}$. 
During the  observations, the system noise temperatures ranged from 240 K to 420 K.
The telescope pointing was checked every hour by observing the SiO maser sources AH Sco, VX Sgr, and W Hya, and was better than 5$\arcsec$ throughout the entire run.  
Our mapping observations were made with the OTF mapping technique. The positions (RA$_{\rm J2000}$, DEC$_{\rm J2000}$) = (16:29:06.4, -24:26:57.0), (16:25:34.5, -24:36:44.0), and (16:26:14.5, -24:02:00.0) were used as off positions. We obtained a series of OTF maps with two different scanning directions along the RA or Dec axes covering a subfield of 6$\arcmin$$\times$6$\arcmin$ each and combined them into a single map to reduce scanning effects as much as possible.  
We smoothed the data spatially with a Gaussian function of $19.5\arcsec$ (FWHM), resulting in an effective beam size of 50$\arcsec$. 
The scanning effects were minimized by combining scans along the RA and Dec directions with the \citet{Emerson88} PLAIT algorithm. 
The 1$\sigma$ noise level of the final data at an effective angular resolution of 50$\arcsec$ and a velocity resolution of 0.1 km s$^{-1}$ is 0.5 K in $T_{\rm MB}$.

\begin{table*}
\centering
\begin{threeparttable}
\tiny
\caption{Physical parameters derived from H$^{13}$CO$^{+}$ (1--0) observations \label{table:param_h13cop}}
\begin{tabular}{|l|ccc|ccc|ccc|ccc|}
\hline
Region &  \multicolumn{3}{|c|}{$V_{\rm peak}$ [km s$^{-1}$] }   & \multicolumn{3}{|c|}{$dV_{\rm FWHM}$ [km s$^{-1}$] } & \multicolumn{3}{|c|}{$N_{\rm H^{13}CO^+}$ [cm$^{-2}$]}  &  \multicolumn{3}{|c|}{$X_{\rm H^{13}CO^+}$}   \\
            & min  & max  & mean &    min  & max  & mean &  min  & max  & mean   &   min  & max  & mean      \\
\hline
Aquila/W40                  &  4.0 & 8.8 & 7.0  & 0.3    &  3.3  & 0.7
                                    	&  4.4$\times$10$^{10}$  &  1.7$\times$10$^{12}$ & 2.9$\times$10$^{10}$   &   2.8$\times$10$^{-12}$ &  1.2$\times$10$^{-11}$ &   5.5$\times$10$^{-11}$  \\
Aquila/Serp S               &  3.8 & 8.9  & 7.3  & 0.3    &  2.3 & 0.8
                                    	&  4.6$\times$10$^{10}$  &  4.1$\times$10$^{12}$ & 3.3$\times$10$^{10}$  &   3.3$\times$10$^{-12}$ &  4.6$\times$10$^{-11}$ &   1.5$\times$10$^{-11}$  \\
Aquila/cold                  &   5.6 & 8.6  & 6.5  & 0.3    &  1.1  & 0.5
                                    	&  7.2$\times$10$^{10}$  &  5.7$\times$10$^{11}$ & 2.4$\times$10$^{11}$   &   9.2$\times$10$^{-12}$ &  4.1$\times$10$^{-11}$ &   2.2$\times$10$^{-11}$  \\
Oph/main (L1688)                     &  2.5 & 4.9  & 3.6  & 0.3    &  2.0  & 0.7
                                    	&   6.4$\times$10$^{11}$  &   2.9$\times$10$^{12}$ &  1.1$\times$10$^{12}$   &   8.0$\times$10$^{-12}$ &   2.0$\times$10$^{-10}$  &   3.3$\times$10$^{-11}$    \\
Oph/cold                     &  2.9 & 3.6  & 3.3  &   0.3  & 0.7  & 0.4
                                    	&  6.3$\times$10$^{11}$  &   1.5$\times$10$^{12}$ &  9.6$\times$10$^{11}$   &   2.2$\times$10$^{-11}$ &   1.2$\times$10$^{-10}$  &   5.5$\times$10$^{-11}$    \\
Orion B/NGC 2023       &   8.1 & 11.7  & 10.0  & 0.2    & 2.3  & 1.0
                                    	&   2.6$\times$10$^{11}$  &  37.8$\times$10$^{11}$ & 15.2$\times$10$^{11}$ &  1.8$\times$10$^{-11}$  &  9.0$\times$10$^{-11}$  &   5.0$\times$10$^{-11}$   \\
Orion B/NGC 2024       &  5.1 & 13.2  & 10.9  & 0.2   &  2.7  & 0.9
                                    	&  2.3$\times$10$^{11}$  &  37.9$\times$10$^{11}$& 14.3$\times$10$^{11}$   &  1.9$\times$10$^{-11}$  &  9.7$\times$10$^{-11}$  &   5.8$\times$10$^{-11}$   \\
Orion B/NGC 2068       & 7.1 & 14.8 & 10.9 & 0.2    & 1.8    & 0.6
                                    	&  2.3$\times$10$^{11}$ &  22.4$\times$10$^{11}$ & 10.2$\times$10$^{11}$   &   1.8$\times$10$^{-11}$  &  6.9$\times$10$^{-11}$  &   4.5$\times$10$^{-11}$   \\
Orion B/NGC 2071       & 6.1 & 12.7  & 9.5  & 0.2    &  2.2   &0.9
                                    	&  2.6$\times$10$^{11}$  & 42.0$\times$10$^{11}$ & 15.6$\times$10$^{11}$   &  1.8$\times$10$^{-11}$  &  1.1$\times$10$^{-10}$  &   4.5$\times$10$^{-11}$   \\
\hline
\end{tabular}
\end{threeparttable}
\end{table*}

  % Table 2
%
% This is produced onTue Jan 31 10:24:54 2017
% using Table_Mvir_h13cop.pro
\begin{table*}
\centering
\scalebox{0.76}[0.76]{
\begin{threeparttable}
\caption{Virial masses from H$^{13}$CO$^+$$^\dag$ \label{table:virial_h13cop}}
\begin{tabular}{|l|cccccc|cccccc|}
\hline
Region & $dV_{\rm H^{13}CO^+}^{\rm detect}$ & $A_{\rm H^{13}CO^+}^{\rm detect}$ & $R_{\rm H^{13}CO^+}^{\rm detect}$ & $M_{\rm VIR,H^{13}CO^+}^{\rm detect}$& $M_{\rm Herschel}^{\rm H^{13}CO^+-detect}$ & $\frac{M_{\rm VIR,H^{13}CO^+}^{\rm detect}}{M_{\rm Herschel}^{\rm H^{13}CO^{+}-detect}}$  & $dV_{\rm H^{13}CO^+}^{A_{\rm V}>8}$ & $A_{\rm H^{13}CO^+}^{A_{\rm V}>8}$ & $R_{\rm H^{13}CO^+}^{A_{\rm V}>8}$ & $M_{\rm VIR,H^{13}CO^+}^{A_{\rm V}>8}$ & $M_{\rm Herschel}^{A_{\rm V}>8}$ & $\frac{M_{\rm VIR,H^{13}CO^+}^{A_{\rm V}>8}}{M_{\rm Herschel}^{A_{\rm V}>8}}$ \\
 & [km s$^{-1}$] & [pc$^2$] & [pc] & [$M_{\odot}$] & [$M_{\odot}$]  & & [km s$^{-1}$] & [pc$^2$] & [pc] & [$M_{\odot}$] &  [$M_{\odot}$]  & \\  
\hline
Aquila/W40&  1.66&  0.90&  0.54& 307.3&   339.1&   0.9&  2.18&  2.68&  0.92&
   917.0&   748.1&   1.2\\
Aquila/Serp S&  1.19&  1.53&  0.70& 205.8&   707.4&   0.3&  1.35&  2.51&  0.89&
   339.0&   954.1&   0.4\\
Aquila/cold&  1.04&  0.36&  0.34&  76.7&    74.2&   1.0&  1.07&  0.40&  0.36&
    85.9&    88.8&   1.0\\
\hdashline
Aquila (total) & --- &--- &--- & 589.8&1120.8&   0.5& ---& --- &---&1341.8&
1790.9&   0.7\\
\hline
Oph/main (L1688)&  1.48&  0.23&  0.27& 122.0&   167.9&   0.7&  2.09&  0.91&
  0.54&   492.0&   416.0&   1.2\\
Oph/cold&  0.40&  0.02&  0.07&   2.5&     6.5&   0.4&  0.55&  0.06&  0.14&
     8.5&    16.2&   0.5\\
\hdashline
Oph (total) & --- &--- &--- & 124.5& 174.4&   0.7& ---& --- &---& 500.5& 432.1&
   1.2\\
\hline
Orion B/NGC2023&  1.50&  0.30&  0.31& 144.9&   175.2&   0.8&  1.89&  0.76&  0.49
&   367.7&   300.1&   1.2\\
Orion B/NGC2024&  1.97&  0.26&  0.29& 236.3&   206.3&   1.1&  2.57&  0.76&  0.49
&   680.8&   335.2&   2.0\\
Orion B/NGC2068&  0.99&  0.23&  0.27&  55.0&   113.7&   0.5&  1.21&  0.52&  0.41
&   125.3&   189.6&   0.7\\
Orion B/NGC2071&  1.23&  0.46&  0.38& 121.1&   262.9&   0.5&  1.47&  0.94&  0.55
&   247.3&   398.7&   0.6\\
\hdashline
Orion B (total) & --- &--- &--- & 557.3& 758.2&   0.7& ---& --- &---&1421.1&
1223.5&   1.2\\
\hline
\end{tabular}
\begin{tablenotes}
\item[$^\dag$] {See Table \ref{list_symbols} for the definition of each notation.}
\end{tablenotes}
\end{threeparttable}
}
\end{table*}

  % Table 3
%
% This is produced onThu Nov 24 10:57:17 2016
% using Table_Mvir_h13cn.pro
\begin{table*}
\scalebox{0.79}[0.79]{
\centering
\begin{threeparttable}
\caption{Virial masses from H$^{13}$CN$^\dag$ \label{table:virial_h13cn}}
\begin{tabular}{|l|cccccc|cccccc|}
\hline
Region & $dV_{\rm H^{13}CN}^{\rm detect}$ & $A_{\rm H^{13}CN}^{\rm detect}$ & $R_{\rm H^{13}CN}^{\rm detect}$ & $M_{\rm VIR,H^{13}CN}^{\rm detect}$& $M_{\rm Herschel}^{\rm H^{13}CN-detect}$ & $\frac{M_{\rm VIR,H^{13}CN}^{\rm detect}}{M_{\rm Herschel}^{\rm H^{13}CN-detect}}$  & $dV_{\rm H^{13}CN}^{A_{\rm V}>8}$ & $A_{\rm Herschel}^{A_{\rm V}>8}$ & $R_{\rm Herschel}^{A_{\rm V}>8}$ & $M_{\rm VIR,H^{13}CN}^{A_{\rm V}>8}$ & $M_{\rm Herschel}^{A_{\rm V}>8}$ & $\frac{M_{\rm VIR,H^{13}CN}^{A_{\rm V}>8}}{M_{\rm Herschel}^{A_{\rm V}>8}}$ 
\\
 & [km s$^{-1}$] & [pc$^2$] & [pc] & [$M_{\odot}$] & [$M_{\odot}$]  &  & [km s$^{-1}$] & [pc$^2$] & [pc] & [$M_{\odot}$] &  [$M_{\odot}$]  & \\  
\hline
Aquila/W40&  1.40&  0.25&  0.28&   115.0&   108.5&  1.1&  2.54&  2.68&  0.92&  1246.8&   748.1&   1.7\\
Aquila/Serp S&  1.17&  0.77&  0.49&   140.7&   445.0&   0.3&  1.57&  2.51&  0.89&   460.6&   954.1&   0.5\\
\hline
Oph/main(L1688)&  1.58&  0.16&  0.23&   117.2&    90.0&   1.3&  2.43&  0.91&  0.54&   665.5&   416.0&   1.6\\
\hline
Orion B/NGC2024&  1.86&  0.20&  0.25&   181.5&   172.2&   1.1&  2.62&  0.76&  0.49&   705.5&   335.2&   2.1\\
Orion B/NGC2068&  1.91&  0.17&  0.23&   179.0&    71.8&   2.5&  2.53&  0.52&  0.41&   546.4&   189.6&   2.9\\
Orion B/NGC2071&  1.90&  0.37&  0.34&   257.1&   203.0&   1.3&  2.40&  0.94&  0.55&   661.7&   398.7&   1.7\\
\hline
\end{tabular}
\begin{tablenotes}
\item[$^\dag$] {See Table \ref{list_symbols} for the definition of each notation.}
\end{tablenotes}
\end{threeparttable}
}
\end{table*}
  % Table 4

\subsection{NRO 45m observations toward the Orion B cloud}
Between 7 May and 21 May 2015, we carried out mapping observations toward a 0.14 deg$^2$ region in the Orion B cloud, {including the four subregions NGC2023, NGC2024, NGC2068, and NGC2071 (see Figs. \ref{fig_ngc2023_maps}--\ref{fig_ngc2071_maps}),} with the TZ receiver on the Nobeyama 45m telescope. All molecular line data were obtained simultaneously.  At 86 GHz, the telescope has a beam size of $19.1\arcsec$ (HPBW) and a main beam efficiency $\eta_{\rm MB}$ of $\sim$50\%.  As backend, we used the SAM45 spectrometer which provides a bandwidth of 31 MHz and a frequency resolution of 7.63 kHz. 
The latter corresponds to a velocity resolution of $\sim$0.02 km s$^{-1}$ at 86 GHz. The standard chopper wheel method was used to convert the observed signal to the antenna temperature $T_{\rm A}^*$ in units of K, corrected for the atmospheric attenuation. 
The data are given in terms of the main beam brightness temperature, $T_{\rm MB}$ = $T_{\rm A}^*$/$\eta_{\rm MB}$. 
During the observations, the system noise temperatures ranged from 140 K to 630 K. 
The telescope pointing was checked every hour by observing the SiO maser source Ori-KL, and was better than 3$\arcsec$ throughout the entire observing run.  
Our mapping observations were made with the OTF mapping technique. 
We chose the positions (RA$_{\rm J2000}$, DEC$_{\rm J2000}$) = (5:39:21.819, -2:9:8.54) and (5:44:41.15, 0:33:4.98) as off positions. 
We obtained OTF maps with two different scanning directions along the RA or Dec axes covering 
a subfield of 6$\arcmin$$\times$6$\arcmin$ and combined them into a single map to reduce the scanning effects as much as possible.  
As a convolution function, we adopted a Gaussian function with a FWHM of half the beam size. The scanning effects were minimized 
by combining scans along the RA and Dec directions with the \citet{Emerson88} PLAIT algorithm. We also applied spatial smoothing 
to the data with a Gaussian function resulting in an effective beam size of 30$\arcsec$. The 1$\sigma$ noise level of the final data at an effective 
resolution of 30$\arcsec$ and a velocity resolution of 0.1 km s$^{-1}$ is 0.28 K in $T_{\rm MB}$.

%%%%%%%%%%%%%%%%%%%%%%%%%%%%%
% Results
%%%%%%%%%%%%%%%%%%%%%%%%%%%%%

\section{Results and Analysis}

%Figure 2
\begin{figure}
\begin{center}
\includegraphics[width=80mm, angle=0]{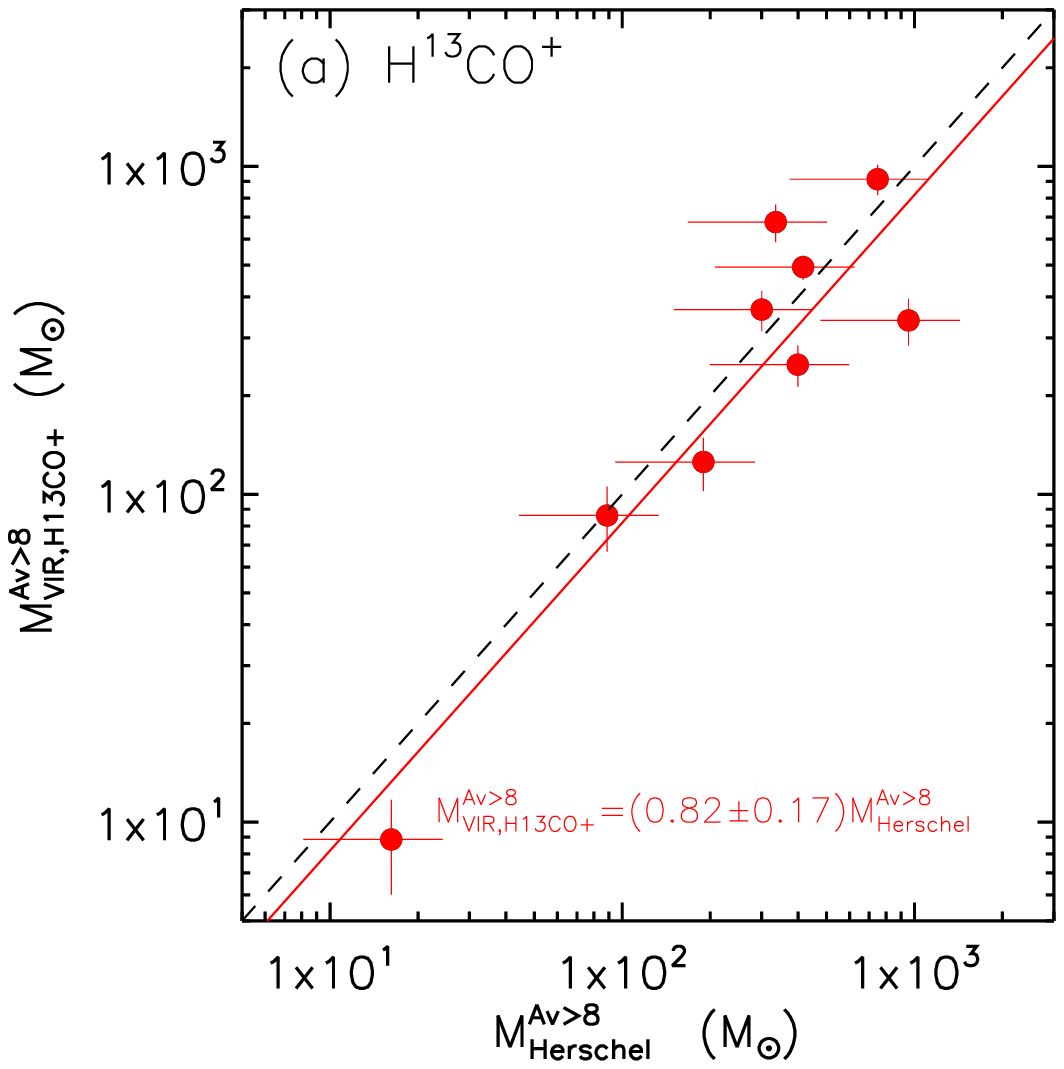} 
\includegraphics[width=80mm, angle=0]{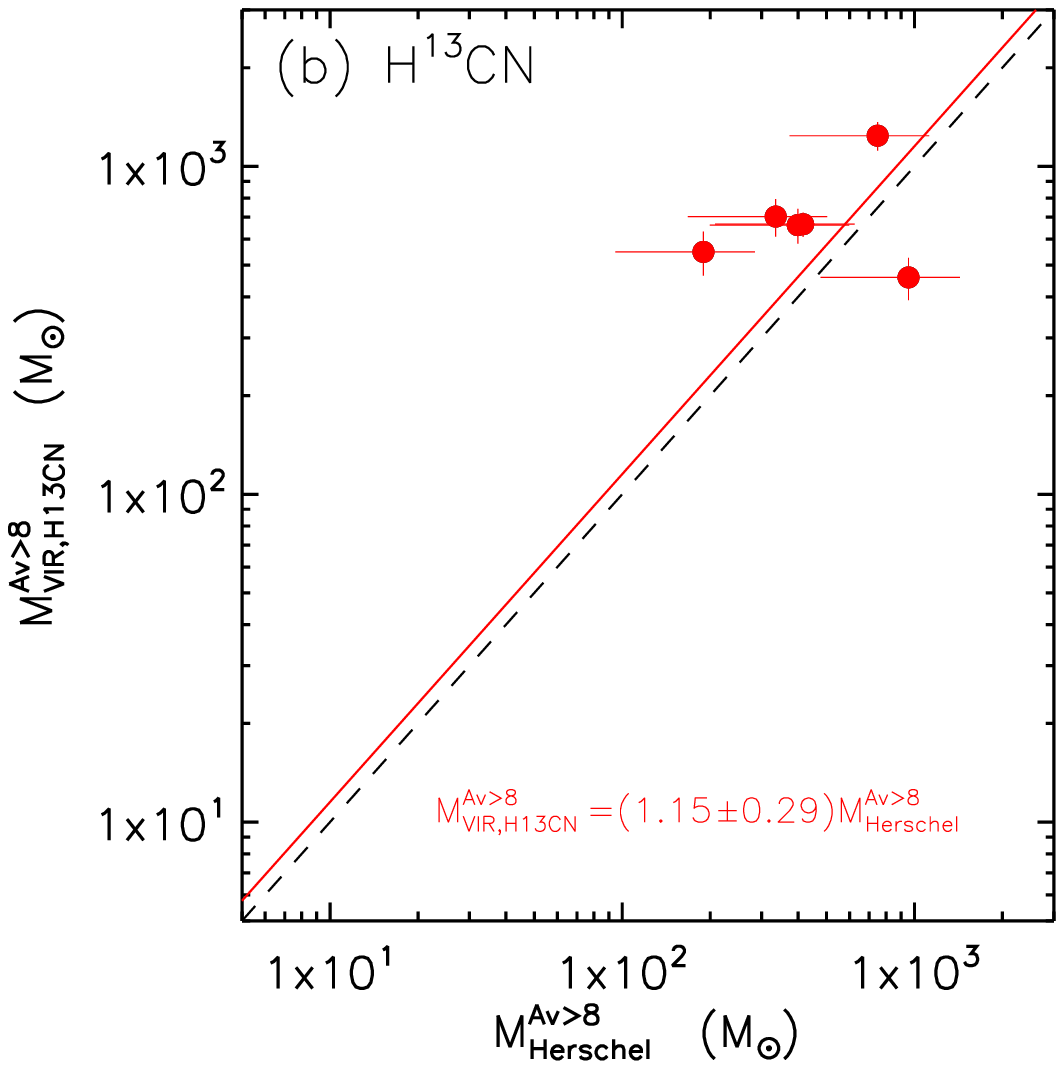} 
\caption{(a) $M_{\rm VIR,H^{13}CO^+}^{A_{\rm V}>8}$ vs. $M_{\rm Herschel}^{A_{\rm V}>8}$ and (b) $M_{\rm VIR,H^{13}CN}^{A_{\rm V}>8}$ vs. $M_{\rm Herschel}^{A_{\rm V}>8}$ relations. The black dashed lines indicate $M_{\rm VIR,mol}^{A_{\rm V}>8}$ = $M_{\rm Herschel}^{A_{\rm V}>8}$ and 
the red lines indicate the best-fit results: 
$M_{\rm VIR,H^{13}CO^+}^{A_{\rm V}>8} = (0.82\pm0.17) \times M_{\rm Herschel}^{A_{\rm V}>8}$ 
and $M_{\rm VIR,H^{13}CN}^{A_{\rm V}>8} = (1.15\pm0.29) \times M_{\rm Herschel}^{A_{\rm V}>8}$. 
(The uncertainty in $M_{\rm VIR,mol}^{A_{\rm V}>8}$ was derived from the uncertainties in $R_{\rm Herschel}^{A_{\rm V}>8}$ and in $dV_{\rm mol}^{A_{\rm V}>8}$ \citep[see][]{Ikeda07, Shimajiri15a}.) {The x- and y-axis ranges are the same in both panels.}}
\label{Fig-Mvir_Mherschel}
\end{center}
\end{figure}

%%%%%%%%%%%%%%%%%%%%%%%%%
% Results of H13CO+ and H13CN
%%%%%%%%%%%%%%%%%%%%%%%%%
\subsection{H$^{13}$CO$^+$(1--0) and H$^{13}$CN(1--0) emission}

Figure \ref{fig1} and Figures \ref{fig1_cold}--\ref{fig_ngc2071_maps} 
compare  the H$^{13}$CO$^+$(1--0), and H$^{13}$CN(1--0) integrated intensity maps 
observed toward Aquila, Ophiuchus, Orion B/NGC2023, Orion B/NGC2024, Orion B/NGC2068, and Orion B/NGC2071 
with the column density maps derived from HGBS data toward the same sub-regions.
Above the $A_{\rm V}$ = 16 contour in the {\it Herschel} column density maps 
\citep[assuming $N_{\rm H_2}$/$A_{\rm V}$ = 0.94$\times$10$^{21}$cm$^{-2}$,][]{Bohlin78}, 
it can be seen that the spatial distribution of the H$^{13}$CO$^+$(1--0) emission is closely correlated
with the texture of the dense gas (e.g. filamentary structure) seen by {\it Herschel}.
In particular, the H$^{13}$CO$^+$(1--0) emission traces the dense ``supercritical'' filaments detected by {\it Herschel} very well. 
Likewise, the spatial distribution of the H$^{13}$CN (1--0) emission is closely correlated with 
the column density distribution above the $A_{\rm V}$ = 20 level, especially in the Aquila and Orion B/NGC 2024 regions.
As apparent in Fig.~\ref{fig1} and Figs.~\ref{fig1_cold}--\ref{fig_ngc2071_maps}, 
and in agreement with the effective excitation densities quoted in Sect.~2, 
the H$^{13}$CN (1--0) emission traces higher column density gas compared to the H$^{13}$CO$^+$ (1--0) emission. 
Figure~\ref{fig_h13cn_spec} displays mean H$^{13}$CN(1--0) spectra, obtained by averaging the data  
over the detected portion of each region and sub-region.

{
The 1 K km s$^{-1}$ level in the H$^{13}$CO$^+$(1-0) integrated intensity map roughly matches the $A_{\rm V}$=30 level 
in the {\it Herschel} column density map (see Fig.~\ref{fig1}f), 
which in turn corresponds to a volume density $\sim  8.6 \times10^{4}$ cm$^{-3}$ 
assuming most of the dense gas is concentrated in $\sim 0.1\,$pc filaments (cf. Sect.~\ref{sect:Intro}). 
This is roughly consistent with the H$^{13}$CO$^{+}$ (1--0) effective excitation  density of $\sim4 \times 10^{4}$ cm$^{-3}$, 
suggesting that local thermodynamical equilibrium (LTE) may not be too bad an approximation for H$^{13}$CO$^+$(1-0).
Under the LTE assumption,} the column density of H$^{13}$CO$^+$ can be {derived} as follows \citep[cf.][]{Tsuboi11}: 

\begin{equation}\label{eq:h13cop_column}
N_{\rm H^{13}CO^+} [{\rm cm^{-2}}] = 5.99 \times 10^{10} T_{\rm ex} \int T_{\rm MB} dv\ [{\rm K\ km\ s^{-1}}].  
\end{equation}

\noindent We further assume that the excitation temperature $T_{\rm ex}$ of the H$^{13}$CO$^+$ (1--0) transition is equal to the dust temperature $T_{\rm dust}$ derived from the HGBS data. 
The dust temperature ranges from 11~K to 46 K.
Table \ref{table:param_h13cop} summarizes the results. 
The mean H$^{13}$CO$^+$ column densities range from 2.9$\times$10$^{10}$ cm$^{-2}$ to 1.6$\times$10$^{12}$ cm$^{-2}$. 
The H$^{13}$CO$^+$ abundances relative to H$_2$,
$X_{\rm H^{13}CO^+} \equiv N_{\rm H^{13}CO^+}/N_{\rm H_2}$, have mean values in the range (1.5--5.8)$\times$10$^{-11}$, 
using  $N_{\rm H^{13}CO^+}$ values estimated from the present data and $N_{\rm H_2}$ values from HGBS data \citep{Andre10,Konyves15}.
These abundance estimates are consistent within a factor of a few 
with the findings of previous studies in other regions  
(1.1$\pm$0.1$\times$10$^{-11}$ in OMC2-FIR4: \citet{Shimajiri15b}, and 1.8$\pm$0.4$\times$10$^{-11}$ in Sagittarius A: \citet{Tsuboi11}). 

Assuming 
spherical shapes and uniform density, 
the virial masses $M_{\rm VIR,mol}^{\rm detect}$ of the detected clumps and structures 
can be estimated as \citep[see][]{Ikeda07, Shimajiri15a}, 

\begin{equation}\label{eq:virial_mass}
\begin{split}
M_{\rm VIR,mol}^{\rm detect} [M_{\odot}] & = \frac{5 R_{\rm mol}^{\rm detect} \sigma^2 }{G} \\
 &  = 209 \left(\frac{R_{\rm mol}^{\rm detect}}{\rm pc} \right) \left(\frac{dV_{\rm mol}^{\rm detect}}{\rm km\ s^{-1}} \right)^2.
\end{split}
\end{equation}

\noindent The radius $R_{\rm mol}^{\rm detect}$ of each clump or cloud was estimated as $R_{\rm mol}^{\rm detect} {\rm \ [pc]} = \sqrt{A_{\rm mol}^{\rm detect}/\pi}$. 
The velocity dispersion $\sigma$ was determined as $\sigma$ = $dV_{\rm mol}^{\rm detect}$/$\sqrt{8\ln{2}}$, 
where $dV_{\rm mol}^{\rm detect}$ is the mean FWHM velocity width among pixels where the emission was detected. 
The derived values $dV_{\rm mol}^{\rm detect}$, $R_{\rm mol}^{\rm detect}$, $A_{\rm mol}^{\rm detect}$, and $M_{\rm VIR,mol}^{\rm detect}$ are 
given in Tables \ref{table:virial_h13cop} and \ref{table:virial_h13cn}.

We also estimated the total virial mass $M_{\rm VIR,mol}^{A_{\rm V}>8}$ of the area above $A_{\rm V}$ = 8 for each cloud by scaling the virial mass derived for the detected subregion using the well-known 
linewidth-size relation  $\sigma_V \propto L^{0.5}$ \citep{Larson81,Heyer09}. 
In practice, we assumed

\begin{equation}
dV_{\rm mol}^{A_{\rm V}>8} \  [{\rm km\ s^{-1}}] = dV_{\rm mol}^{\rm detect} \left( \frac{R_{\rm Herschel}^{A_{\rm V}>8}}{R_{\rm mol}^{\rm detect}} \right)^{0.5},
\end{equation}

\noindent and we estimated the cloud radius $R_{\rm Herschel}^{A_{\rm V}>8}$  
as $R_{\rm Herschel}^{A_{\rm V}>8}${\rm \ [pc]} = $\sqrt{A_{\rm Herschel}^{A_{\rm V}>8}/\pi}$, 
where $A_{\rm Herschel}^{A_{\rm V}>8}$ is the projected area of each observed cloud above $A_{\rm V}$ = 8. 
The total virial mass, $M_{\rm VIR,H^{13}CO^+}^{A_{\rm V}>8}$, was then estimated from the scaled velocity width 
($dV_{\rm VIR,mol}^{A_{\rm V}>8}$) using Eq.~(\ref{eq:virial_mass}).

The mass of each cloud was also estimated from HGBS data as

\begin{equation}
M_{\rm X} [M_{\odot}] = \Sigma_{\rm Herschel} A_{\rm Y}, \\
\end{equation}

\begin{equation}
\Sigma_{\rm Herschel} = N({\rm H_2}) m_{\rm H} \mu_{\rm H_2},
\end{equation}

\noindent where $M_{\rm X}$ (either $M_{\rm Herschel}^{\rm mol-detect}$ or $M_{\rm Herschel}^{A_{\rm V}>8}$)
is the mass integrated over the area of the corresponding {\it Herschel} column density map 
where (H$^{13}$CO$^+$ or H$^{13}$CN) line emission was detected or where $A_{\rm V} > 8$. 
$A_{\rm Y}$ is the surface area (either $A_{\rm mol}^{\rm detect}$ or $A_{\rm Herschel}^{A_{\rm V}>8}$), $m_{\rm H}$  is the hydrogen atom mass, 
and $\mu_{\rm H_2}$ = 2.8 is the mean molecular weight per ${\rm H_2}$ molecule. 
The uncertainties in $M_{\rm Herschel}^{\rm mol-detect}$ and $M_{\rm Herschel}^{A_{\rm V}>8}$ are typically a factor 2, 
mainly due to uncertainties in the dust opacity \citep[cf.][]{Roy14}. 
The total gas masses derived from {\it Herschel}, $M_{\rm Herschel}^{\rm H^{13}CO^+-detect}$ and $M_{\rm Herschel}^{A_{\rm V}>8}$, 
range from 6.5 $M_{\odot}$ to 707 $M_{\odot}$ and from 16 $M_{\odot}$ to 954 $M_{\odot}$, respectively.

Table \ref{table:virial_h13cop} and Table \ref{table:virial_h13cn} 
also include estimates of 
the virial mass ratios, $\mathcal{R}_{\rm VIR,mol}^{\rm detect}$ and $\mathcal{R}_{\rm VIR,mol}^{A_{\rm V}>8}$, defined as

\begin{equation}
\mathcal{R}_{\rm VIR,mol}^{\rm detect} = \frac{M_{\rm VIR,mol}^{\rm detect}}{M_{\rm Herschel}^{\rm mol-detect}},  
\end{equation}

\noindent and

\begin{equation}
\mathcal{R}_{\rm VIR,mol}^{A_{\rm V}>8} = \frac{M_{\rm VIR,mol}^{A_{\rm V}>8}}{M_{\rm Herschel}^{A_{\rm V}>8}}.  
\end{equation}

\noindent 
The $\mathcal{R}_{\rm VIR,H^{13}CO^{+}}^{\rm detect}$ and $\mathcal{R}_{\rm VIR,H^{13}CO^+}^{A_{\rm V}>8}$ ratios 
range from $\sim$0.30 to $\sim$1.1 and from $\sim$0.40 to $\sim$2.0, respectively, 
suggesting that {all the dense clumps we observed} are gravitationally bound, especially the portions detected in H$^{13}$CO$^+$. 
Figure~\ref{Fig-Mvir_Mherschel}~(a) plots $M_{\rm VIR,H^{13}CO^+}^{A_{\rm V}>8}$ against $M_{\rm Herschel}^{A_{\rm V}>8}$, 
and shows that these two estimates of the mass of dense ($A_{\rm V} > 8$) gas agree to within better than 50\% in each region 
(see also Table \ref{table:virial_h13cop}).

We also estimated $M_{\rm VIR,H^{13}CN}^{\rm detect}$ and $M_{\rm VIR,H^{13}CN}^{A_{\rm V}>8}$ for the regions 
where H$^{13}$CN emission was detected, i.e.,  Aquila/W40, Aquila/Serp. South, Oph/main,  Orion B/NGC2024, Orion B/NGC2068, and Orion B/NGC2071. 
The virial mass ratios $\mathcal{R}_{\rm VIR,H^{13}CN}^{\rm detect}$ and $\mathcal{R}_{\rm VIR,H^{13}CN}^{A_{\rm V}>8}$ range  
from 0.3 to 2.5 and from 0.5 to 2.9, respectively (see also Fig. \ref{Fig-Mvir_Mherschel}(b)).

%%%%%%%%%%%%%%%%%%%%%%%%%
%   Result of HCN & HCO+
%%%%%%%%%%%%%%%%%%%%%%%%%

%Figure 3
\begin{figure*}
\begin{center}
\includegraphics[width=160mm, angle=0]{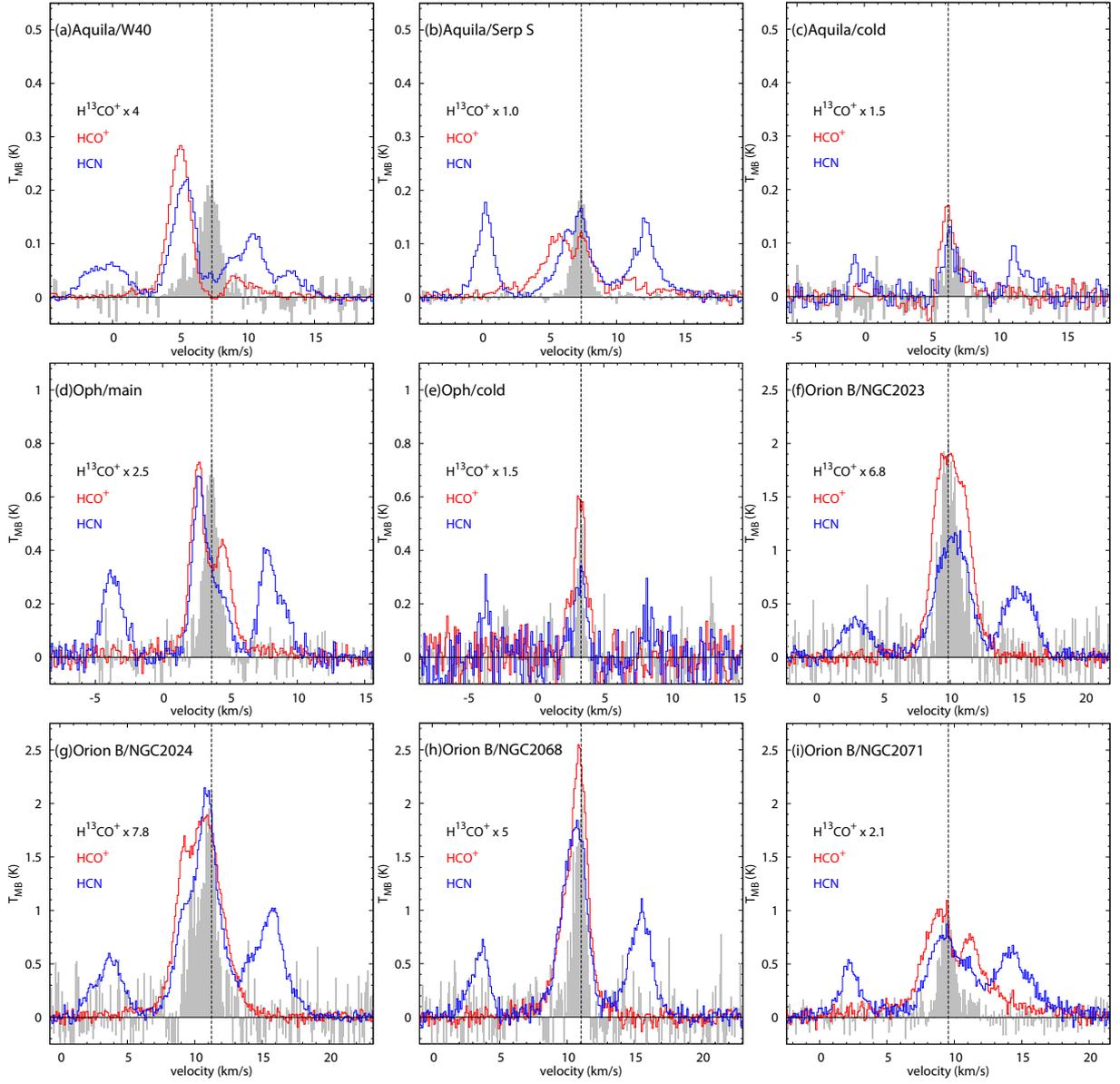}
\caption{Comparison of the HCN, HCO$^+$, and H$^{13}$CO$^+$(1-0) spectra averaged over the observed area in (a) Aquila/W40, (b) Aquila/Serp--South, (c) Aquila/cold, (d) Oph/main, (e) Oph/cold, (f) Orion B/NGC2023, (g) Orion B/NGC2024, (h) Orion B/NGC2068, and (i) Orion B/NGC2071. In each panel, red, blue, and gray lines show the mean HCN, HCO$^+$, and H$^{13}$CO$^+$(1-0) spectra in the corresponding subregion; the vertical dashed line marks the peak velocity of the H$^{13}$CO$^+$(1-0) line.}
\label{fig_spectra}
\end{center}
\end{figure*}

%
% This is produced onWed Nov 23 14:31:41 2016
% using Table_HCN-HCOp_ratio.pro
%./HCN-HCO+/aquila_5sigma.txt
%./HCN-HCO+/oph_5sigma.txt
%./HCN-HCO+/ngc2023_5sigma.txt
%./HCN-HCO+/ngc2024_5sigma.txt
%./HCN-HCO+/ngc2068_5sigma.txt
%./HCN-HCO+/ngc2071_5sigma.txt
\begin{table}
\centering
\begin{threeparttable}
%\caption{Intensity Ratio of HCN (1--0) to HCO$^+$ (1--0)}  \label{table:hcn_hcop}
\caption{HCN (1--0) to HCO$^+$ (1--0) line intensity ratios}  \label{table:hcn_hcop}
\begin{tabular}{lcc}
\hline
Region  & Median & Standard Deviation    \\
\hline
Aquila&   1.6&   1.5\\
Oph&   1.3&   0.9\\
Orion B/NGC2023&   0.7&   0.2\\
Orion B/NGC2024&   1.1&   0.3\\
Orion B/NGC2068&   1.0&   0.3\\
Orion B/NGC2071&   0.9&   0.2\\
\hline
\end{tabular}
\end{threeparttable}
\end{table}
 %Table 5

\subsection{HCN (1--0) \& HCO$^+$ (1--0) Emission}\label{sect:HCN-HCO+}

In contrast to the H$^{13}$CO$^+$ emission, the spatial distributions of HCN(1--0) and HCO$^{+}$(1--0) emission differ significantly from 
the column density distribution derived from {\it Herschel} data, as shown in Fig. \ref{fig1} and Figs. \ref{fig1_cold}--\ref{fig_ngc2071_maps}. 
The HCN(1--0) and HCO$^+$(1--0) maps appear to trace more extended regions than the denser filaments traced in 
the {\it Herschel} column density maps. In the Aquila cloud, the HCN(1--0) and HCO$^+$ (1--0) integrated intensities are strongest 
toward the W40 H{\small II} subregion, while column density is highest in the Serpens South subregion.  
The W40 H{\small II} region is known to be excited by the luminous stars IRS 1A North of spectral type O9.5 and IRS 1A South of spectral type B1V \citep{Shuping12}. 
In the Ophiuchus cloud, the HCN(1--0) and HCO$^+$(1--0) integrated intensities tend to be strong around the compact H{\small II} region excited by the B3 star S1 \citep{Grasdalen73,Andre88}. These findings suggest that  the HCN(1--0) and HCO$^+$(1--0) intensities depend on the strength of the local far-UV (FUV) 
radiation field.

Figure \ref{fig_spectra} shows comparisons of the mean H$^{13}$CO$^+$(1--0), HCO$^{+}$(1--0) and HCN(1--0) spectra observed toward each region. 
Clear dips in the HCN(1--0) and HCO$^{+}$(1--0) spectra can be seen at $V_{\rm LSR}$ $\sim$7, 4, 10 km s$^{-1}$ in Aquila/W40, Oph/main, 
and Orion B/NGC2071, respectively.
Furthermore, the velocities of these dips coincide with the peak velocities of the H$^{13}$CO$^+$(1--0) spectra. 
This suggests that the HCN(1--0) and HCO$^{+}$(1--0) spectra are strongly affected by self-absorption effects. 
In these subregions, the blueshifted components of the HCN(1--0) and HCO$^+$(1--0) spectra 
are stronger than the redshifted components. 
This type of asymmetric spectral shape, known as blue-skewed asymmetry, 
suggests the presence of infalling motions in the clouds \citep[cf.][]{Myers96,Schneider10}. 
Thus, the Aquila/W40, Oph/main, and Orion B/NGC2071 clumps may be undergoing large-scale collapse.

{Assuming the same excitation temperature for the two isotopic species and a $^{12}$C/$^{13}$C isotopic ratio 
$R_{\rm i} = 62$ 
\citep{Langer93}, we also estimate the optical depth of HCN(1--0) and HCO$^{+}$(1--0) as follows:}

{
\begin{equation}\label{eq:optical_depth}
\frac{T_{\rm peak}({\rm i})}{T_{\rm peak}({\rm n})} = \frac{1-e^{-\tau({\rm n})}/R_{\rm i}}{1-e^{-\tau({\rm n})}}, 
\end{equation}
}

{\noindent where $T_{\rm peak}({\rm i})$ is the peak intensity of the rare isotopic species [H$^{13}$CN(1--0) or H$^{13}$CO$^{+}$(1--0)] derived from mean spectra averaged over the observed area (see Fig. \ref{fig_spectra}), whereas $T_{\rm peak}({\rm n)}$ and $\tau({\rm n})$ are the peak intensity and optical depth of the normal species [HCN(1--0) or HCO$^{+}$(1--0)] 
at the peak velocity of the rare isotopic species in the averaged spectra. 
In all observed regions, the HCN(1--0) and HCO$^+$(1--0) lines are optically thick (see also Table~\ref{table:opacity}).

Recently, \citet{Brain16} observed a significant dependence of the $I_{\rm HCN}/I_{\rm HCO^+}$ intensity ratio on metallicity, 
with $I_{\rm HCN}/I_{\rm HCO^+}$ increasing from $\sim$1/4 at 0.3 solar metallicity to $\sim$1 at solar metallicity 
among galaxies of the local group.
In the nearby clouds observed here, the median $I_{\rm HCN}/I_{\rm HCO^+}$ intensity ratio ranges 
from 0.7 to 1.6 (see Table \ref{table:hcn_hcop}), consistent with the solar or near solar metallicity of these clouds.

%%%%%%%%%%%%%%%%%%%%%%%%%%%%%%%%
%   3.3 FUV 
%%%%%%%%%%%%%%%%%%%%%%%%%%%%%%%%
\subsection{Estimating the strength of the FUV radiation field }\label{ap:FUV}

 The FUV field strength, $G_0$, can be derived from {\it Herschel} 70 $\mu$m and 100 $\mu$m photometric data using the following equations
\citep{Kramer08, Schneider16}: 

\begin{equation}\label{eq:G0_FIR}
G_0 = \frac{4\pi I_{\rm FIR}} {1.6\times 10^{-3} [{\rm erg\ cm^{-2}\ s^{-1}}]} \ [{\rm in\ Habing\ units}]
\end{equation}

\begin{equation}\label{eq:G0_IFIR}
\begin{split}
I_{\rm FIR} &= \left(\frac{F_{\rm 70 \mu m}}{{[{\rm erg\ cm^{-2}\ s^{-1} Hz^{-1} sr^{-1}}}]} \times \frac{B_{\rm 60 \mu m - 80 \mu m}}{{\rm [Hz]}}  \right) \\ 
&+ \left(\frac{F_{\rm 100 \mu m}}{{[{\rm erg\ cm^{-2}\ s^{-1} Hz^{-1} sr^{-1}}]}} \times \frac{B_{\rm 80 \mu m - 125 \mu m}}{{\rm [Hz]}}  \right) \ [{\rm erg\ cm^{-2}\ s^{-1} sr^{-1}}],
\end{split}
\end{equation}

\noindent where $I_{\rm FIR}$ is the far-infrared (FIR) intensity, and $B_{\rm 60 \mu m - 80 \mu m}$ and $B_{\rm 80 \mu m - 125 \mu m}$ are
the bandwidths of the {\it Herschel}/PACS broad-band filters at 70 $\mu$m and 100 $\mu$m, respectively. 
{The $G_0$ values are in units of the local interstellar radiation field \citep{Habing68}.}

According to \citet{Hollenbach91}, $G_0$ can also be estimated from 
the dust temperature $T_{\rm dust}$ based on the following relations:

\begin{equation}\label{eq:Td_G0}
\begin{split}
T_{\rm d} & = (8.9 \times 10^{-11} \nu_0 G_0 \exp(-1.8 A_{\rm V}) \\
&\quad + 2.7^5 + 3.4 \times 10^{-2} [0.42 - \ln(3.5 \times 10^{-2} \tau_{100} T_0])\\
&\quad \times \tau_{100} T_0^6)^{0.2}\  [{\rm K}]
\end{split}
\end{equation}

\begin{equation}\label{eq:T_-}
T_0 = 12.2 G_0^{0.2} \   [{\rm K}]
\end{equation}

\begin{equation}\label{eq_tau}
\tau_{100} = \frac{2.7 \times 10^2}{T_0^5} G_0
\end{equation}

\noindent where, $\tau_{100}$, $T_{0}$, and $\nu_0$ are the effective optical depth at 100$\mu$m, the dust temperature of the cloud surface, and the frequency at 0.1$\mu$m.

Figure~\ref{fig:Td-G0} shows the pixel-to-pixel correlation between the $G_0$ values estimated 
in our target clouds from {\it Herschel} 70 $\mu$m and 100 $\mu$m data using Eqs.~(\ref{eq:G0_FIR}--\ref{eq:G0_IFIR}) 
and the $G_0$ values estimated from the  {\it Herschel} $T_{\rm dust}$ maps 
using Eqs.~(\ref{eq:Td_G0}--\ref{eq_tau}). 
{
The best-fit results are $G_0$($T_{\rm dust}$) = 1.38$\times$$G_0$(70,100$\mu$m) for Aquila,  $G_0$($T_{\rm dust}$) = 0.62$\times$$G_0$(70,100$\mu$m) for Oph, and  $G_0$($T_{\rm dust}$) = 2.60$\times$$G_0$(70,100$\mu$m) for Orion B.  In Orion B, the highest $G_0$ values come from pixels in NGC2024 and affect the best-fit results. 
The best-fit results for pixels with $G_0$(70,100$\mu$m) < 100 is $G_0$($T_{\rm dust}$) = 1.04$\times$$G_0$(70,100$\mu$m).  
In summary, our two estimates of $G_0$ generally agree to within a factor 2 to 3. }
{\citet{Pety16} also estimated the strength of the FUV radiation field 
toward NGC 2023/2024 in Orion B 
using Eq. (\ref{eq:T_-}).  {Their estimate agrees with our $G_0$(70,100$\mu$m) value  within 30\%}.}

%%%%%%%%%%%%%%%%%%%%%%%%%
%  Discussion
%%%%%%%%%%%%%%%%%%%%%%%%%

%Figure 4
\begin{figure}
\begin{center}
\includegraphics[width=90mm, angle=0]{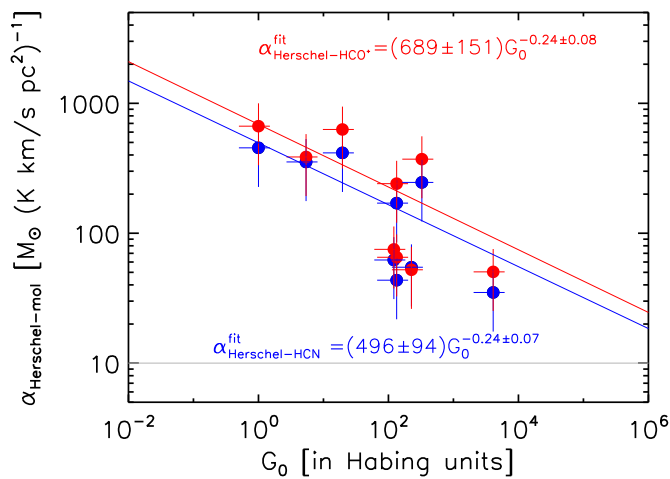}
\caption{Correlations between $\alpha_{\rm Herschel-HCN}$ and $G_0$ (blue line and filled circles) and 
between $\alpha_{\rm Herschel-HCO^+}$ and $G_0$ (red line and filled circles).  
The blue and red lines correspond to the best-fit relations: 
$\alpha_{\rm Herschel-HCN}^{\rm fit}$ = (496$\pm$94)$\times$$G_0^{-0.24\pm0.07}$ [$M_{\odot}$  (K km s$^{-1}$ pc$^{2}$)$^{-1}$] 
and $\alpha_{\rm Herschel-HCO^+}^{\rm fit}$ = (689$\pm$151)$\times$$G_0^{-0.24\pm0.08}$ [$M_{\odot}$  (K km s$^{-1}$ pc$^{2}$)$^{-1}$].  
{The horizontal line marks $\alpha_{\rm GS04-HCN}=10$ [$M_{\odot}$  (K km s$^{-1}$ pc$^{2}$)$^{-1}$].}
The uncertainties in $\alpha_{\rm Herschel-HCN}$ and $\alpha_{\rm Herschel-HCO^+}$ as estimated from the $M_{\rm Herschel}^{A_{\rm V}>8}$ uncertainties 
are a factor of 2 \citep{Roy14}. The uncertainties in $G_0$ are also a factor of 2 (see Sect. \ref{ap:FUV}).
}
\label{fig_alpha-G0}
\end{center}
\end{figure}

%Figure 5
\begin{figure}
\begin{center}
\includegraphics[width=80mm, angle=0]{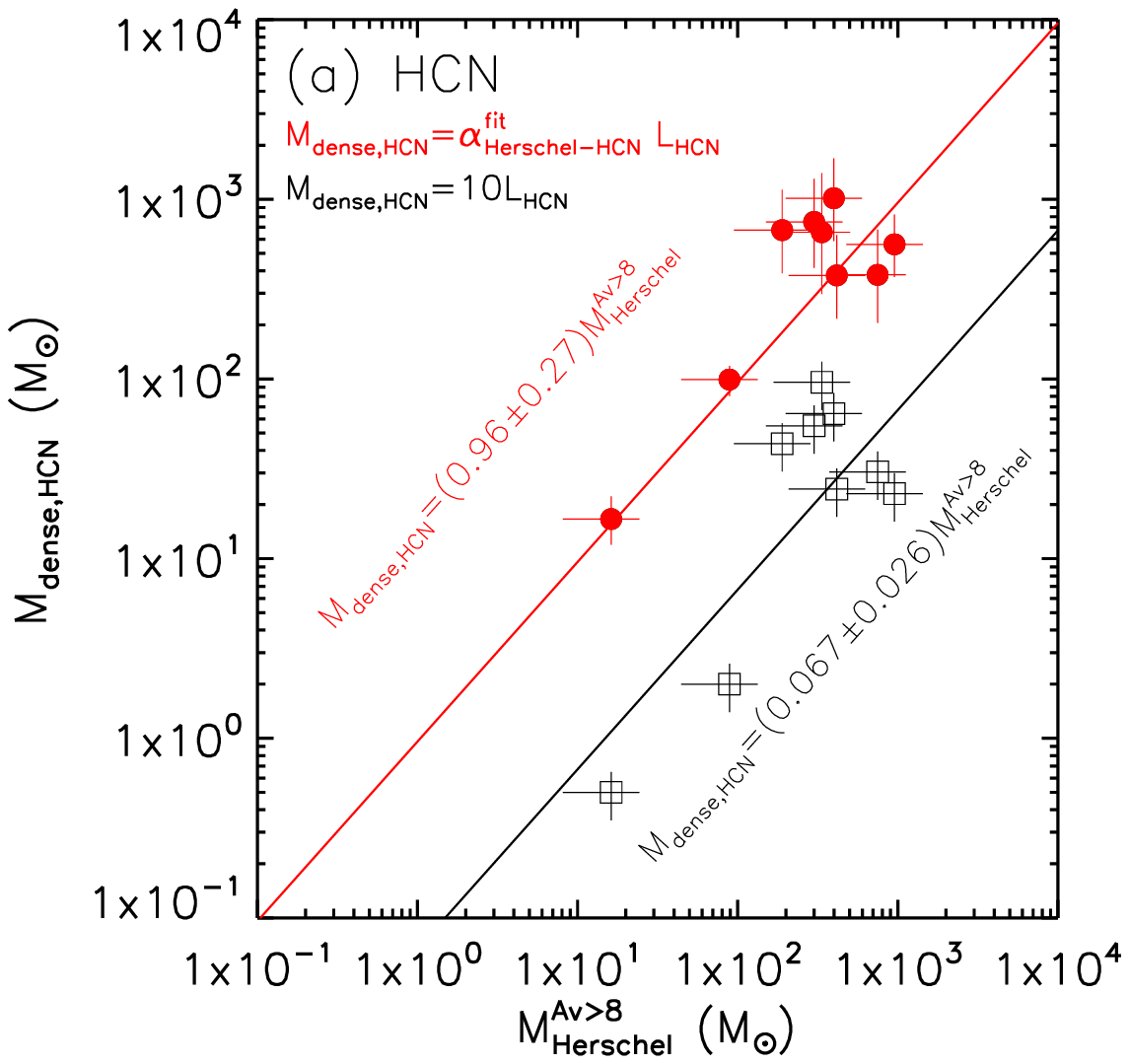} 
\includegraphics[width=80mm, angle=0]{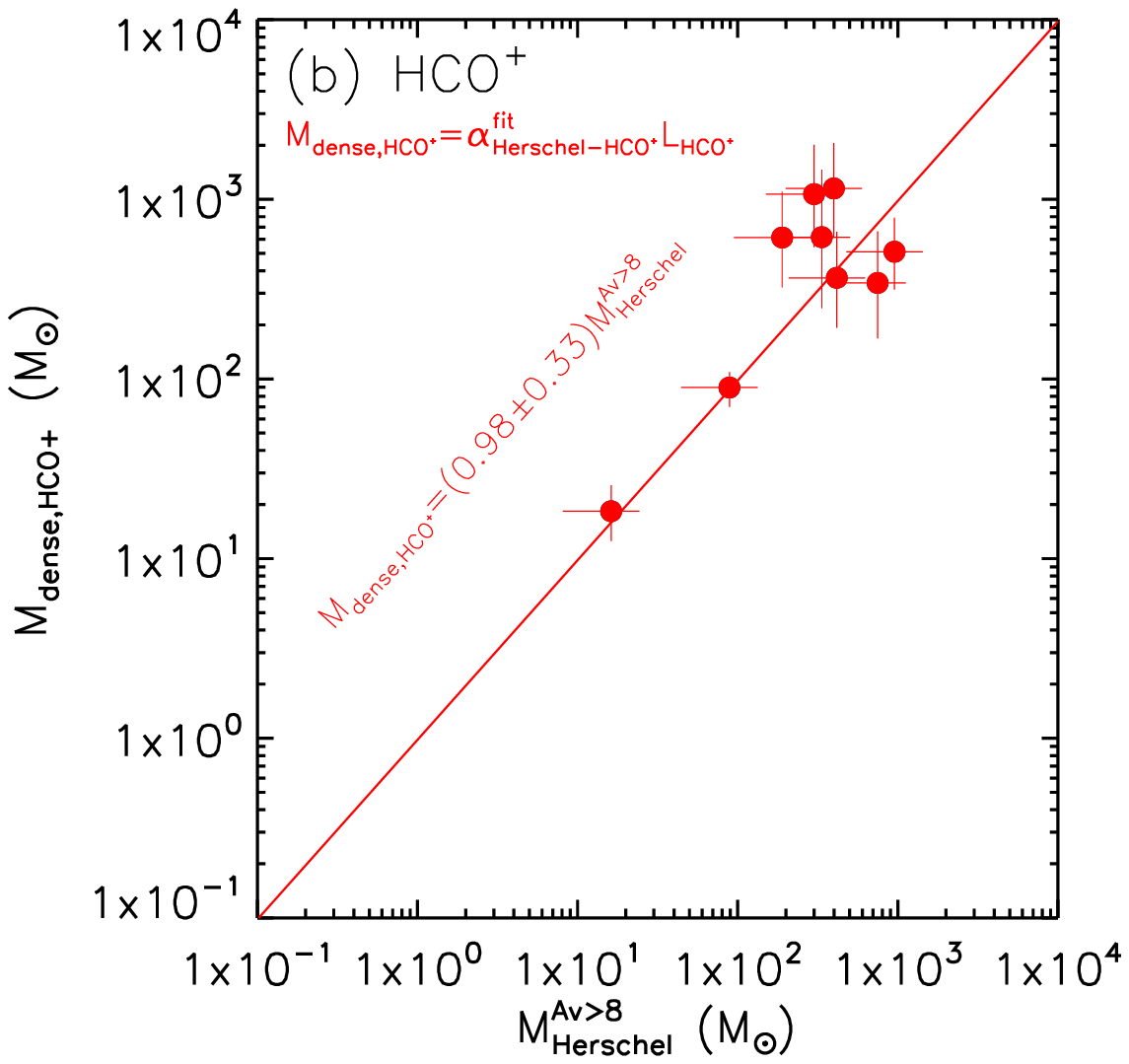} 
\caption{Plots of  (a) $M_{\rm dense,HCN}$ against $M_{\rm Herschel}^{A_{\rm V}>8}$, and (b) $M_{\rm dense,HCO^+}$ against $M_{\rm Herschel}^{A_{\rm V}>8}$. 
Red lines correspond to best-fit linear relations to the red data points: 
{$M_{\rm dense,HCN} = (0.96\pm0.27) \times M_{\rm Herschel}^{A_{\rm V}>8}$ and  $M_{\rm dense,HCO^+} = (0.98\pm0.33) \times M_{\rm Herschel}^{A_{\rm V}>8}$.} 
The red filled circles use $M_{\rm dense,mol}$ values derived from the relation $M_{\rm dense,mol} = \alpha_{\rm Herschel-mol}^{\rm fit} \times L_{\rm mol}$. 
In panel (a), the black open squares use $M_{\rm dense, HCN}$ values derived from the ``extragalactic'' 
relation $M_{\rm dense,HCN}$ = $\alpha_{\rm GS04-HCN} \times L_{\rm HCN}$, where $\alpha_{\rm GS04-HCN}$ = 10 $M_{\odot}$  (K km s$^{-1}$ pc$^{2}$)$^{-1}$; 
the black line corresponds to the best-fit linear relation to the open squares: 
$M_{\rm dense,HCN}$ ($\equiv$ $\alpha_{\rm GS04-HCN}$ $L_{\rm HCN}) = (0.067\pm0.026) \times M_{\rm Herschel}^{A_{\rm V}>8}$.} 
\label{Fig-Mvir_Mherschel2}
\end{center}
\end{figure}

\section{Discussion}

\begin{table*}
\centering
\scalebox{0.87}[0.87]{
\tiny
\begin{threeparttable}
%\caption{Parameters for the area where $A_{\rm V}$ is more than 8mag  \label{table:table_8mag}}
\caption{Derived parameters for the dense portions of the target nearby clouds where $A_{\rm V} > 8$ mag   \label{table:table_8mag}}
\begin{tabular}{lccccccccc}
\hline
Region & $M_{\rm Herschel}^{\rm map>8mag}$$^\dag$ & $G_0$  &  $A_{\rm Herschel}^{\rm map>8mag}$$^\dag$ & $I_{\rm HCN}$ & $L_{\rm HCN}$ & $\alpha_{\rm Herschel-HCN}$$^\dag$ & $I_{\rm HCO^+}$ & $L_{\rm HCO^+}$ & $\alpha_{\rm Herschel-HCO^+}$$^\dag$\\
 & [$M_{\odot}$] & & [pc$^2$] & [K km s$^{-1}$] & [K km s$^{-1}$ pc$^2$]& [$M_{\odot}$ (K km s$^{-1}$ pc$^2$)$^{-1}$] & [K km s$^{-1}$] & [K km s$^{-1}$ pc$^2$]& [$M_{\odot}$ (K km s$^{-1}$ pc$^2$)$^{-1}$] \\
\hline
Aquila/W40&   748 &   327&    2.68&    1.13&    3.04&   246&    0.75&
    2.01&   372\\
Aquila/Serp S&   954&    20&    2.51&    0.91&    2.30&   416&    0.60&
    1.52&   629\\
Aquila/cold&    89&     1&    0.40&    0.48&    0.20&   454&    0.33&
    0.13&   667\\
\hline
Oph/main (L1688)&   416&   134&    0.91&    2.68&    2.44&   170&    1.90&    1.73
&   241\\
Oph/cold&    16&     5&    0.06&    0.79&    0.05&   354&    0.73&    0.04
&   386\\
\hline
Orion B/NGC2023&   300&   226&    0.76&    7.21&    5.49&    55&    7.53&
    5.74&    52\\
Orion B/NGC2024&   335&  4091&    0.76&   12.57&    9.59&    35&    8.72&
    6.65&    50\\
Orion B/NGC2068&   190&   134&    0.52&    8.32&    4.36&    44&    5.54&
    2.90&    65\\
Orion B/NGC2071&   399 &   121 &    0.94&    6.79&    6.41&    62&    5.62&
    5.31&    75\\
\hline
\end{tabular}
\begin{tablenotes}
\item[$^\dag$] {See Table \ref{list_symbols} for the definition of each notation.}
\end{tablenotes}
\end{threeparttable}
}
\end{table*}
 %Table 6

%Figure 6
\begin{figure*}
\begin{center}
\includegraphics[width=120mm, angle=270]{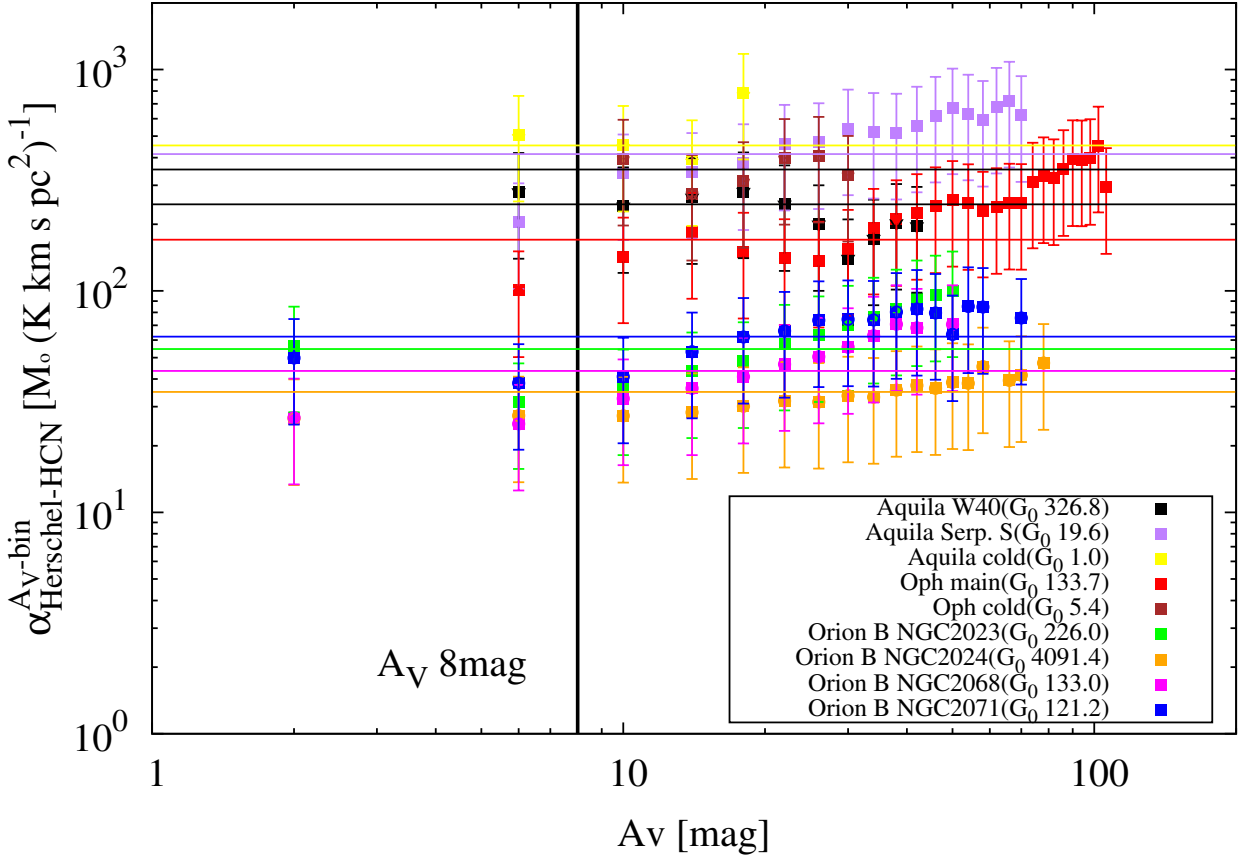}
\caption{Plot of the empirical conversion factor {$\alpha_{\rm Herschel-HCN}^{\rm A_V-bin} \equiv M_{\rm Herschel}^{\rm A_V-bin} \times L_{\rm HCN}^{\rm A_V-bin}$ 
against column density (expressed in $A_{\rm V}$ units) for the sub-regions of our sample. 
$M_{\rm Herschel}^{\rm A_V-bin}$ and $L_{\rm HCN}^{\rm A_V-bin}$ represent the mass derived from $Herschel$ column density data 
and the HCN luminosity for the portion of each sub-region corresponding to a given column density  bin. 
The width of each column density bin is 4 mag in $A_{\rm V}$ units. 
The horizontal lines indicate $\alpha_{\rm Herschel-HCN}$ for the various regions (see Table \ref{table:table_8mag}).}
}
\label{fig_alpha-Av}
\end{center}
\end{figure*}

\subsection{Evidence of large variations in the $\alpha_{\rm HCN}$ and $\alpha_{\rm HCO^+}$ conversion factors} \label{alpha_HCN}

In many extragalactic studies \citep[e.g.][]{Gao04b}, the mass of dense gas $M_{\rm dense}$ is estimated from the HCN(1--0) luminosity 
$L_{\rm HCN}$ using the relation $M_{\rm dense}$ = $\alpha_{\rm HCN}$$L_{\rm HCN}$
and assuming a fixed conversion factor $\alpha_{\rm HCN}$. 
Gao \& Solomon derived a simple formula for the conversion factor, namely 
$\alpha_{\rm GS04-HCN}$ = 2.1 $\sqrt{n({\rm H_2})}$/$T_{\rm b}$ = 10 $M_{\odot}$ (K km s$^{-1}$ pc$^2$)$^{-1}$, 
under the assumption that the HCN(1--0) emission originates from gravitationally-bound ``cores'' or clumps with  
volume-averaged density $n({\rm H_2}$)$\sim$3$\times$10$^4$ cm$^{-3}$ and brightness temperature $T_{\rm b}$$\sim$35 K. 
On this basis, they adopted the single value $\alpha_{\rm GS04-HCN}$ = 10 $M_{\odot}$ (K km s$^{-1}$ pc$^2$)$^{-1}$ 
in their seminal HCN study of galaxies \citep[][]{Gao04b}.
Clearly, however, if the brightness temperature of the HCN emitting clumps is larger than 35~K or if their volume-averaged density
is less than 3$\times$10$^4$ cm$^{-3}$, the $\alpha_{\rm GS04-HCN}$ factor can become smaller than 10 $M_{\odot}$ (K km s$^{-1}$ pc$^2$)$^{-1}$. 
Calibrating the conversion factor $\alpha_{\rm HCN}$ in Galactic clouds and assessing its potential variations is thus of crucial importance.
For a sample of massive Galactic clumps, \citet{Wu05} investigated the relationship between virial mass (estimated from C$^{34}$S observations)
and HCN luminosity 
and found a logarithmic mean value $\alpha_{\rm Wu05-HCN}$ = 7$\pm$2 $M_{\odot}$ (K km s$^{-1}$ pc$^2$)$^{-1}$ for the 
conversion factor. The fact that the $\alpha_{\rm Wu05-HCN}$ and $\alpha_{\rm GS04-HCN}$ values differ by only $\sim 30\%$ 
is very encouraging for extragalactic studies, but the HCN excitation conditions in the massive clumps studied by \citet{Wu05} 
are not necessarily representative of the bulk of the HCN-emitting dense gas in galaxies.

Here, we have both high-resolution HCN data and independent estimates of the mass of dense gas (from {\it Herschel} data) 
for a sample of nearby clouds/clumps spanning a broad range of radiation-field conditions, and are thus in a good position 
to calibrate the $\alpha_{\rm HCN}$ factor. 
To do so, we used the mass estimates derived from the {\it Herschel} column density maps,  $M_{\rm Herschel}^{A_{\rm V}>8}$,  
and the HCN(1--0) luminosities from the present observations to compute a 
$\alpha_{\rm Herschel-HCN} \equiv M_{\rm Herschel}^{A_{\rm V}>8}/L_{\rm HCN}$ factor for each cloud in our sample. 
As explained in Sect.~\ref{sect:Intro}, because most of the dense gas is distributed in filaments of $\sim 0.1\,$pc width, 
the $A_{\rm V} > 8$ level in {\it Herschel} column density maps of nearby molecular clouds is an excellent tracer of the 
gas denser than $n_{\rm H_2} \sim 2.3 \times10^4\, \rm{cm}^{-3}$, corresponding to supercritical, star-forming filaments \citep[cf.][]{Andre14}.  
The ${\rm H_2}$ volume density of $2 \times10^4\, \rm{cm}^{-3}$ is also very close to the typical gas density of $\simgt 3 \times 10^4 /\tau \, \rm{cm}^{-3} $ 
traced by the HCN(1--0) line in normal spiral galaxies \citep[][where $\tau \simgt 1$ is the optical depth of the line]{Gao04b} 
and lies between the effective excitation density (8.4$\times$10$^{3}$ cm$^{-3}$) and the critical density ($\simgt 3 \times$ 10$^{5}$ cm$^{-3}$) 
of HCN(1--0) \citep{Shirley15}. 
The masses $M_{\rm Herschel}^{A_{\rm V}>8}$ derived from {\it Herschel} data therefore provide good reference estimates 
of the mass of dense gas in nearby clouds. 
The estimated values of $\alpha_{\rm Herschel-HCN}$ 
range from $\sim$35 to $\sim$454 $M_{\odot}$ (K km s$^{-1}$ pc$^2$)$^{-1}$ 
(see Table \ref{table:table_8mag}). Clearly, large variations in $\alpha_{\rm Herschel-HCN}$ are present.

As described in Sect. \ref{sect:HCN-HCO+}, 
the HCN emission tends to be strong 
in areas where the FUV radiation field is strong. 
\citet{Meijerink07} demonstrated that the HCN 
emission is stronger in photon-dominated regions (PDRs) 
by a factor of two for densities larger than 10$^5$ cm$^{-3}$. Therefore, the variations we observe in $\alpha_{\rm Herschel-HCN}$ 
may be due to variations in the strength of the FUV field among the sub-regions.

The blue filled circles in 
Fig.~\ref{fig_alpha-G0} show a correlation plot between $\alpha_{\rm Herschel-HCN}$ and 
the mean FUV radiation field strength, $G_0$, 
estimated from {\it Herschel} 70 $\mu$m and 100 $\mu$m data (cf. Sect.~3.3). 
The correlation coefficient between the two variables is $-0.82$, showing the presence of a clear anti-correlation. 
Quantitatively, $\alpha_{\rm Herschel-HCN}$ decreases as $G_0$ increases according to the following empirical relation: 

\begin{equation}\label{Eq:alpha}
\alpha_{\rm Herschel-HCN}^{\rm fit} = (496\pm94) \times G_{\rm 0}^{-0.24\pm0.07} \ [M_{\odot} {\rm (K\ km\ s^{-1}\ pc^{2})^{-1}]}.
\end{equation}

Figure~\ref{Fig-Mvir_Mherschel2}(a) plots the mass of dense gas $M_{\rm dense,HCN}$ estimated from HCN for each cloud in our sample,
using both the standard extragalactic conversion factor $\alpha_{\rm GS04-HCN}$ [= 10 (K km s$^{-1}$ pc$^2$)$^{-1}$]  (black open squares)
and the conversion factor $\alpha_{\rm Herschel-HCN}^{\rm fit}$ from Eq. (\ref{Eq:alpha}) (red filled circles), 
as a function of the reference mass estimate $M_{\rm Herschel}^{A_{\rm V}>8}$. 
As can be seen, the $M_{\rm dense,HCN}$ values obtained with the $\alpha_{\rm GS04-HCN}$ conversion factor 
underestimate the reference masses $M_{\rm Herschel}^{A_{\rm V}>8}$ by an order of magnitude on average 
in nearby clouds. In contrast, the $M_{\rm dense,HCN}$ estimates using the $G_0$-dependent conversion factor 
$\alpha_{\rm Herschel-HCN}^{\rm fit}$ agree well with the reference dense gas mass estimates $M_{\rm Herschel}^{A_{\rm V}>8}$.

The HCO$^{+}$(1--0) line is another tracer of dense gas which can be used in extragalactic studies \citep[][]{Brain16}. 
Like the HCN emission, the HCO$^{+}$ emission tends to be strong in areas where the FUV radiation field is strong (see  Sect.~\ref{sect:HCN-HCO+}).
The red filled circles in Fig.~\ref{fig_alpha-G0} show a correlation plot between $\alpha_{\rm Herschel-HCO^+}$ and $G_0$. 
Here again, a clear anti-correlation is present with a correlation coefficient of $-0.77$. 
The $\alpha_{\rm Herschel-HCO^+}^{\rm fit}$ values are slightly larger than the $\alpha_{\rm Herschel-HCN}^{\rm fit}$ values, but 
the basic trend as a function of $G_{0}$ is the same. 
The $\alpha_{\rm Herschel-HCO^+}$ conversion factor decreases as $G_{0}$ increases according to the following best-fit relation:

\begin{equation}\label{Eq:alpha-hcop}
\alpha_{\rm Herschel-HCO^+}^{\rm fit} = (689\pm151) \times G_{0}^{-0.24\pm0.08} \ [M_{\odot} {\rm (K\ km\ s^{-1}\ pc^{2})^{-1}]}.
\end{equation}

\noindent 
Figure \ref{Fig-Mvir_Mherschel2} (b) plots the mass of dense gas $M_{\rm dense,HCO^+}$ derived from the HCO$^+$ luminosity 
using the conversion factor $\alpha_{\rm Herschel-HCO^+}^{\rm fit}$ from Eq. (\ref{Eq:alpha-hcop}) 
as a function of the reference mass estimate $M_{\rm Herschel}^{A_{\rm V}>8}$. 
As can be seen, the $M_{\rm dense,HCO^+}$ estimates using 
$\alpha_{\rm Herschel-HCO^+}^{\rm fit}$ 
agree well with the reference dense gas mass estimates $M_{\rm Herschel}^{A_{\rm V}>8}$.

We conclude that both the HCN(1--0) luminosity and the HCO$^{+}$(1--0) luminosity can be used as reasonably good tracers 
of the total mass of dense gas down to molecular cloud scales $\simgt 1\,$pc, 
{\it provided} that appropriate $G_0$-dependent conversion factors are employed (and the strength of the radiation field can be estimated). 
This is true despite the fact that the HCN(1--0) and HCO$^{+}$(1--0) lines are optically thick and do not trace well the details 
of the filamentary structure of the dense molecular gas (cf. Sect.~\ref{sect:HCN-HCO+}).

%%%%%%%%%%%%%%%%%%%%%%%%%%%%%%%%
% 4.2 Relationship between SFR and Mdense
%%%%%%%%%%%%%%%%%%%%%%%%%%%%%%%%

\subsection{Relationship between star formation rate and dense gas mass} \label{SFR-Mdense}

%
% This is produced onTue Nov 22 14:14:39 2016
% using Table_classII_SFR.pro
\begin{table}
\centering
\begin{threeparttable}
\caption{Number of {\bf YSOs} and star formation rate in each {\bf subregion}  \label{table:sfr}}
\begin{tabular}{lcc}
\hline
Region        & $N$({\bf YSOs})\tnote{$\dagger$} & SFR$_{\rm YSO}$ [$M_{\odot}$ yr$^{-1}$]\tnote{$\ddagger$}    \\
\hline
Aquila/W40&      88--115&  (22.0--28.8)$\times$10$^{-6}$ \\
Aquila/Serp S&      57--139&  (14.2--34.8)$\times$10$^{-6}$ \\
Aquila/cold&      11--15&   (2.8--3.8)$\times$10$^{-6}$ \\
\hdashline
Aquila (total)&     156--269&  (39.0--67.3)$\times$10$^{-6}$ \\
\hline
Oph/main (L1688)&      60--112&  (15.0--28.0)$\times$10$^{-6}$ \\
Oph/cold&       0--0&---- \\
\hdashline
Oph (total)&      60--112&  (15.0--28.0)$\times$10$^{-6}$ \\
\hline
Orion B/NGC2023&       {9--15}&   {(2.2--3.8)}$\times$10$^{-6}$ \\
Orion B/NGC2024&      {37--55}&   {(9.2--13.8)}$\times$10$^{-6}$ \\
Orion B/NGC2068&       {6--12}&   {(1.5--3.0)}$\times$10$^{-6}$ \\
Orion B/NGC2071&      26--43&   (6.5--10.8)$\times$10$^{-6}$ \\
\hdashline
Orion B (total)&      {78--125}&  {(19.5--31.3)}$\times$10$^{-6}$ \\
\hline
\end{tabular}
\begin{tablenotes}
\item[$\dagger$] {Number of YSOs in each subregion observed 
in HCN, HCO$^{+}$, H$^{13}$CN, and H$^{13}$CO$^{+}$.
The lower value gives the number of Class II YSOs, the upper value 
%gives  
the combined number of  Class 0/I, Flat, and Class II YSOs.}
\item[$\ddagger$] The uncertainties are $\sqrt{N}$ statistical uncertainties, where $N$ is the number count.
\end{tablenotes}
\end{threeparttable}
\end{table}
 %Table 7

In this subsection, we use our results on nearby clouds, e.g., our finding of a $G_0$-dependent conversion factor $\alpha_{\rm Herschel}(G_0)$,  
to make a new detailed comparison between the star formation efficiency within dense molecular gas 
found in nearby star-forming complexes on one hand and in external galaxies on the other hand.

%%%%%%%%%%%%%%%%%%%%%%%%%%%%%%%%
% 4.2.1 SFR estimate
%%%%%%%%%%%%%%%%%%%%%%%%%%%%%%%%

\subsubsection{Estimates of the star formation rate and star formation efficiency in the observed clouds} \label{estimateSFR}

We first estimated the star formation rate (SFR) in each of our target regions/sub-regions 
from direct counting of young stellar objects (YSOs) using  
the available {\it Spitzer} census of YSOs in nearby clouds \citep{Evans09}. 
Under the assumption that the median lifetime of Class~II YSOs is 2~Myr \citep{Evans09, Covey10, Lada10,Dunham15} 
and that their {mean} mass is 0.5 $M_{\odot}$ \citep{Muench07}, 
the SFR can be derived from the number of YSOs observed with Spitzer as follows:

\begin{equation}
{\rm SFR_{\rm YSO}} = 0.25 \times N({\rm YSOs}) \times 10^{-6} \  [M_{\odot} {\rm yr}^{-1}].
\end{equation}

To evaluate the number of  {YSOs} in each subregion observed here, 
we used the {\it Spitzer} catalog of \citet{Dunham15} for Aquila and Ophiuchus, and the catalog of \citet{Megeath12} for Orion B.  
{To count YSOs at each evolutionary stage from these catalogs, we selected objects with an infrared spectral 
index\footnote{The spectral index $a_{\rm IR}$ is 
defined as the slope of the near-/mid-infrared spectral energy distribution (SED), $a_{\rm IR}$ = $d \log(\lambda S_{\lambda})/d \log(\lambda)$, 
where $\lambda$ and $S_{\lambda}$ denote wavelength and flux density at that wavelength, respectively.}  $a_{\rm IR}$ in the 
ranges $0.3 \le a_{\rm IR}$ for Class 0/I YSOs, $-0.3 \le a_{\rm IR} < 0.3$ for Flat Spectrum sources, and $-1.6 \le a_{\rm IR} < -0.3$ 
for Class II objects in agreement with 
%{\bf 
standard YSO classification criteria \citep{Greene94}. }

\citet{Dunham15} used 2MASS and {\it Spitzer} data between 2 $\mu$m and 24 $\mu$m. 
\citet{Megeath12} used {\it Spitzer} data between 4.5 $\mu$m and 24 $\mu$m. 
While \citet{Dunham15} published $a_{\rm IR}$ values both before and after correction for extinction, 
\citet{Megeath12} used uncorrected $a_{\rm IR}$ values. 
For the sake of consistency, we used uncorrected $a_{\rm IR}$ values to select Class~II YSOs in all regions. 
The {resulting YSO number counts} and corresponding SFRs are reported in Table \ref{table:sfr} for each region/sub-region observed in molecular lines.
The blue filled circles in Fig.~\ref{fig_SFR-Mdense} (a) show the correlation between SFR$_{\rm YSO}$ and $M_{\rm Herschel}^{A_{\rm V}>8}$ for the clouds of our sample, 
which can be expressed as {SFR$_{\rm YSO}$ = (2.1--5.0)$\times$10$^{-8}$ $M_{\odot}$ yr$^{-1}$ $\times$($M_{\rm Herschel}^{A_{\rm V}>8}$/$M_{\odot}$)}.
This relationship is in good agreement with previous studies of nearby Galactic clouds using extinction maps to estimate the mass of dense gas 
\citep[][]{Lada10,Lada12, Evans14}.

%%%%%%%%%%%%%%%%%%%%%%%%%%%%%%%%%%%%%%%%%%%%%%%%%%%%
% 4.2.2 Calibration of the dense gas mass in external galaxies
%%%%%%%%%%%%%%%%%%%%%%%%%%%%%%%%%%%%%%%%%%%%%%%%%%%%
\subsubsection{Calibration of the dense gas mass in external galaxies }\label{Sect:calibration}

In Sect.~\ref{alpha_HCN}, we showed that reliable HCN-based estimates of the dense gas mass in nearby clouds could be obtained 
using the $G_0$-dependent conversion factor $\alpha_{\rm Herschel-HCN}^{\rm fit}(G_0)$ given by Eq.~(\ref{Eq:alpha}). 
In this subsection, we try to account for this $G_0$ dependence in galaxies with published HCN data 
in an effort to improve the current estimates of their dense gas masses. 
Before doing so, it is important to notice that the $\alpha_{\rm Herschel-HCN}^{\rm fit}(G_0)$ values derived for the nearby clouds 
of our sample are a factor of $\simgt 3$ to $\sim 50$ higher than the standard extragalactic conversion factor 
$\alpha_{\rm GS04-HCN}$ = 10 $M_{\odot}$ (K km s$^{-1}$ pc$^2$)$^{-1}$ (see Fig.~\ref{fig_alpha-G0}). 
Since the mean $G_0$ value in a typical galaxy cannot be much higher than the highest $G_0$ found in our regions ($G_0 \sim 4000$ for NGC2024),
another effect besides the $G_0$ dependence must explain this large difference in conversion factor. 

While our HCN/HCO$^+$ observations were specifically focused on the densest ($A_{\rm V}$ > 8) portions of the target clouds, 
they also cover a small fraction of the lower density gas in these clouds, and our data clearly show that the HCN(1--0) 
and HCO$^{+}$(1--0) lines are tracing molecular gas down to much lower column densities than the $A_{\rm V} = 8$ fiducial limit. 
In a recent  independent molecular line study of the southern part of the Orion~B cloud (including NGC2023 and NGC2024) with the IRAM 30m telescope, 
\citet{Pety16} found that 36\% of the HCN(1--0) line flux was emitted from low extinction (2 $\le$ $A_{\rm V}$ $<$ 6) areas,  
with $> 98\% $ of the line flux coming from $A_{\rm V}$ $\ge$ 2 areas. {Likewise, the results of the CHaMP census of molecular clumps in the southern Milky Way with the MOPRA telescope \citep[e.g.][]{Barnes2016} demonstrate that the HCO$^{+}$(1--0) line emission is generally tracing gas down to volume densities of $\sim 10^3\, {\rm cm}^{-3}$ or less.}
Dividing the regions and sub-regions observed in our study in various column density bins, we can investigate possible variations 
of the empirical conversion factor $\alpha_{\rm Herschel-HCN}^{\rm A_V-bin} \equiv M_{\rm Herschel}^{\rm A_V-bin} \times L_{\rm HCN}^{\rm A_V-bin}$ 
with column density (see Fig.~\ref{fig_alpha-Av}). 
Significant variations in $\alpha_{\rm Herschel-HCN}^{\rm A_V-bin}$ from region to region can be seen,  
in agreement with the dependence of $\alpha_{\rm Herschel-HCN}$ on the strength of the FUV radiation field discussed in Sect.~\ref{alpha_HCN}. 
Figure~\ref{fig_alpha-Av} nevertheless suggests that, in any given sub-region (i.e., for a given $G_0$ value), the conversion between gas mass 
and HCN(1--0) luminosity remains approximately constant, irrespective of gas (column) density or $A_{\rm V}$, down to $A_{\rm V} \sim 2$. 
This means that, in addition to the dense molecular gas $n_{\rm H_2} \simgt  2 \times10^4\, \rm{cm}^{-3}$ (corresponding to $A_{\rm V} \simgt 8$ in resolved nearby clouds),  
extragalactic HCN(1--0) studies with spatial resolutions from $\sim 8\,$pc to $\sim 36\,$kpc  \citep{Chin97, Chin98,Gao04a,Brouillet05, Buchbender13,Chen15, Chen16} 
must also probe essentially all of the molecular gas corresponding to $2 <  A_{\rm V} < 8$ in the observed galaxies. 
Using typical column density probability density functions (PDFs) observed toward Galactic molecular clouds \citep[e.g.][]{Schneider13}, 
we can estimate the relative amount of molecular gas at $A_{\rm V} > 8$ and $2 <  A_{\rm V} \le 8$ in GMCs, $M(A_{\rm V} >2)/M(A_{\rm V} >8)$.
In practice, we adopt the combined column density PDF derived from {\it Herschel} HGBS data in the Aquila and Orion~B complexes, 
which is well described by a power law, $dN/d{\rm log}N_{\rm H_2} \propto N_{\rm H_2}^{-2.7 \pm 0.2}$ \citep[][; K\"onyves et al., in prep.]{Konyves15}, 
as our template. Such a column density PDF implies that $M(A_{\rm V} >2)/M(A_{\rm V} >8) = (2/8)^{-1.7 \pm 0.2} \approx 10 $, at least in molecular clouds such as Aquila and Orion~B. 
Clearly, if extragalactic HCN measurements of dense gas include low-density gas down to $A_{\rm V} \sim  2$, 
then the effective $\alpha_{\rm HCN}$ conversion factor for these measurements must be an order of magnitude lower 
than the $\alpha_{\rm Herschel-HCN}^{\rm fit}(G_0)$ factor given by Eq.~(\ref{Eq:alpha}) and thus within a factor of a few 
of the standard extragalactic conversion factor $\alpha_{\rm GS04-HCN}$ = 10 $M_{\odot}$ (K km s$^{-1}$ pc$^2$)$^{-1}$.

%Figure 7
\begin{figure*}
\begin{center}
\includegraphics[width=91mm, angle=0]{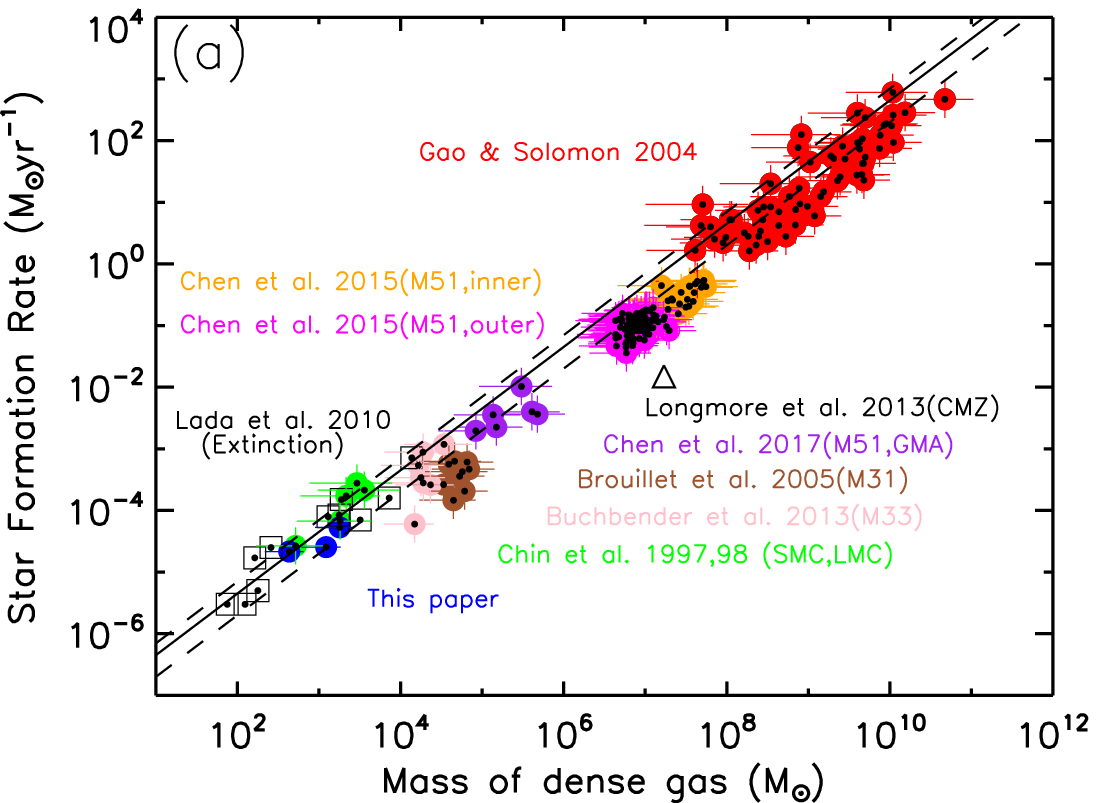}
\includegraphics[width=91mm, angle=0]{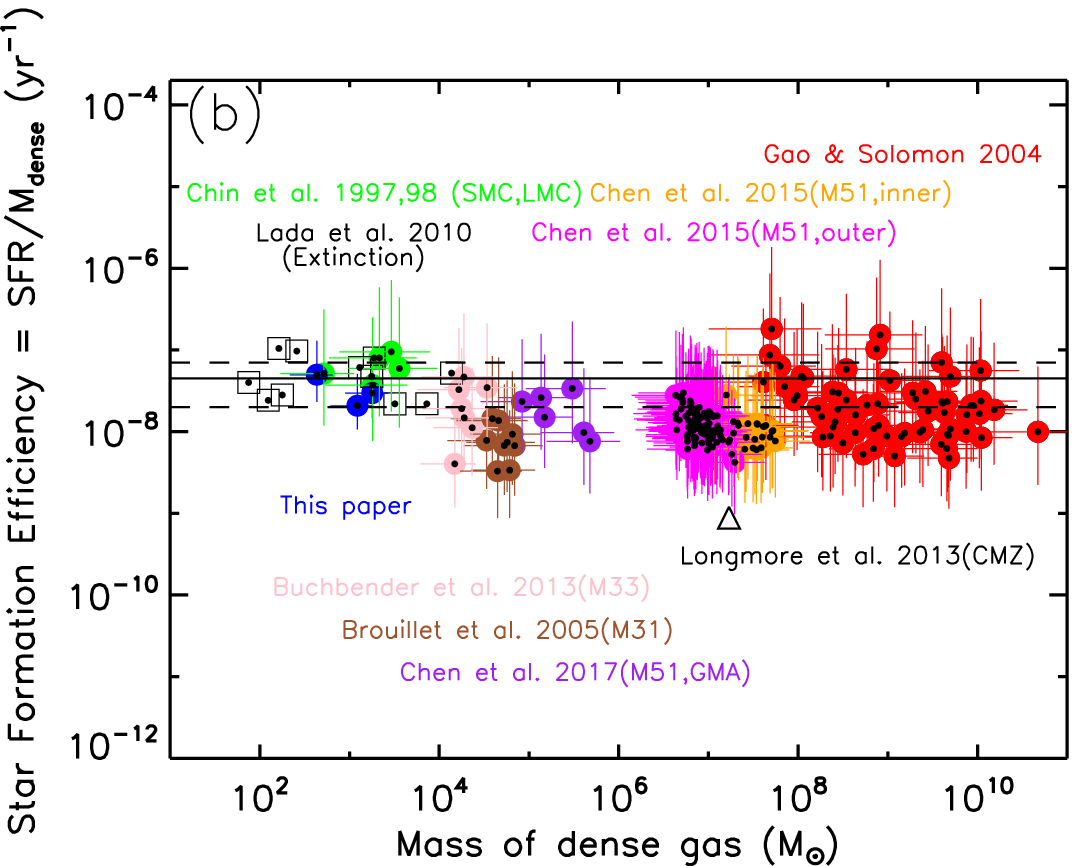}
\caption{Plots of (a) SFR against $M_{\rm dense}$ and (b) SFE against $M_{\rm dense}$. 
Blue filled squares indicate the SFR--$M_{\rm dense}$(= $M_{\rm Herschel}^{A_{\rm V}>8}$) and  SFE(= SFR/$M_{\rm dense}$) -- $M_{\rm dense}$ 
relations for the regions observed in this paper.
Black open circles indicate the relation between SFR and $M_{\rm dense}$ in nearby star-forming clouds \citep{Lada10}. 
Red filled circles  
indicate the SFR-$M_{\rm dense}$ and SFE -- $M_{\rm dense}$ relations obtained for external galaxies \citep{Gao04a}, 
assuming $M_{\rm dense}$ = $\alpha_{\rm HCN}$$\times$$L_{\rm HCN}$ with the corrected $\alpha_{\rm HCN}$ values from our Eq.~(\ref{Eq:alpha2}).  
Green, brown, pink, purple, magenta, and orange filled circles indicate the SFR-$M_{\rm dense}$ and  SFE -- $M_{\rm dense}$ relations in the SMC, LMC, M31, M33, and M51 galaxies \citep{Chin97,Chin98,Brouillet05,Buchbender13,Chen15,Chen16} using our relation of Eq.~(\ref{Eq:alpha2}).  
The uncertainties in the extragalactic SFR values are a factor of $\sim$2 \citep[cf.][]{Bell03}.
The black triangle indicates the position of the central molecular zone (CMZ: |$l$| < 1 deg, |$b$| < 0.5 deg) in each plot, 
where SFR and $M_{\rm dense}$ were estimated from the number count of massive YSOs and $Herschel$ column density, respectively \citep{Longmore13}.
In both panels, the black solid line and black dashed lines display the simple empirical  
relation expected from the ``microphysics" of prestellar core formation within filaments 
described in Sect.~\ref{Sect:origin} \citep[see also][]{Andre14,Konyves15}:
${\rm SFR} = (4.5\pm2.5)\times10^{-8}\,  M_\odot \, {\rm yr}^{-1}\, \times \left(M_{\rm dense}/M_\odot \right) $,  
i.e., SFE = (4.5$\pm$2.5)$\times$10$^{-8}\, {\rm yr}^{-1} $.
}
\label{fig_SFR-Mdense}
\end{center}
\end{figure*}

To go further and to improve current extragalactic $M_{\rm dense,HCN}$ estimates for radiation-field effects,  
we also need to account for the fact that the minimum column density or extinction $A_{V, {\rm min}}$(HCN) down to which significant HCN(1--0) line flux 
is emitted itself depends on $G_0$. 
The values of the effective excitation density of HCN(1--0) calculated by \citet{Shirley15} scale approximately as $n_{\rm eff}^{\rm HCN} \propto T_{\rm gas}^{-0.65}$, 
where $T_{\rm gas}$ is the gas kinetic temperature. 
Assuming $T_{\rm gas} \approx T_{\rm dust} $ and using the simplified relation between $T_{\rm dust} $ and $G_0$ given by Eq.~(\ref{eq:T_-}), 
this means that $n_{\rm eff}^{\rm HCN} \propto G_0^{-0.13}$. 
For both a spherical and a cylindrical cloud with a density gradient $\rho \propto r^{-\alpha} $, column density scales as  $r^{1-\alpha} $ or $\rho^{1-\frac{1}{\alpha}} $. 
Adopting $\alpha = 1.75 \pm 0.25$, a value intermediate between 1.5 (free-falling cloud onto a point mass) and 2 (isothermal spheres) 
which is also consistent with the logarithmic slope of $-2.7$ for the column density PDF \citep[cf.][]{Konyves15}, 
we may thus expect the minimum column density probed by HCN(1--0) to scale as $A_{V, {\rm min}}({\rm HCN}) \propto (n_{\rm eff}^{\rm HCN})^{0.43} \propto G_0^{-0.056 \pm 0.012}$. 
Normalizing this scaling relation using the fact that  $A_{V, {\rm min}}({\rm HCN}) \approx 2$ for NGC2023/2024 in Orion~B  where $G_0 \approx 20$ \citep{Pety16}, 
we obtain the following $\alpha_{\rm Herschel-HCN}^{\rm fit^\prime}$ conversion factor for external galaxies, under the assumption that extragalactic HCN observations 
sample all of the molecular gas down to $A_{V, {\rm min}}$(HCN):

\begin{equation}\label{Eq:alpha2}
\alpha_{\rm Herschel-HCN}^{\rm fit ^\prime} \approx 0.13^{+0.06}_{-0.06} \times G_{\rm 0}^{-0.095\pm 0.02}  \times \alpha_{\rm Herschel-HCN}^{\rm fit}  \propto G_{\rm 0}^{-0.34\pm0.08} 
\end{equation}

Using the total infrared luminosities $L_{\rm IR}$ listed in the \citet{Gao04a} paper  
and adopting $L_{\rm FIR}$ = $L_{\rm IR}/1.3$ \citep{Gracia08}}, 
we can derive 
relevant $G_{0}$ values from 
{Eq. (\ref{eq:G0_FIR}) and 
{$I_{\rm FIR}=L_{\rm FIR}/4\pi d^2(\frac{\pi \theta_{\rm beam}^2}{4\ln2})$, where $d$ and $\theta_{\rm beam}$ 
are the distance to each galaxy and the telescope beam size of the corresponding HCN measurement respectively
\citep[cf. ][]{Buchbender13}.
We can then estimate an effective conversion factor $\alpha_{\rm Herschel-HCN}^{\rm fit^\prime}(G_0)$ for each galaxy in the Gao \& Solomon sample.
The resulting $G_0$ and $\alpha_{\rm Herschel-HCN}^{\rm fit^\prime}$ values range 
from 44 to 3.9$\times$10$^{4}$ (mean: 1.7$\times$10$^{3}$) and from 1.8 $M_{\odot}$ (K km s$^{-1}$ pc$^2$)$^{-1}$ 
to 17.7 $M_{\odot}$ (K km s$^{-1}$ pc$^2$)$^{-1}$ (mean: 10.2 $M_{\odot}$ (K km s$^{-1}$ pc$^2$)$^{-1}$), respectively.
Using these $\alpha_{\rm Herschel-HCN}^{\rm fit^\prime}$ values for the HCN conversion factor, 
the corrected masses of dense gas range from 4.3$\times$10$^{7}$ $M_{\odot}$ to 4.9$\times$10$^{10}$ $M_{\odot}$. 
Under the assumption that the average gas densities of normal galaxies and extreme starbursts 
are 2$\times$10$^2$ cm$^{-3}$ and 1$\times$10$^5$ cm$^{-3}$, respectively, 
\citet{Garcia12} advocated revised conversion factors 
$\alpha_{\rm GB12-HCN}$$\sim$3--4 $M_{\odot}$(K km s$^{-1}$ pc$^2$)$^{-1}$ 
for normal galaxies ($L_{\rm IR}$ < 10$^{11}$ $L_{\odot}$ corresponding to SFR $<$ 2$\times$10 $M_{\odot}$yr$^{-1}$) 
and $\alpha_{\rm GB12-HCN}$$\sim$1--2 $M_{\odot}$(K km s$^{-1}$ pc$^2$)$^{-1}$ for extreme starbursts ($L_{\rm IR}$ > 10$^{11}$ 
$L_{\odot}$  corresponding to SFR $>$ 2$\times$10 $M_{\odot}$yr$^{-1}$).
The revised conversion factors $\alpha_{\rm GB12-HCN}$ of \citet{Garcia12} are roughly consistent with our suggested
$\alpha_{\rm Herschel-HCN}^{\rm fit^\prime}$ values of 1.8--17.7 $M_{\odot}$(K km s$^{-1}$ pc$^2$)$^{-1}$ for the galaxies sampled 
by \citet{Gao04a} ($L_{\rm IR}$$\sim$10$^{10-12}$ $L_{\odot}$). 
This agreement suggests that the dependence of the $\alpha_{\rm Herschel-HCN}$ conversion factor 
on the strength of the FUV radiation field is not restricted to Galactic clouds and also applies to external galaxies.
Figure~\ref{fig_SFR-Mdense} (a) compiles observations of the SFR--$M_{\rm dense}$ relation from the present study, \citet{Lada10}, and \citet{Gao04a}. 
The SFR-$M_{\rm dense}$ relation found in our study is consistent with that of \citet{Lada10}, as expected since both are focused on nearby Galactic clouds.
The initial SFR-$M_{\rm dense}$ relation from \citet{Gao04a} lies a factor of $\sim \,$2--3 below the best-fit relation found for nearby star-forming regions, 
namely SFR = 4.6$\times$10$^{-8}$ $M_{\rm dense}$ \citep{Lada10}.
Conversely, the corrected SFR-$M_{\rm dense}$ relation for the \citet{Gao04a} sample 
lies above the nearby-cloud relation by a factor of $\sim$3.
We also compiled data points for nearby galaxies of the local group, namely the Small Magellanic Cloud (SMC), the Large Magellanic Cloud (LMC), 
M31, M33, and M51 from \citet{Chin97,Chin98}, \citet{Brouillet05}, \citet{Buchbender13}, and \citet{Chen15,Chen16}. 
The SFR for each galaxy was estimated using SFR = 2.0$\times$10$^{-10}$($L_{\rm IR}$/$L_{\odot}$) $M_{\odot}$yr$^{-1}$  following \citet{Gao04a}. 
As can be seen in Fig.~\ref{fig_SFR-Mdense} (a), the observed trend is basically a linear relation 
between star formation rate and dense gas mass, and the nearby-cloud relation provides a good overall fit to most data points.

Figure~\ref{fig_SFR-Mdense} (b) provides a blow-up view of the scatter around the nearby-cloud star formation law
by plotting the star formation efficiency SFE$_{\rm dense}$ $\equiv$ SFR/$M_{\rm dense}$ against $M_{\rm dense}$ for the same objects as in Fig.~\ref{fig_SFR-Mdense} (a).
As can be seen, SFE$_{\rm dense}$ remains roughly constant within a scatter of less than 
1.5 orders of magnitude
over 8 orders of magnitude in $M_{\rm dense}$ from $\sim10^2\, M_{\odot}$ to $10^{10}\, M_{\odot}$. 
This scatter in SFE$_{\rm dense}$ is significantly smaller than the scatter in SFE$_{\rm total}$,  
defined as SFR divided by total molecular gas mass $M_{\rm total}$, observed 
among nearby clouds and massive Galactic clouds, which exceeds 2 to 3 orders of magnitude 
over only 4 orders of magnitude in gas mass 
\citep[cf.][]{Lada10,Vutisalchavakul16}.
{The mean and standard deviation of (the logs of) all the data points in Fig.~\ref{fig_SFR-Mdense}b (including the CMZ point) are $\langle$log SFE$_{\rm dense}$ (yr$^{-1}$)$\rangle$=-7.85$\pm$0.31. This is in excellent agreement with the results of \citet{Vutisalchavakul16} who found  $\langle$log SFE$_{\rm dense}$ (yr$^{-1}$)$\rangle$ =-7.74$\pm$0.50 toward star-forming regions in the Galactic plane using masses derived from the Bolocam Galactic Plane Survey (BGPS) data at 1.1 mm and SFR values estimated from $WISE$ mid-infrared data. }

The vertical scatter in Fig.~\ref{fig_SFR-Mdense}b can almost entirely be attributed to uncertainties 
in the $\alpha_{\rm HCN}$ conversion factor used to estimate the dense gas mass for galaxies. 
In nearby clouds, for which assumptions about this conversion factor are not needed, the observed scatter in SFE$_{\rm dense}$
is less than a factor of $\sim 3$.
To conclude, our results support the view that the relationship between star formation rate and dense gas mass, i.e., 
the star formation efficiency in the dense molecular gas of galaxies, is quasi-universal on a wide range of 
scales from $\sim 1$--10 pc to $>$ 10 kpc.

\subsection{Origin of the quasi-universal star formation efficiency in dense molecular gas}\label{Sect:origin}

On the grounds that filaments dominate the mass budget of dense molecular gas in GMCs and that most, if not all, prestellar cores form in filaments (see Sect.~1), 
\citet{Andre14} proposed that the quasi-universal star formation efficiency in dense gas discussed above and summarized in Fig.~\ref{fig_SFR-Mdense} 
originates from the physics of filament evolution and core formation along filaments.
Following \citet{Andre14}, we suggest that the star formation efficiency in dense gas, SFE$_{\rm dense} \equiv  {\rm SFR}/M_{\rm dense}$, 
is primarily set by three parameters characterizing dense cores along filaments, namely the typical fraction of dense filament gas in the form of prestellar cores
(or ``core formation efficiency''), 
$f_{\rm pre}$, 
the typical lifetime of prestellar cores, $t_{\rm pre}$, and the efficiency of the conversion from prestellar core mass to stellar mass (or stellar system mass), 
$\epsilon_{\rm core}$, according to the simple relation: SFE$_{\rm dense} = f_{\rm pre} \times \epsilon_{\rm core} / t_{\rm pre}$. 
This relation assumes that, in steady state, each supercritical filament 
has converted a fraction $f_{\rm pre}$ of its mass into prestellar cores, and that each prestellar core converts a fraction 
$\epsilon_{\rm core}$ of its mass into either a single star or a small stellar system on a timescale $t_{\rm pre}$. 
The latter assumption is supported by the similarity between the prestellar core mass function (CMF) and the stellar initial mass function (IMF) \cite[e.g.][]{Motte98, Enoch08,Andre10}.
Observationally, it seems that 
the three parameters $f_{\rm pre}$, $t_{\rm pre}$, $\epsilon_{\rm core}$ 
have reasonably constant values with little variations from cloud to cloud, at least in Gould Belt regions. 
Based on the results of the {\it Herschel} Gould Belt survey in the Aquila, Ophiuchus, Orion~B, and Taurus/L1495 clouds, the prestellar core formation efficiency 
in the dense ($A_{\rm V} > 8$) 
gas of supercritical filaments is estimated to be 
{$f_{\rm pre} = 15\%^{+5\%}_{-5\%}$} (e.g. \citealp{Konyves15}; K\"onyves et al.  in prep.; Ladjelate al. in prep.).
The typical lifetime of low- to intermediate-mass prestellar cores is known to be $t_{\rm pre} = 1^{+0.5}_{-0.3}\, $Myr \citep[e.g.][]{Lee99, Jessop00, Konyves15}. 
In reality, $t_{\rm pre} $ likely depends on both core density \citep{Jessop00} and core mass \citep{Hatchell08}, 
but what matters here is the characteristic lifetime of the bulk of prestellar cores near the peak of the CMF and forming in filaments just above 
the critical line mass $M_{\rm line, crit}$. 
The efficiency of the conversion process from core mass to stellar mass, as estimated from the global shift between the CMF and the IMF 
is believed to be $\epsilon_{\rm core} = 30\%^{+20\%}_{-10\%} $ \citep[e.g.][]{Alves07,Nutter07,Konyves15}. 
Combining these estimates leads to the following ``prediction'' for SFE$_{\rm dense}$ 
from the ``microphysics'' of star formation in filaments:\\
SFE$_{\rm dense}^{\rm pre} \equiv f_{\rm pre} \times \epsilon_{\rm core} / t_{\rm pre} = (4.5\pm2.5)\times 10^{-8}\, {\rm yr}^{-1} $,\\ 
which is plotted as a solid straight line/strip in Fig.~\ref{fig_SFR-Mdense}. 
As can be seen, SFE$_{\rm dense}^{\rm pre}$ provides a reasonably good fit to both the Galactic and extragalactic data points of Fig.~\ref{fig_SFR-Mdense} (b).

%%%%%%%%%%%%%%%%%%%%%%%%%%%%%%%%%%%%
% 4.3 Interpolation into external galaxies
%%%%%%%%%%%%%%%%%%%%%%%%%%%%%%%%%%%%
\subsection{Comments on apparent SFE variations in resolved galactic disks}

Recently, \citet{Usero15}, \citet{Chen15}, and \citet{Bigiel16} reported significant variations 
in the apparent star formation efficiency in dense gas, SFE$^{\rm extragal}_{\rm dense} \equiv L_{\rm FIR}/L_{\rm HCN}$, 
across the spatially-resolved disks of several nearby galaxies such as M51.
In particular, the $L_{\rm FIR}$/$L_{\rm HCN}$ ratio was observed to decrease as $L_{\rm FIR}$ increases from the outskirts 
to the center of the M51 disk \citep[see for instance the map of the $L_{\rm IR}$/$L_{\rm HCN}$ ratio presented by][]{Chen15}.
One plausible interpretation of this trend was a decrease in the star formation efficiency 
SFE$_{\rm dense}$ from the outer disk to the center of the M51 galaxy \citep[e.g.][]{Bigiel16}. 
We note, however, that since the FUV radiation field is significantly stronger near the center of the disk compared to the outer parts of the galaxy, 
and since $L_{\rm IR}/L_{\rm HCN}$ scales as $ \frac{\rm SFR}{M_{\rm dense}} \times \alpha_{\rm HCN} = {\rm SFE_{dense}} \times \alpha_{\rm HCN}$, 
the trend observed in M51 can at least partly originate from the expected decrease in the $\alpha_{\rm HCN} $ factor 
toward the center of the disk according to Eq.~(\ref{Eq:alpha}), with the efficiency  ${\rm SFE_{dense}}$ remaining approximately constant.

We can further quantify this statement using the detailed results published by \citet{Usero15} for 29 nearby disk galaxies. 
These authors found an anti-correlation between the apparent local star formation efficiency in dense gas SFE$_{\rm Usero-dense} \equiv I_{\rm FIR}/I_{\rm HCN}$
and the mass surface density of stars  $\Sigma_{\rm star}$,  expressed as 
SFE$_{\rm Usero-dense}$ (Myr$^{-1}$) = 10$^{-0.78\pm0.30}$ $\Sigma_{\rm Usero-star}^{-0.37\pm0.11}$($M_{\odot}$pc$^{-2}$)  \citep[Fig.~2 of ][]{Usero15}.  
\citet{Usero15} evaluated the mass surface density of dense gas  $\Sigma_{\rm dense}$ as $\alpha_{\rm Usero-HCN} \times I_{\rm HCN}$
and the mass surface density of old stars $\Sigma_{\rm Usero-star}$ values from the $3.6 \mu$m intensity $I_{\rm 3.6 \mu m}$ 
(with $\Sigma_{\rm Usero-star} \propto I_{\rm 3.6 \mu m}$). 
Since the $I_{\rm 3.6 \mu m}$ intensity is directly proportional to the FUV intensity $I_{\rm FUV}$ across the disk of nearby galaxies \citep[e.g.][]{Ford13}, 
the anti-correlation SFE$_{\rm Usero-dense} \propto  \Sigma_{\rm Usero-star}^{-0.37\pm0.11}$ 
can be rewritten as SFE$_{\rm dense} \times \alpha_{\rm Usero-HCN}  \propto I_{\rm FUV}^{-0.37\pm0.11}$. 
The dependence of $\alpha_{\rm HCN}$ on $G_0$ found in Sect.~\ref{alpha_HCN} and \ref{Sect:calibration} and expressed by Eq.~(\ref{Eq:alpha2}), i.e., 
$\alpha_{\rm Herschel-HCN}^{\rm fit^\prime} \propto G_0^{-0.34\pm0.08}$, can largely account for the \citet{Usero15} anti-correlation, 
and suggests that SFE$_{\rm dense}$ actually depends at most very weakly on $\Sigma_\star$ 
as SFE$_{\rm dense} \propto \Sigma_\star^{-0.03\pm0.13}$.

Significant variations in SFE$_{\rm dense}$ may still exist in the most extreme star-forming environments 
like the central molecular zone (CMZ) of our Milky Way 
{\citep[e.g.][see also Fig. \ref{fig_SFR-Mdense}]{Longmore13}} or extreme starburst galaxies \citep[e.g.][]{Garcia12}, 
especially at high redshift \citep[e.g.][]{Gao07}. 
By and large, however, the results summarized in Fig.~\ref{fig_SFR-Mdense} (b) suggest that SFE$_{\rm dense}$ 
is remarkably constant over a wide range of scales and environments.

%%%%%%%%%%%%%%%%%%%%%%%%%
% Conclusions
%%%%%%%%%%%%%%%%%%%%%%%%%

\section{Summary and conclusions}
In an effort to calibrate dense gas tracers commonly used in extragalactic studies, 
we carried out wide-field mapping observations at a spatial resolution of $\sim$0.04 pc in HCN ($J$=1--0), H$^{13}$CN ($J$=1--0),  HCO$^+$ ($J$=1--0), 
and H$^{13}$CO$^+$ ($J$=1--0) toward the nearby molecular clouds in Ophiuchus, Aquila, and Orion~B  
using the MOPRA 22m, IRAM 30m, and Nobeyama 45m telescopes. 
Our main results can be summarized as follows:

\begin{enumerate}
\item The spatial distributions of the H$^{13}$CO$^+$(1--0) and H$^{13}$CN(1--0) emission are tightly correlated 
with the filamentary texture of the dense gas seen in {\it Herschel} column density maps, 
showing that the H$^{13}$CO$^+$(1--0) and H$^{13}$CN(1--0) lines are  
good tracers of the densest (``supercritical'') filaments seen in {\it Herschel} submillimeter continuum images. 
Quantitatively, H$^{13}$CO$^+$(1--0) and H$^{13}$CN(1--0) trace {\it Herschel} filaments very well 
above $A_{\rm V} > 16$ (i.e., $M_{\rm line} \simgt 30\, M_\odot \, {\rm pc}^{-1} $) 
and $A_{\rm V} > 20$ (i.e., $M_{\rm line} \simgt 40\, M_\odot \, {\rm pc}^{-1} $) respectively. 
Moreover, the virial mass estimates derived from the H$^{13}$CO$^+$(1--0) and H$^{13}$CN(1--0) velocity 
dispersions agree well with the dense gas mass estimates derived from {\it Herschel} data for the same sub-regions.

\item In contrast, the spatial distributions of the HCN(1--0) and HCO$^+$(1--0) emission differ significantly 
from the column density distribution derived from {\it Herschel} data. The HCN(1--0) and HCO$^+$ (1--0) lines 
are only poor tracers of the supercritical filaments seen with {\it Herschel} and tend to be stronger 
around H${\rm II}$ regions. 
Based on a detailed comparison of the HCN(1--0) and HCO$^+$(1--0) integrated intensity maps
with the {\it Herschel} 70/100 $\mu$m and dust temperature maps, it appears that 
the HCN(1--0) and HCO$^+$(1--0) integrated intensities are strongly correlated with the strength of the local FUV radiation field ($G_0$). 
The luminosities of the HCN and HCO$^+$ lines can nevertheless be used to derive reasonably good estimates of the 
masses of dense gas in the nearby clouds we observed, {\it provided} that appropriate $G_0$-dependent conversion factors
$\alpha_{\rm Herschel-HCN}(G_0)$ and $\alpha_{\rm Herschel-HCO^+}(G_0)$ are adopted 
when converting $L_{\rm HCN}$ and $L_{\rm HCO^+}$ to $M_{\rm dense}$.

\item 
Using the masses $M_{\rm Herschel}^{A_{\rm V}>8}$ derived from {\it Herschel} column density maps above $A_{\rm V} > 8$ as reference estimates of the mass of dense gas in each nearby cloud, we found that  the conversion factors $\alpha_{\rm HCN}$ and $\alpha_{\rm HCO^+}$ are anti-correlated with the strength of the local FUV radiation field according to the following best-fit empirical relations:
$\alpha_{\rm Herschel-HCN}^{\rm fit}$ = $(496\pm94)$$\times$$G_0$$^{-0.24\pm0.07}$ $M_{\odot}$ (K km s$^{-1}$ pc$^2$)$^{-1}$ and $\alpha_{\rm Herschel-HCO^+}^{\rm fit}$ = $(689\pm151)$$\times$$G_0$$^{-0.24\pm0.08}$ $M_{\odot}$ (K km s$^{-1}$ pc$^2$)$^{-1}$.

\item The relation between the star formation rate, estimated from direct counting of {\it Spitzer} {YSOs}, and the mass of dense gas, derived from {\it Herschel} data above $A_{\rm V} > 8$, for the nearby clouds/clumps of our sample can be expressed as  {SFR$_{\rm YSO}$ =  (2.1--5.0)$\times$10$^{-8}$ $M_{\odot}$ yr$^{-1}$$\times$($M_{\rm Herschel}^{A_{\rm V}>8}$/$M_{\odot}$)}. This is consistent within errors with the relation found by Lada and collaborators for another and broader sample of nearby star-forming clouds based on {\it Spitzer} and near-infrared extinction data \citep[SFR = (4.6$\pm$2.6)$\times$10$^{-8}$ $M_{\odot}$ yr$^{-1}$$\times$($M_{\rm dense}$/$M_{\odot}$),][]{Lada10,Lada12}. 

\item 
In nearby molecular clouds, the optically thick HCN(1--0) and HCO$^+$(1--0) lines are tracing both moderate column density areas ($2 \le A_{\rm V} \le 8$) and high column density areas ($A_{\rm V}$ > 8) \citep[see also][]{Pety16}. Therefore, published extragalactic HCN(1--0) studies, which have spatial resolutions between $\sim 10\,$pc and $\sim 50\,$kpc, must be tracing all of the moderate density gas down to $n_{\rm H_2}  \simlt 10^3\, \rm{cm}^{-3}$ in the observed galaxies, in addition to the dense gas with $n_{\rm H_2}  > 10^4\, \rm{cm}^{-3}$. Estimating the contribution of this moderate density gas from the typical column density PDFs observed in nearby clouds, we obtained the following effective HCN conversion factor for external galaxies:  
$\alpha_{\rm Herschel-HCN}^{\rm fit^\prime} = 0.13^{+0.06}_{-0.06}\times$ $G_{0}^{-0.095\pm0.02}$$\times$$\alpha_{\rm Herschel-HCN}^{\rm fit}(G_0)$ $\propto$ $G_{0}^{-0.34\pm0.08}$.

\item
Using this $G_0$-dependent conversion factor $\alpha_{\rm Herschel-HCN}^{\rm fit^\prime}(G_0)$ to improve the dense gas mass estimates of external galaxies with published HCN data, we found that the star formation efficiency in dense molecular gas, SFE$_{\rm dense} \equiv $ SFR/$M_{\rm dense}$, is remarkably constant, with a scatter of less than 1.5 orders of magnitude around the nearby-cloud value of $4.5 \times 10^{-8}\, {\rm yr}^{-1} $, over 8 orders of magnitude in $M_{\rm dense}$ from $\sim10^2\, M_{\odot}$ to $10^{10}\, M_{\odot}$. This suggests that the star formation efficiency in the dense molecular gas of galaxies is quasi-universal on a wide range of scales from $\sim 1$--10 pc to $> 10\,$kpc.

\item Following \citet{Andre14}, we suggest that  SFE$_{\rm dense}$ is primarily set by three parameters characterizing the ``microphysics'' of core/star formation along molecular filaments, 
namely the typical fraction of dense filament gas in the form of prestellar cores, $f_{\rm pre} = 15\%^{+10\%}_{-5\%}$, the typical lifetime of solar-type prestellar cores, $t_{\rm pre} = 1^{+0.5}_{-0.3}\, $Myr, and the efficiency of the conversion from prestellar core mass to stellar mass, $\epsilon_{\rm core} = 30\%^{+20\%}_{-10\%}$, according to the simple relation: SFE$_{\rm dense}^{\rm pre} = f_{\rm pre} \times \epsilon_{\rm core} / t_{\rm pre} = (4.5\pm2.5)\times10^{-8}\, {\rm yr}^{-1} $.

\end{enumerate}

\begin{acknowledgements}
We thank the referee for useful suggestions which improved the clarity of the paper. 
The Mopra radio telescope is part of the Australia Telescope National Facility which is funded by the Australian Government for operation as a National Facility managed by CSIRO. Based on observations carried out with the IRAM 30m Telescope. IRAM is supported by INSU/CNRS (France), MPG (Germany) and IGN (Spain). The 45-m radio telescope is operated by Nobeyama Radio Observatory, a branch of National Astronomical Observatory of Japan. 
This research has made use of data from the Herschel Gould Belt survey (HGBS) project (http://gouldbelt-herschel.cea.fr). The HGBS is a Herschel Key Programme jointly carried out by SPIRE Specialist Astronomy Group 3 (SAG 3), scientists of several institutes in the PACS Consortium (CEA Saclay, INAF-IFSI Rome and INAF-Arcetri, KU Leuven, MPIA Heidelberg), and scientists of the Herschel Science Center (HSC).
Our work was supported by the ANR-11-BS56-010 project ``STARFICH" and the European Research Council under the European Union's Seventh Framework Programme (ERC Advanced Grant Agreement no. 291294 --  `ORISTARS'). N.S. acknowledges support from the DFG through project number Os 177/2-1 and 177/2-2, and central funds of the DFG-priority program 1573 (ISM-SPP). Y.G. acknowledges support from the National Natural Science Foundation of China (grants 11390373 and 11420101002) and the Chinese Academy of Sciences’ Key Research Program of Frontier Sciences.

\end{acknowledgements}

%-------------------------------------------------------------------

\bibliographystyle{aa}
%\bibliography{HCN}

%%%%%%%%%%%%%%%%%%%%

%%%%%%%%%%%%%%%%%%%%%%%%%%
% Appendix
%%%%%%%%%%%%%%%%%%%%%%%%%%

%%%%%%%%%%%%%%%%%%%%%%%%%%%%%%%%%%%%
%  Appendix A: Complementary figures
%%%%%%%%%%%%%%%%%%%%%%%%%%%%%%%%%%%%
\appendix{}
\section{Complementary figures and tables}

Figures \ref{fig1_cold} -- \ref{fig:spec_comp_av} are complementary figures. {The optical depths of the HCN (1--0) and HCO$^+$ (1--0) lines in each $A_{\rm V}$ bin are listed in Table \ref{table:opacity}. Table \ref{list_symbols} summarizes the definition of each notation used in this paper.}

%Figure A.1
\begin{figure*}
\begin{center}
\includegraphics[width=190mm, angle=0]{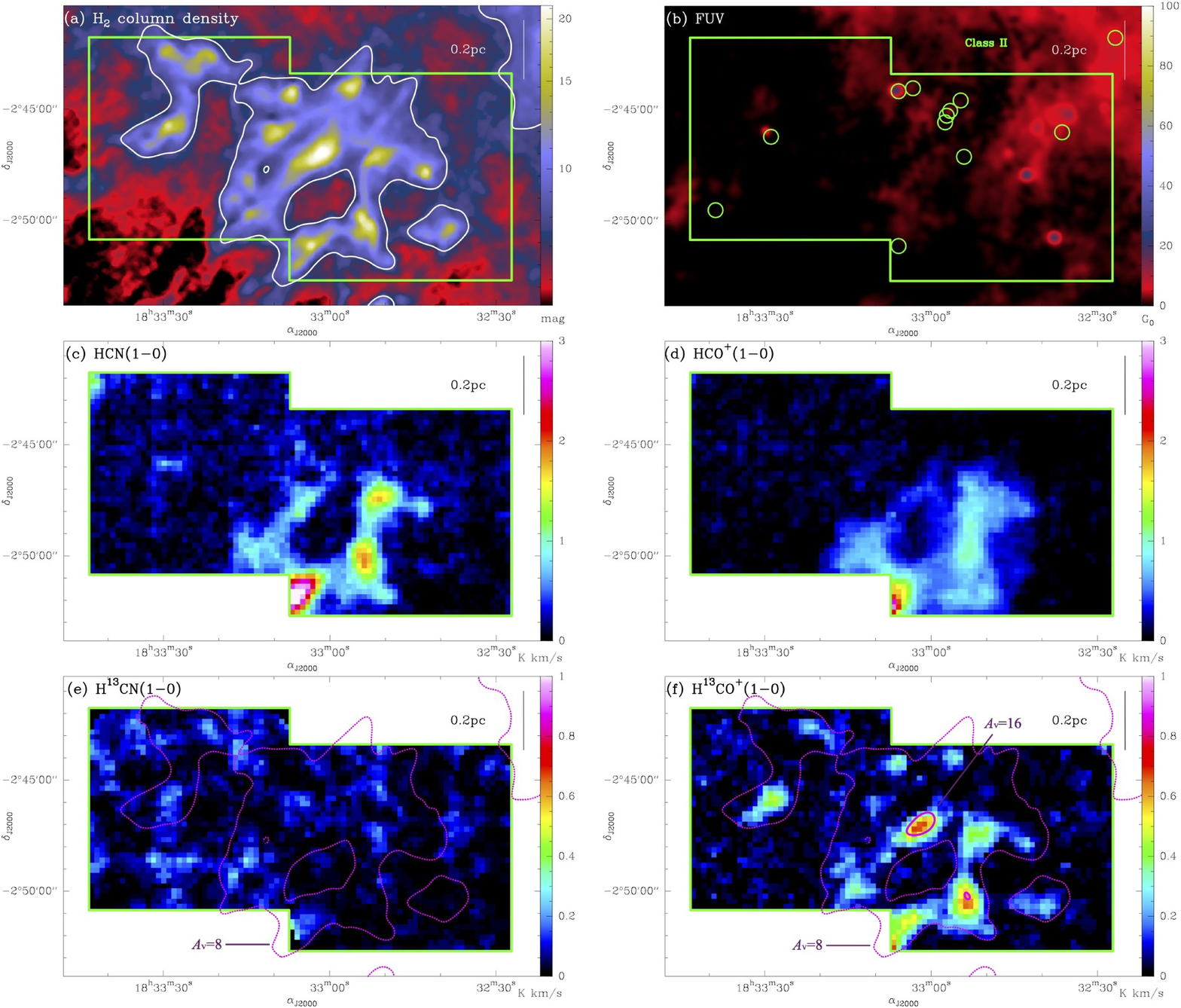}
\caption{Same as Fig. \ref{fig1}, but for Aquila/cold. The H$_2$ column density map of the Aquila region is from {\it Herschel} Gould Belt Survey (HGBS) data \citep{Andre10, Konyves15}.  {The white contour in panel (a) and magenta dotted contours in panels (e) and (f) indicate} the $A_{\rm V}$ = 8  level derived from the {\it Herschel} column density map smoothed to 40$\arcsec$ resolution. 
{In panel (f), the magenta solid contour indicates the rough $A_{\rm V}$ column density level above which significant H$^{13}$CO$^+$ (1--0) emission is detected, 
i.e.,  $A_{\rm V}$ = 16.}}
\label{fig1_cold}
\end{center}
\end{figure*}

%Figure A.2
\begin{figure*}
\begin{center}
\includegraphics[width=190mm, angle=0]{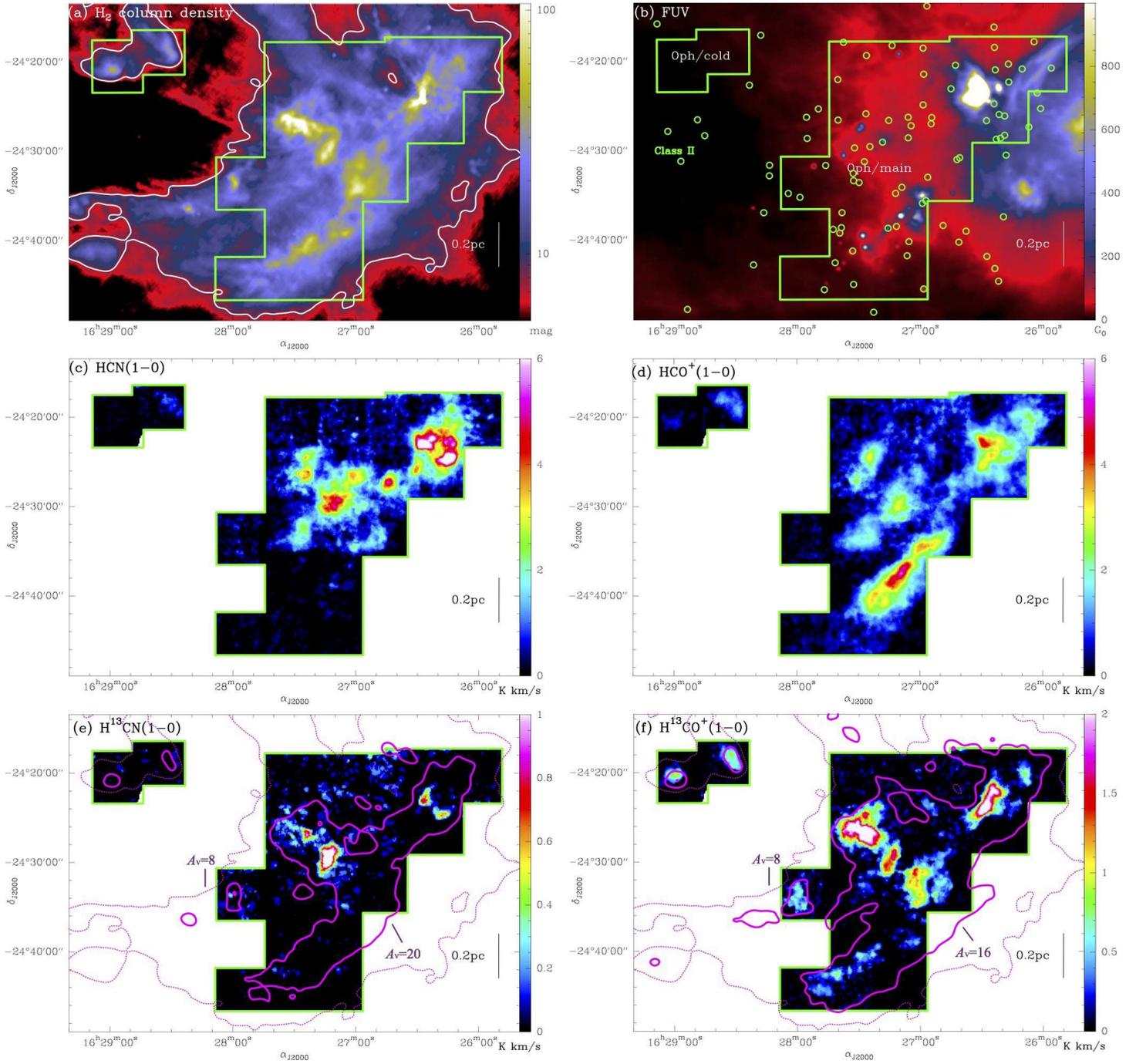}
\caption{Same as Fig. \ref{fig1}, but for Ophiuchus. The H$_2$ column density map of the Ophiuchus region is from HGBS data (Ladjelate et al. in prep.). The angular resolutions of HCN, HCO$^{+}$, and H$^{13}$CO$^+$, and H$^{13}$CN maps are 50$\arcsec$, 50$\arcsec$, 50$\arcsec$, and 60$\arcsec$. {The white contour in panel (a) and magenta dotted contours in panels (e) and (f) indicate} the $A_{\rm V}$ = 8 level derived from the {\it Herschel} column density map smoothed to 50$\arcsec$ resolution. {In panel (e) and (f), the magenta solid contour indicates the rough $A_{\rm V}$ column density level above which significant line emission is detected, i.e., $A_{\rm V}$ = 20 for H$^{13}$CN (1--0) and  $A_{\rm V}$ = 16 for H$^{13}$CO$^+$ (1--0).}}
\label{fig_oph_maps}
\end{center}
\end{figure*}

%Figure A.3
\begin{figure*}
\begin{center}
\includegraphics[width=190mm, angle=0]{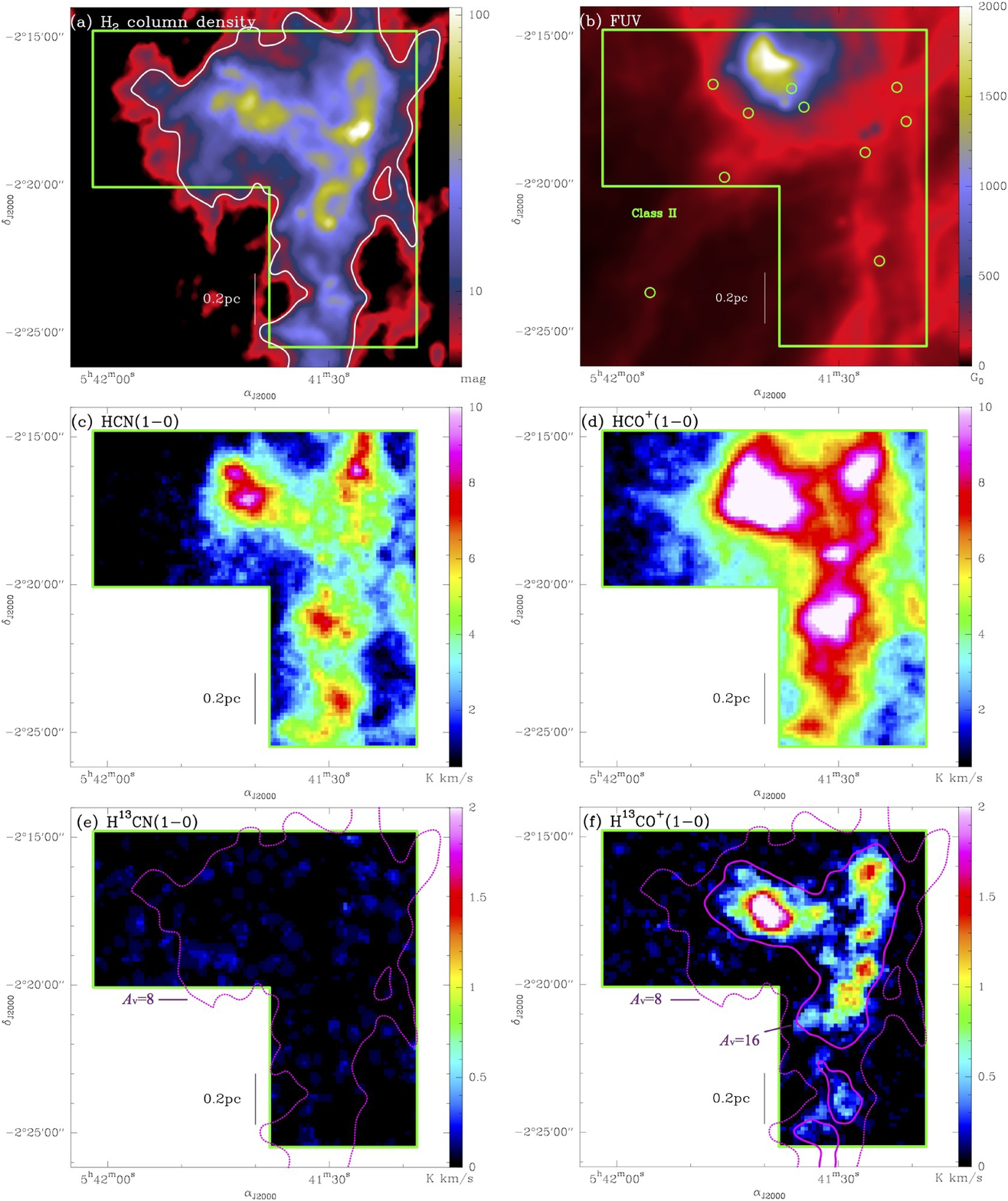}
\caption{Same as Fig. \ref{fig1}, but for NGC2023 in Orion B. The H$_2$ column density map of the Orion B region is from HGBS data {\citep[K\"onyves et al. in prep., see also][]{Schneider13}}. The angular resolutions of HCN, HCO$^{+}$, and H$^{13}$CO$^+$, and H$^{13}$CN maps are 30$\arcsec$, 30$\arcsec$, 30$\arcsec$, and 40$\arcsec$. {The white contour in panel (a) and magenta dotted contours in panels (e) and (f) indicate} the $A_{\rm V}$ = 8 level derived from the {\it Herschel} column density map smoothed to 30$\arcsec$ resolution. {In panel (f), the magenta solid contour indicates the rough $A_{\rm V}$ column density level 
above which significant H$^{13}$CO$^+$ (1--0) emission is detected, 
i.e., $A_{\rm V}$ = 16.}}
\label{fig_ngc2023_maps}
\end{center}
\end{figure*}

%Figure A.4
\begin{figure*}
\begin{center}
\includegraphics[width=150mm, angle=0]{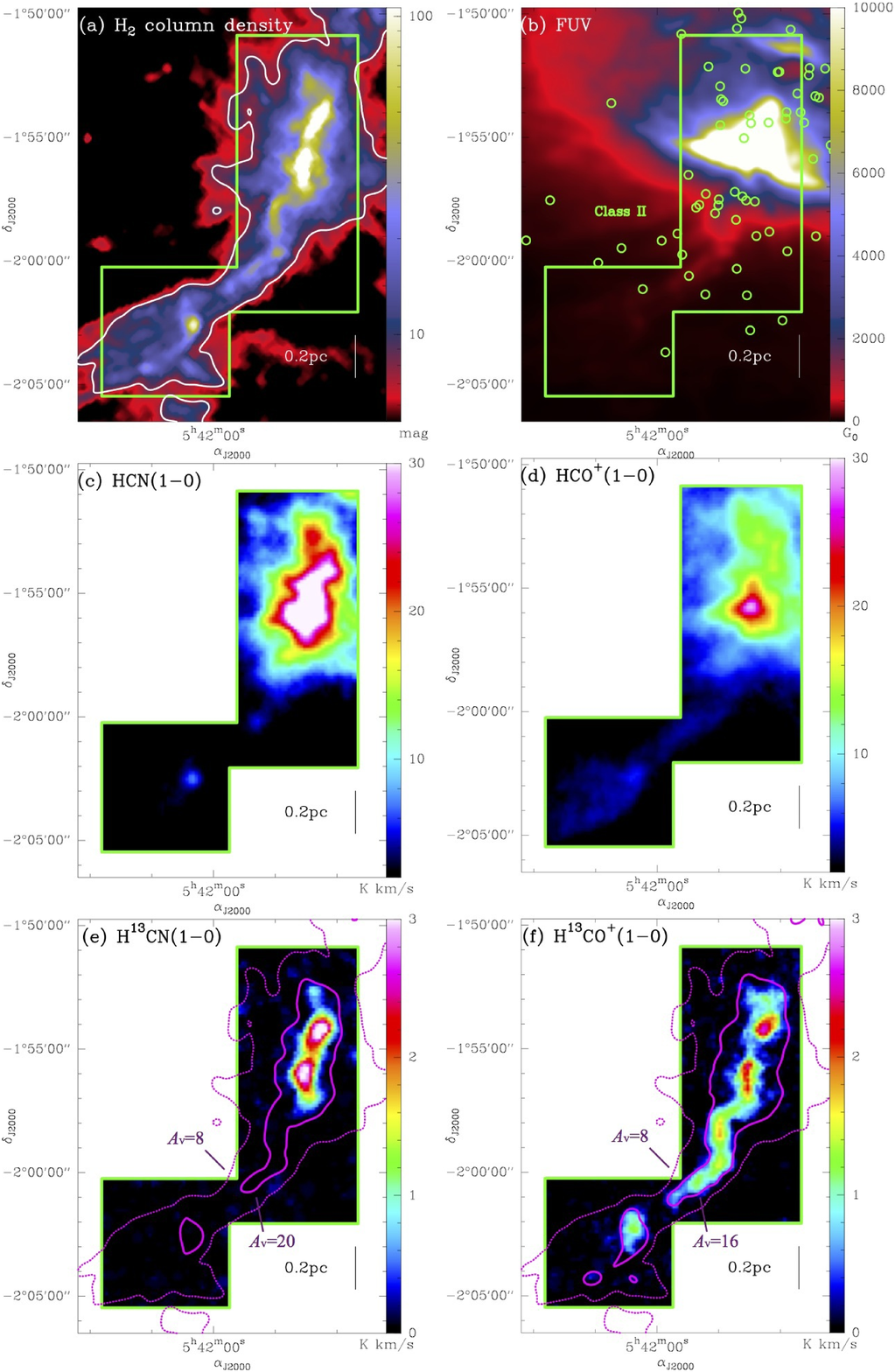}
\caption{Same as Fig. \ref{fig1}, but for NGC2024 in Orion B. The H$_2$ column density map of the Orion B region is from HGBS data {\citep[K\"onyves et al. in prep., see also][]{Schneider13}}. The angular resolutions of HCN, HCO$^{+}$, and H$^{13}$CO$^+$, and H$^{13}$CN maps are 30$\arcsec$, 30$\arcsec$, 30$\arcsec$, and 40$\arcsec$.  {The white contour in panel (a) and magenta dashed contours in panels (e) and (f) indicate} the $A_{\rm V}$ = 8 level derived from the {\it Herschel} column density map smoothed to 30$\arcsec$ resolution. 
{In panel (e) and (f), the magenta solid contour indicates the rough $A_{\rm V}$ column density level 
above which significant line emission is detected, 
i.e., $A_{\rm V}$ = 20 for H$^{13}$CN (1--0) and  $A_{\rm V}$ = 16 for H$^{13}$CO$^+$ (1--0).}}
\label{fig_ngc2024_maps}
\end{center}
\end{figure*}

%Figure A.5
\begin{figure*}
\begin{center}
\includegraphics[width=190mm, angle=0]{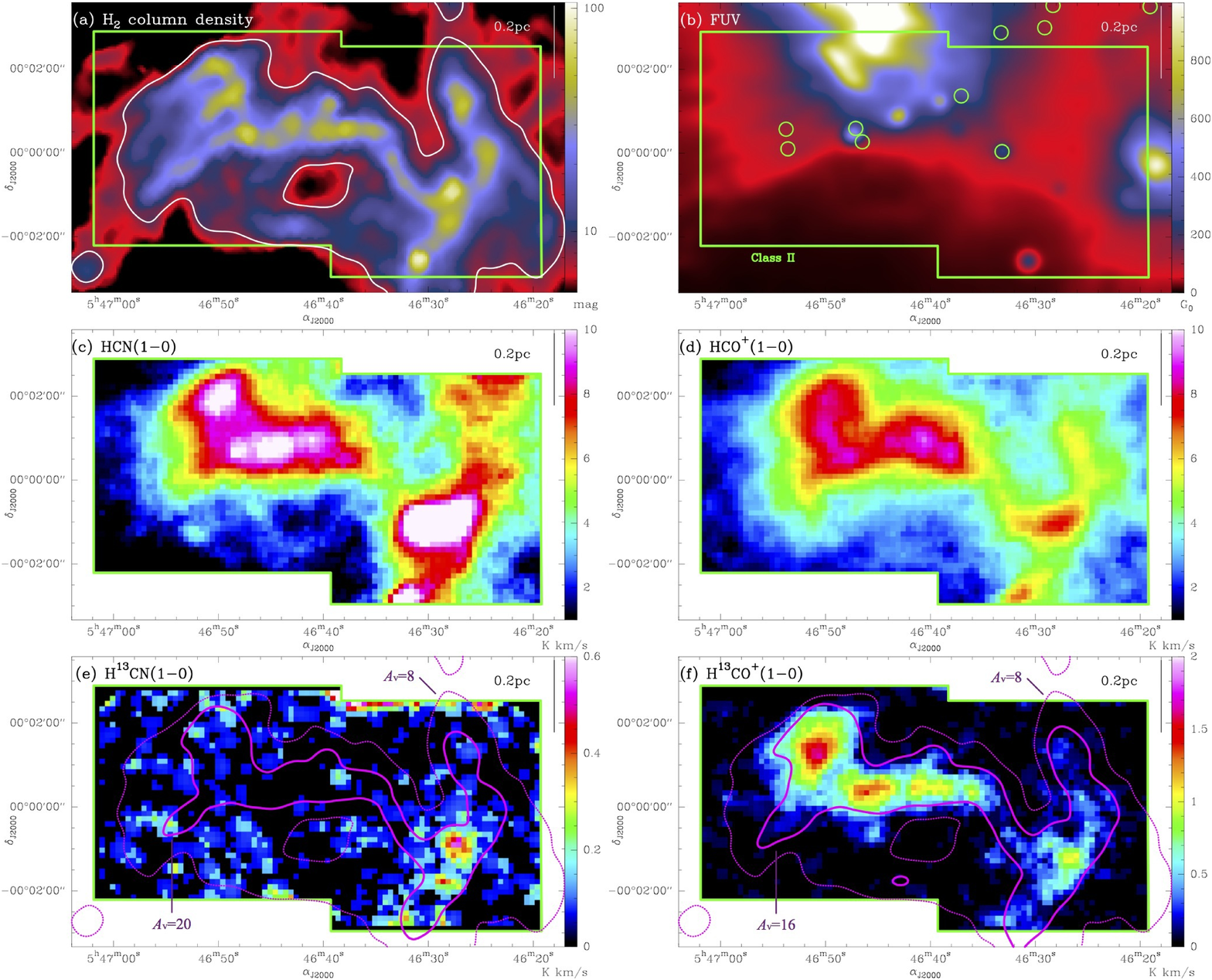}
\caption{Same as Fig. \ref{fig1}, but for NGC2068 in Orion B. The H$_2$ column density map of the Orion B region is from HGBS data {\citep[K\"onyves et al. in prep., see also][]{Schneider13}}. The angular resolutions of HCN, HCO$^{+}$, and H$^{13}$CO$^+$, and H$^{13}$CN maps are 30$\arcsec$, 30$\arcsec$, 30$\arcsec$, and 40$\arcsec$.  {The white contour in panel (a) and magenta dashed contours in panels (e) and (f) indicate} the $A_{\rm V}$ = 8 level derived from the {\it Herschel} column density map smoothed to 30$\arcsec$ resolution.  
{In panel (e) and (f), the magenta solid contour indicates the rough $A_{\rm V}$ column density level 
above which significant line emission is detected, 
i.e., $A_{\rm V}$ = 20 for H$^{13}$CN (1--0) and  $A_{\rm V}$ = 16 for H$^{13}$CO$^+$ (1--0).}}
\label{fig_ngc2068_maps}
\end{center}
\end{figure*}

%Figure A.6
\begin{figure*}
\begin{center}
\includegraphics[width=190mm, angle=0]{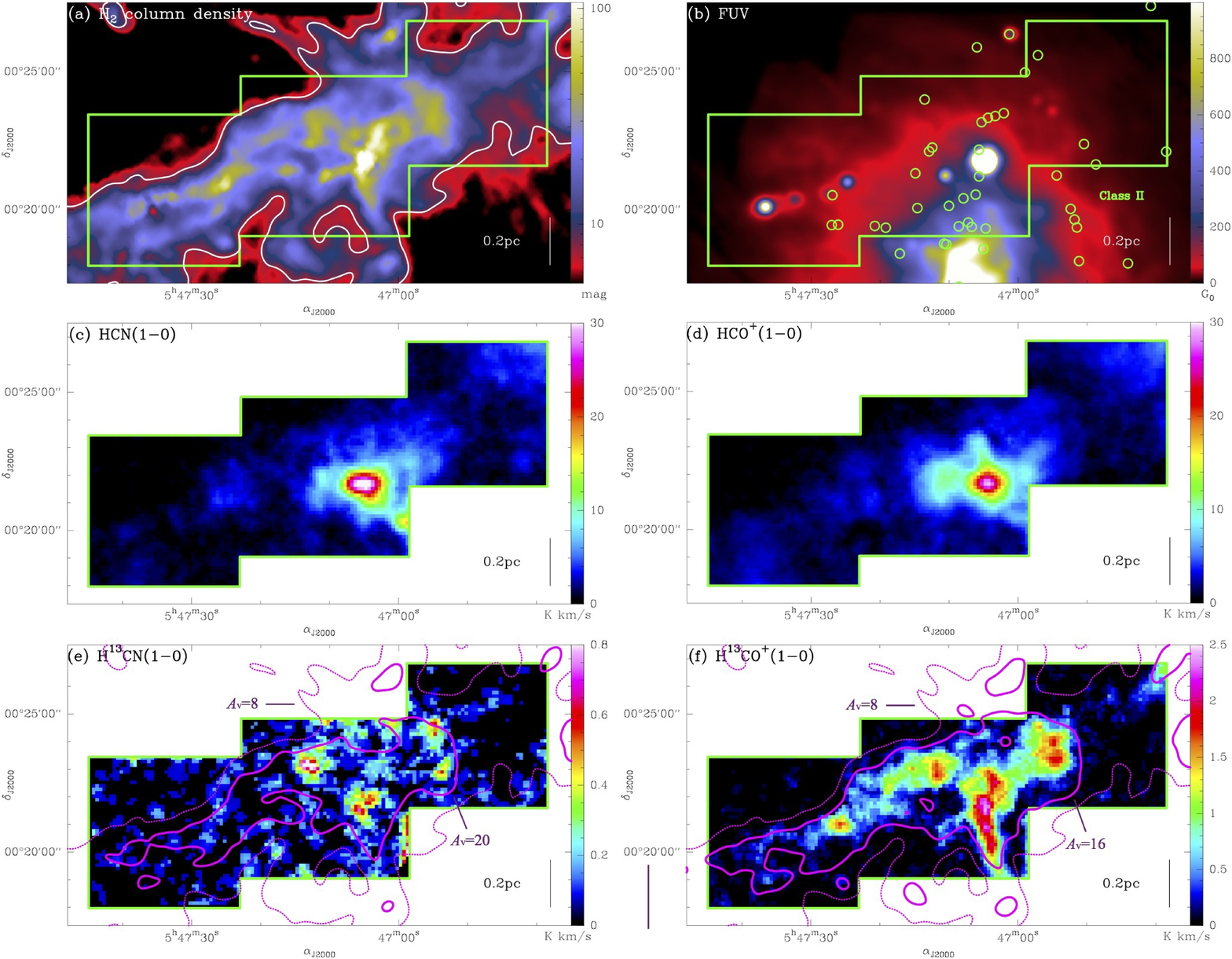}
\caption{Same as Fig. \ref{fig1}, but for NGC2071 in Orion B. The H$_2$ column density map of the Orion B region is from HGBS data {\citep[K\''onyves et al. in prep., see also][]{Schneider13}}. The angular resolutions of HCN, HCO$^{+}$, and H$^{13}$CO$^+$, and H$^{13}$CN maps are 30$\arcsec$, 30$\arcsec$, 30$\arcsec$, and 40$\arcsec$.  {The white contour in panel (a) and magenta dashed contours in panels (e) and (f) indicate} the $A_{\rm V}$ = 8 level derived from the {\it Herschel} column density map smoothed to 30$\arcsec$ resolution.  
{In panel (e) and (f), the magenta solid contour indicates the rough $A_{\rm V}$ column density level 
above which significant line emission is detected, 
i.e., $A_{\rm V}$ = 20 for H$^{13}$CN (1--0) and  $A_{\rm V}$ = 16 for H$^{13}$CO$^+$ (1--0).}}
\label{fig_ngc2071_maps}
\end{center}
\end{figure*}

%Figure A.7
\begin{figure*}
\begin{center}
\includegraphics[width=140mm, angle=0]{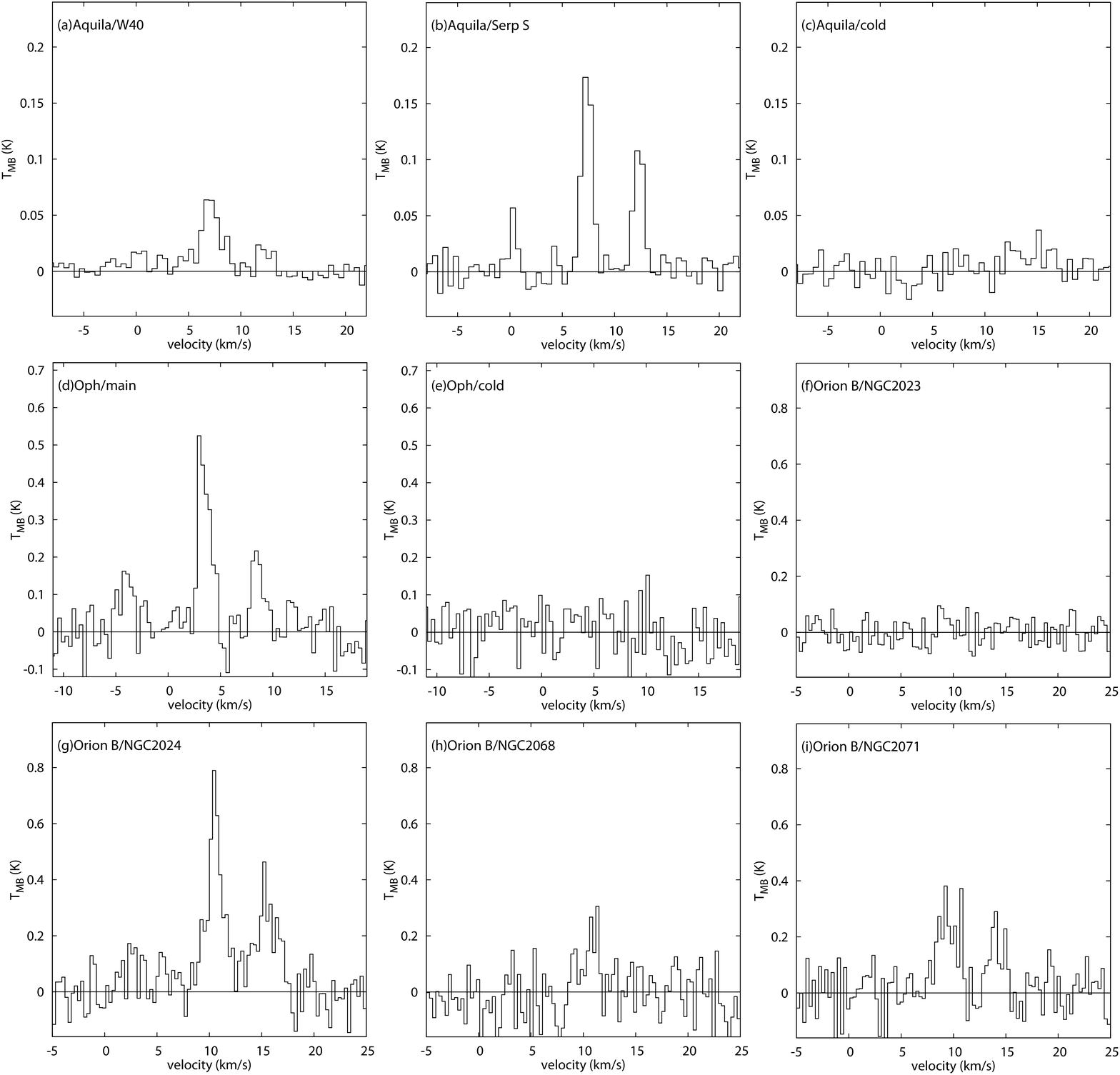}
\caption{Comparison of H$^{13}$CN spectra averaged over pixels where has $\ge$ 5$\sigma$ emission at (a) Oph/main, (b) Oph/cold, (c) Aquila/W40, (d) Aquila/Serp. South, (e) Aquila/cold, (f) NGC2023, (g) NGC2024, (h) NGC2068, and (i) NGC2071. For Oph/cold, Aquila/cold, and NGC2023, the H$^{13}$CN emission is not detected. Thus, we show the spectra averaged over observing area.}
\label{fig_h13cn_spec}
\end{center}
\end{figure*}

%Figure A.8
\begin{figure*}
\begin{center}
\includegraphics[width=55mm, angle=0]{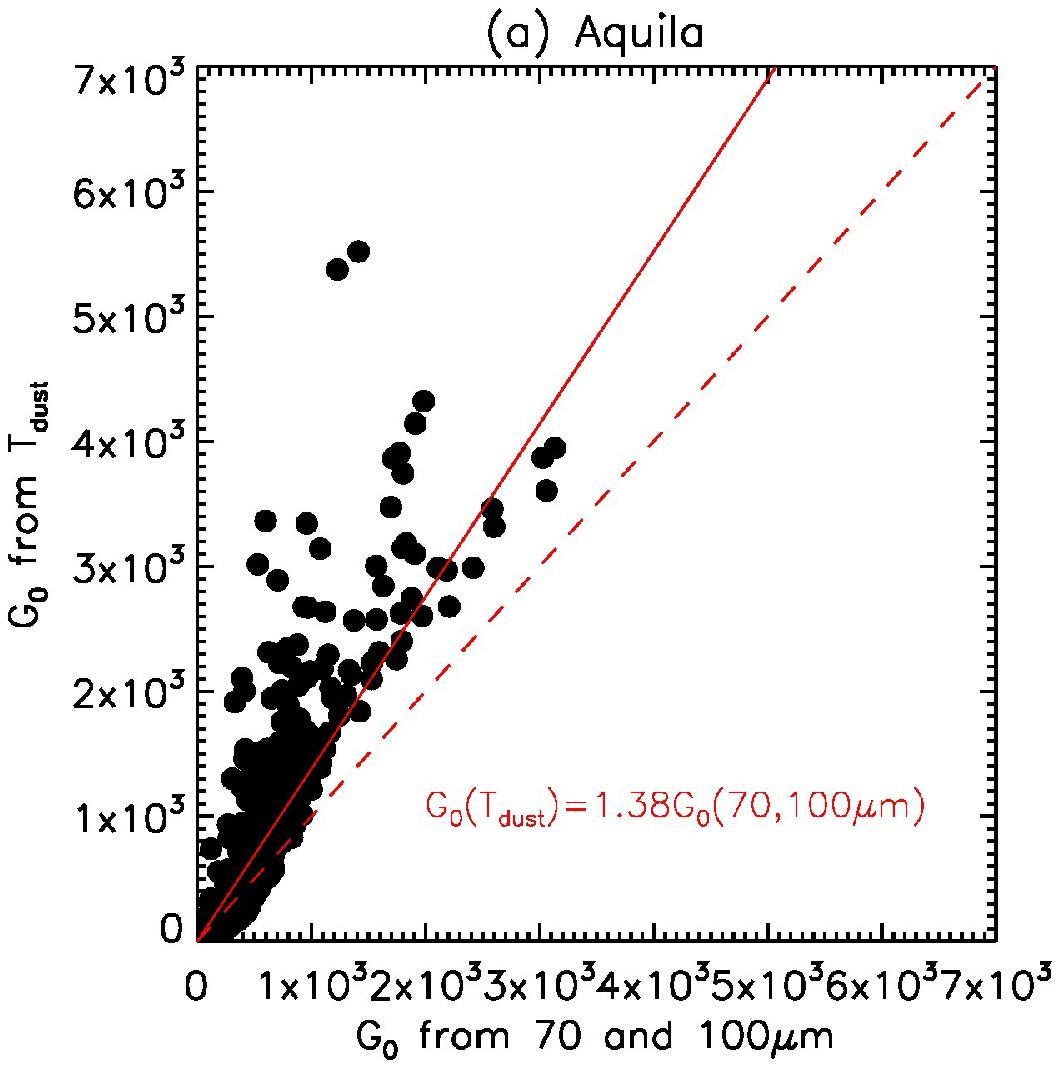}
\includegraphics[width=55mm, angle=0]{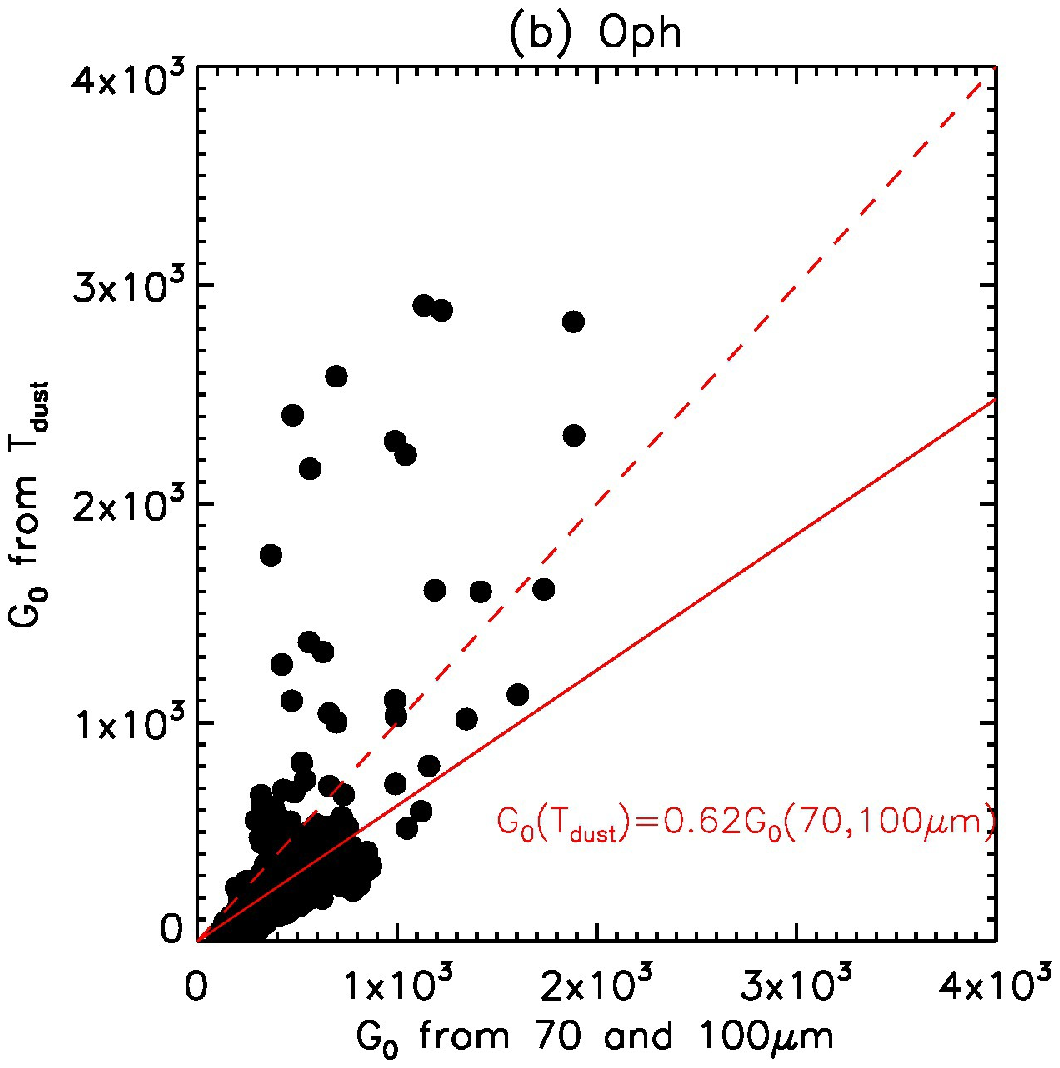}
\includegraphics[width=55mm, angle=0]{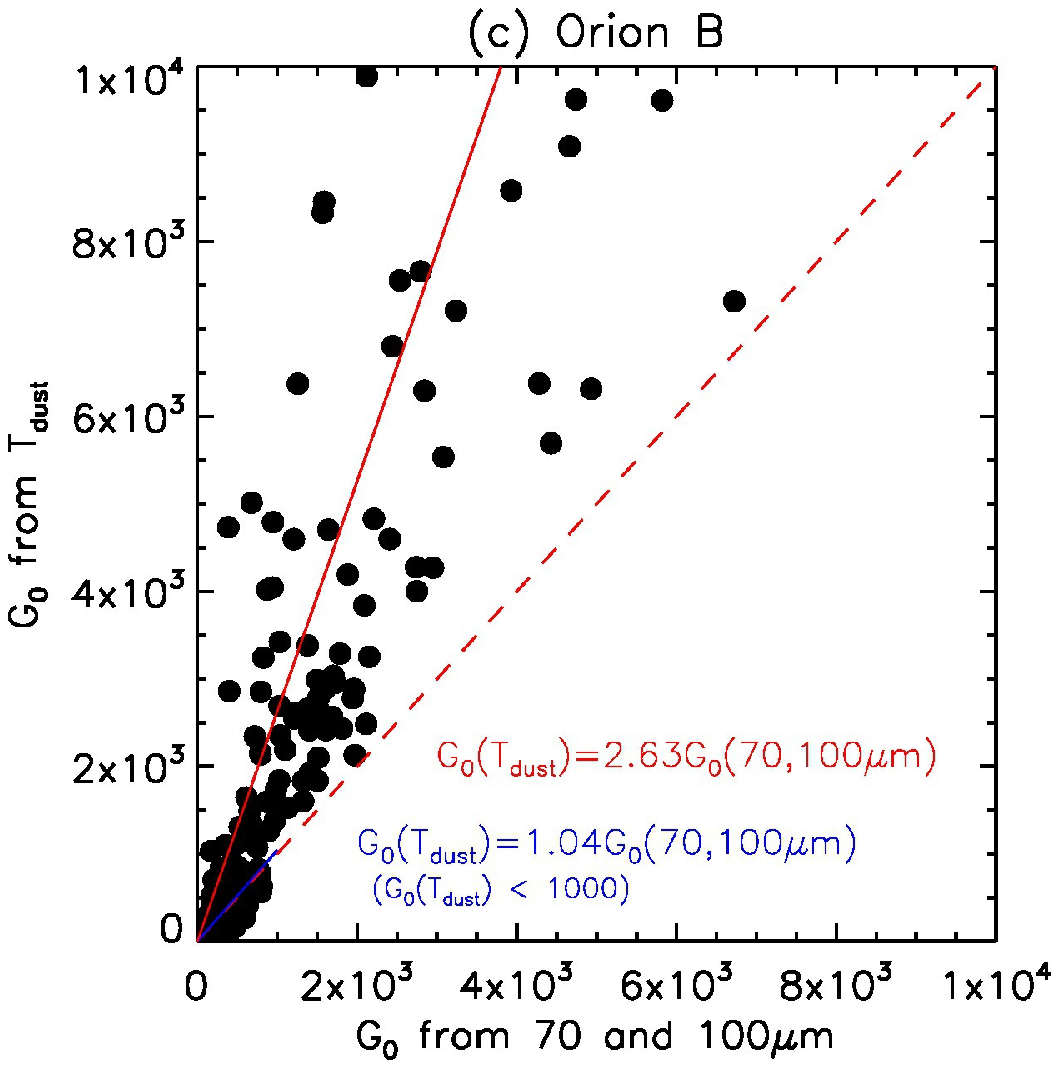}
\caption{{Pixel to pixel correlation between the $G_0$ value estimated from $T_{\rm dust}$, $G_0$($T_{\rm dust}$), using Eqs. (\ref{eq:Td_G0}-\ref{eq_tau}) map and $G_0$ values estimated from $Herschel$ 70 and 100 $\mu$m data, $G_0$(70,100$\mu$m), using Eqs (\ref{eq:G0_FIR},\ref{eq:G0_IFIR}) toward (a) Aquila, (b) Ophiuchus,  (c) Orion B. The dashed red lines indicate $G_0$($T_{\rm dust}$)=$G_0$(70,100$\mu$m). The {solid} red lines indicate the best-fit results: 
$G_0$($T_{\rm dust}$)=(1.38$\pm$0.01)$\times$$G_0$(70,100$\mu$m) for Aquila, 
$G_0$($T_{\rm dust}$)=(0.62$\pm$0.01)$\times$$G_0$(70,100$\mu$m) for Ophiuchus, 
and $G_0$($T_{\rm dust}$)=(2.63$\pm$0.09)$\times$$G_0$(70,100$\mu$m) for Orion B. 
The blue lines in panel (c) indicate the best-fit result for pixels with $G_0$($T_{\rm dust}$) < 1000: $G_0$($T_{\rm dust}$)=(1.04$\pm$0.02)$\times$$G_0$(70,100$\mu$m).}}
\label{fig:Td-G0}
\end{center}
\end{figure*}

%Figure A.10 ? 
\begin{figure*}
\begin{center}
\includegraphics[width=180mm, angle=0]{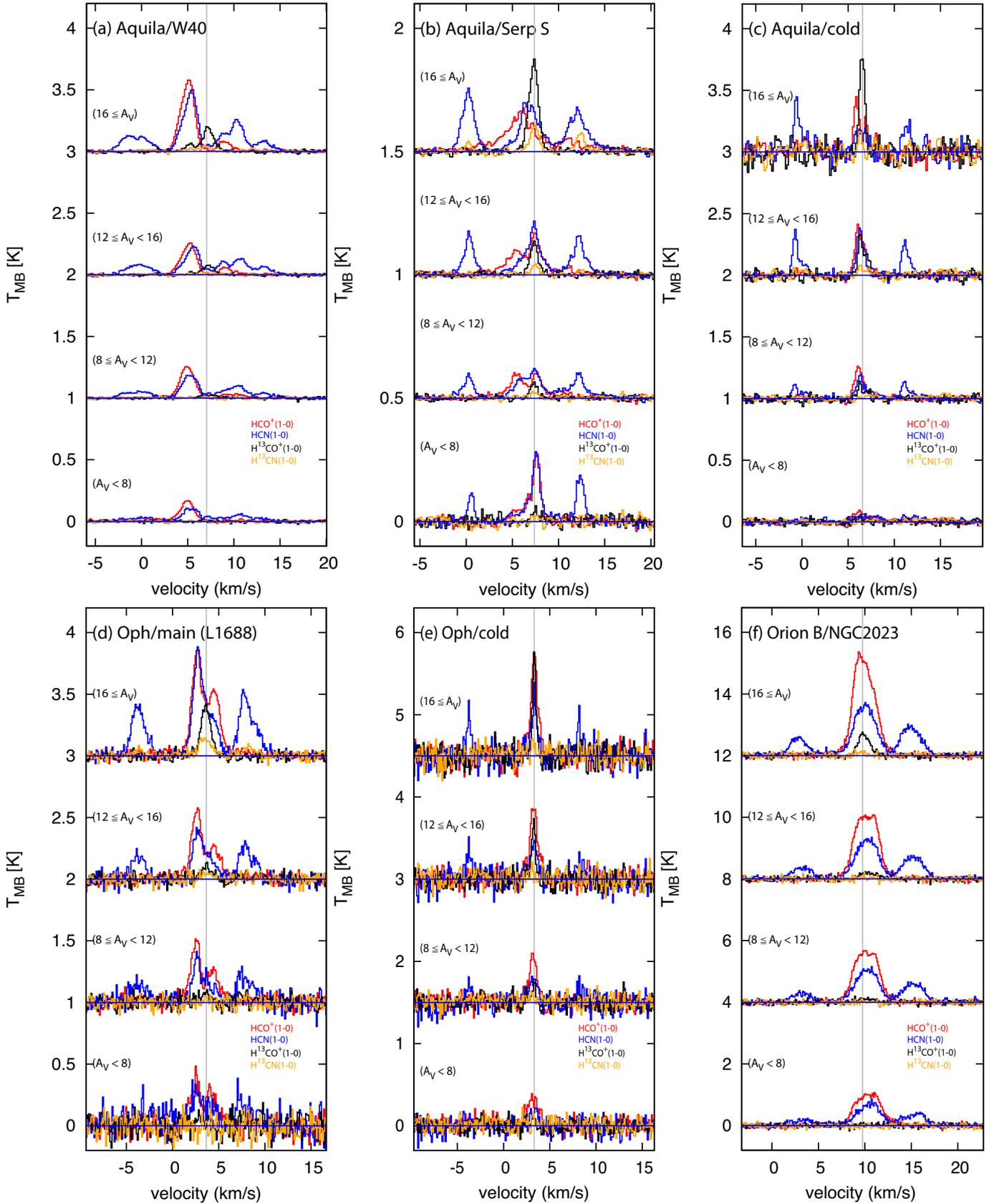}
\caption{Comparison of the HCO$^{+}$ (1--0, red), HCN (1--0, blue), H$^{13}$CO$^+$ (1--0, black), and H$^{13}$CN (1--0, orange) spectra averaged over the observed area in each $A_{\rm V}$ range for (a) Aquila/W40, (b) Aquila/Serp-South, (c) Aquila/cold, (d) Oph/main, (e) Oph/cold, (f) Orion B/NGC2023, (g) Orion B/NGC2024, (h) Orion B/NGC2068, and (i) Orion B/NGC2071. In each panel, the vertical grey line marks the peak velocity of the H$^{13}$CO$^+$ line toward the 16 $\le$ $A_{\rm V}$ area, and the spectra shown from top to bottom correspond to the 16 $\le$ $A_{\rm V}$, 12 $\le$  $A_{\rm V}$ < 16, 8 $\le$ $A_{\rm V}$ < 12, and $A_{\rm V}$ < 8  areas, respectively. }
\label{fig:spec_comp_av}
\end{center}
\end{figure*}

\begin{figure*}
\ContinuedFloat
\begin{center}
\includegraphics[width=180mm, angle=0]{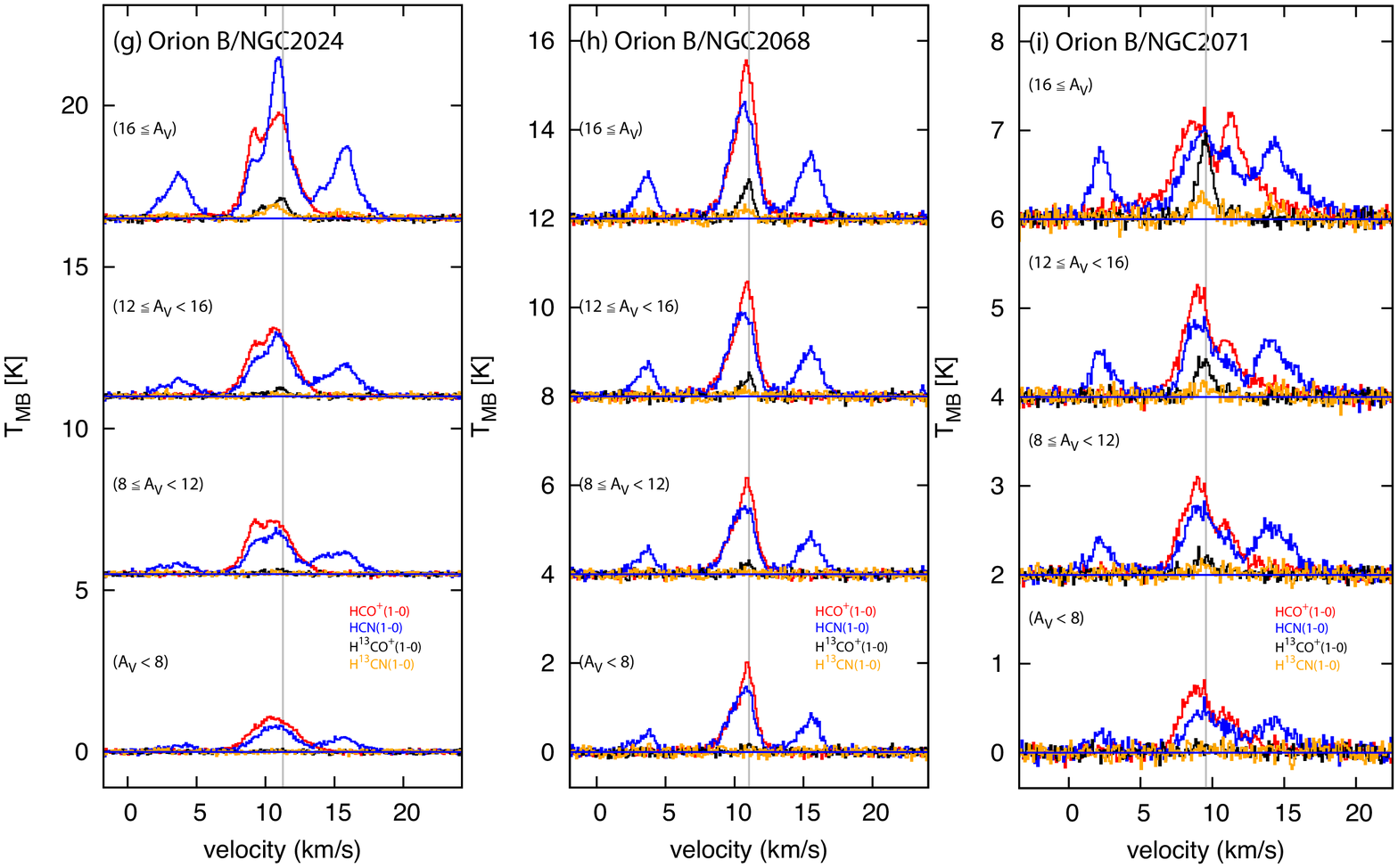}
\caption{--$continued$}
\end{center}
\end{figure*}

\begin{table*}
\centering
\begin{threeparttable}
\caption{Optical depths of the HCN (1--0) and HCO$^{+}$ (1--0) lines $^\dag$  \label{table:opacity}}
\begin{tabular}{lccccc}
\hline
Region    &  molecule    &  $A_{\rm V}$  < 8   & 8 $\le$ $A_{\rm V}$ < 12  & 12 $\le$ $A_{\rm V}$ < 16 & 16 $\le$ $A_{\rm V}$   \\
\hline
Aquial/W40   &  HCN(1--0)  &  ---$^\ddag$     & ---$^\ddag$   &  35 &  >>1\tnote{$SA$}   \\
                     &  HCO$^+$(1--0)  &  ---$^\ddag$     &  159 &  >>1\tnote{$SA$}  &   >>1\tnote{$SA$}  \\
\hline
Aquila/Serp S& HCN(1--0)    &    ---$^\ddag$   & ---$^\ddag$   &  17  &  103   \\ 
                     &  HCO$^+$(1--0)  &   17    &  67 & 102  &   >>1\tnote{$SA$}  \\
\hline
Aquila/cold& HCN(1--0)    &    ---$^\ddag$   & ---$^\ddag$  & 16   &  135  \\
                     &  HCO$^+$(1--0)  &  ---$^\ddag$     &  51 &  124 &  >>1\tnote{$SA$}   \\
\hline
Oph/main (L1688)&  HCN(1--0)   &   ---$^\ddag$    &  ---$^\ddag$  &  ---$^\ddag$ &  14   \\
                     &  HCO$^+$(1--0)  &   ---$^\ddag$    &  ---$^\ddag$ &  80 &   >>1\tnote{$SA$}   \\
\hline
Oph/cold& HCN(1--0)    &    ----$^\ddag$   &   ---$^\ddag$ & $^\ddag$$^\ddag$  &  ---$^\ddag$   \\
                     &  HCO$^+$(1--0)  &    ---$^\ddag$   & ---$^\ddag$  & 128   &  >>1\tnote{$SA$}    \\
\hline
Orion B/NGC2023&  HCN(1--0)   &  ---$^\ddag$     &  ---$^\ddag$  &  112  &  12   \\
                     &  HCO$^+$(1--0)  &   ---$^\ddag$    &  10  &  8  &    18 \\
\hline
Orion B/NGC2024& HCN(1--0)    &  ---$^\ddag$     &  ---$^\ddag$  &   ---$^\ddag$ &  6  \\
                     &  HCO$^+$(1--0)  &  ---$^\ddag$     &  7  &  10  &   14  \\
\hline
Orion B/NGC2068& HCN(1--0)    &   ---$^\ddag$    & ---$^\ddag$   & ---$^\ddag$   &  8   \\
                     &  HCO$^+$(1--0)  &   ---$^\ddag$    & 11   & 16   &  20  \\
\hline
Orion B/NGC2071& HCN(1--0)    &  ---$^\ddag$      &  ---$^\ddag$  &  ---$^\ddag$  & 24    \\
                     &  HCO$^+$(1--0)  & ---$^\ddag$      & 21  & 29   &   >>1\tnote{$SA$}  \\
\hline
\end{tabular}
\begin{tablenotes}
\item[$^\dag$] The spectra in Fig. \ref{fig:spec_comp_av} are used to derive the intensities.
\item[$^\ddag$] The emission of the rare species is not detected. 
\item[$SA$] The intensity of the normal species is weaker than that of the rare species due to the self-absorption effect.
\end{tablenotes}
\end{threeparttable}
\end{table*}

\begin{table*}
\tiny
\scalebox{0.92}[0.92]{
\centering
\begin{threeparttable}
\caption{Definition of each notation used in the paper\label{list_symbols}}
\begin{tabular}{|l|l|l|}
\hline
Quantity & Notation & Meaning \\
\hline
\multirow{2}{*}{Surface area}  & $A_{\rm mol}^{\rm detect}$ & Area where the H$^{13}$CO$^+$ or H$^{13}$CN emission has been detected.\\
                                 & $A_{\rm Herschel}^{A_{\rm V}>8}$ & Area above $A_{\rm V}$=8 according to the $Herschel$ column density map in each observed cloud \\
\hline
\multirow{2}{*}{Cloud radius}  & $R_{\rm mol}^{\rm detect}$ & Equivalent radius of the area where H$^{13}$CO$^+$ or H$^{13}$CN emission has been detected in each cloud.\\
                                 & $R_{\rm Herschel}^{A_{\rm V}>8}$ & Equivalent radius of the area above $A_{\rm V}$=8 in each observed cloud (according to the $Herschel$ column density map). \\
\hline
\multirow{3}{*}{Velocity width}  & $dV_{\rm FWHM}$ & FWHM velocity width at each pixel\\
                         & $dV_{\rm mol}^{\rm detect}$ & FWHM velocity width estimated from the spectrum averaged over the area where H$^{13}$CO$^+$ or H$^{13}$CN emission has been detected.  \\
                         & $dV_{\rm mol}^{A_{\rm V}>8}$ & Scaled FWHM velocity width using $dV_{\rm mol}^{A_{\rm V}>8}$=$dV_{\rm mol}^{\rm detect}$ $\left(\frac{R_{\rm Herschel}^{A_{\rm V}>8}}{R_{\rm mol}^{\rm detect}}\right)^{0.5}$ \\
\hline
\multirow{5}{*}{Mass} & $M_{\rm VIR,mol}^{\rm detect}$           &  Virial mass of the portion of each cloud where H$^{13}$CO$^+$ or H$^{13}$CN emission has been detected. \\
         & $M_{\rm VIR,mol}^{A_{\rm V}>8}$  &   Virial mass of the region above $A_{\rm V}$=8 in each observed cloud.   \\
         & $M_{\rm Herschel}^{\rm mol-detect}$  &  Dense gas mass estimated from the Herschel column density map of the area where H$^{13}$CO$^+$ or H$^{13}$CN emission has been detected.  \\
         & $M_{\rm Herschel}^{A_{\rm V}>8}$ &  Dense gas mass estimated from the Herschel column density map for the area above $A_{\rm V}$=8 in each observed cloud.   \\
         & $M_{\rm dense,mol}$  & Dense gas mass estimated from molecular luminosity conversion factor ($M_{\rm dense,mol}$=$\alpha_{\rm Herschel-mol}^{\rm fit}$$L_{\rm mol}$). \\
\hline
\multirow{2}{*}{Virial ratio} & $\mathcal{R}_{\rm VIR,mol}^{\rm detect,}$           &  Virial mass ratio of the portion of the cloud where the H$^{13}$CO$^+$ or H$^{13}$CN emission has been detected $(\mathcal{R}_{\rm VIR,mol}^{\rm detect}={M_{\rm VIR,mol}^{\rm detect}}/{M_{\rm Herschel}^{\rm mol-detect}})$. \\
                                         & $\mathcal{R}_{\rm VIR,mol}^{A_{\rm V}>8}$    &  Virial mass ratio of the region above $A_{\rm V}$=8 in each observed cloud $(\mathcal{R}_{\rm VIR,mol}^{A_{\rm V}>8}={M_{\rm VIR,mol}^{A_{\rm V}>8}}/{M_{\rm Herschel}^{A_{\rm V}>8}})$.  \\
\hline
\multirow{2}{*}{Star formation rate}  &  SFR$_{\rm YSO}$       &  Star formation rate estimated from the number count of YSOs. \\
                              &  SFR$_{\rm prestellar}$   & Star formation rate estimated from the number count of prestellar cores.  \\
\hline
\multirow{7}{*}{HCN conversion factor}  & $\alpha_{\rm HCN}$  & Conversion factor from HCN(1--0) luminosity to dense gas mass. \\
& $\alpha_{\rm Herschel-HCN}$  & Empirical $\alpha_{\rm HCN}$  factor derived for the target nearby clouds using {\it Herschel} mass estimates as references ($\alpha_{\rm Herschel-HCN}$=$M_{\rm Herschel}^{A_{\rm V}>8}$/$L_{\rm HCN}$). \\
                             & $\alpha_{\rm Herschel-HCN}^{\rm fit}$  & $\alpha_{\rm HCN}$ conversion factor obtained from the relation $\alpha_{\rm Herschel-HCN}^{\rm fit}$=496$\times$ $G_{\rm 0}^{-0.24}$.   \\
                             & $\alpha_{\rm Herschel-HCN}^{\rm fit ^\prime}$  & $\alpha_{\rm HCN}$ conversion factor obtained from the relation $\alpha_{\rm Herschel-HCN}^{\rm fit^{\prime}}$=0.13$\times G_{\rm 0}^{-0.095} \times \alpha_{\rm Herschel-HCN}^{\rm fit}$.   \\
                             & $\alpha_{\rm GS04-HCN}$  & $\alpha_{\rm HCN}$ conversion factor assumed in \citet{Gao04b}.  \\
                             & $\alpha_{\rm Wu05-HCN}$  & $\alpha_{\rm HCN}$ conversion factor obtained in \citet{Wu05}.  \\
                             & $\alpha_{\rm GB12-HCN}$  & $\alpha_{\rm HCN}$ conversion factor obtained in \citet{Garcia12}.  \\
                             & $\alpha_{\rm Usero-HCN}$  & $\alpha_{\rm HCN}$ conversion factor obtained in \citet{Usero15}.  \\
\hline
\multirow{2}{*}{${\rm HCO^+}$ conversion factor}                             & $\alpha_{\rm Herschel-HCO^+}$  & $\alpha_{\rm HCO^+}$ conversion factor obtained toward each observed sub region in the present study ($\alpha_{\rm Herschel-HCO^+}$=$M_{\rm Herschel}^{A_{\rm V}>8}$/$L_{\rm HCO^+}$). \\
                             & $\alpha_{\rm Herschel-HCO^+}^{\rm fit}$  & $\alpha_{\rm HCO^+}$ conversion factor obtained by the relation $\alpha_{\rm Herschel-HCO^+}^{\rm fit}$=689$\times$ $G_{\rm 0}^{-0.24}$.   \\
\hline
\end{tabular}
\end{threeparttable}
}
\end{table*}

\end{document}